\documentclass[10pt]{article}
    \usepackage{amsmath,amssymb,bm,epsf,epsfig,graphicx}
    \input epsf.sty
    \usepackage[dvipsnames]{xcolor}
    \usepackage[utf8]{inputenc}
    \usepackage[english]{babel}
    \usepackage{amsmath}
    \usepackage{latexsym}
    \usepackage{amsfonts}
    \usepackage{mathtools}
    \usepackage{amssymb}
    \usepackage{scalerel}
    \usepackage{extpfeil}
    \usepackage[left=3cm,right=3cm,top=3.1cm,bottom=3.1cm]{geometry}
    \usepackage{tensor}
    \usepackage{tikz}
    \usepackage{tikz-cd}
    \usepackage{tikz-feynman}
    \usepackage{adjustbox}
    \usepackage{multirow}
    \tikzfeynmanset{compat=1.1.0}
    
    \usepackage[mathscr]{euscript}
    \usetikzlibrary{arrows.meta, bending}
    \tikzset{C/.style={circle, minimum size=8mm,
    		node contents={},
    		append after command={\pgfextra{%
    				\draw[-{Straight Barb[flex']}](\tikzlastnode.150) arc (450:110:2.8mm);}
    	}}
    }

    \usepackage{physics}
    \usepackage{bbold}
    \usepackage[pdfencoding=auto, psdextra]{hyperref}
    \usepackage{ dsfont }
    \usepackage{stackengine}
    \usepackage{ mathrsfs }
    \usepackage{capt-of}
    \usepackage{amsthm}
    \usepackage{float}
    \usepackage{titlesec}

    \usepackage{cite}
    \usepackage{csquotes}
  
    \hypersetup{
        linktocpage,
        colorlinks  = true, %Colours links instead of ugly boxes
        urlcolor    = blue, %Colour for external hyperlinks
        linkcolor   = blue, %Colour of internal links
        citecolor   = blue %Colour of citations
    }

    \titleformat{\paragraph}
    {\normalfont\normalsize\bfseries}{\theparagraph}{1em}{}
    \titlespacing*{\paragraph}
    {0pt}{3.25ex plus 1ex minus .2ex}{1.5ex plus .2ex}

    \numberwithin{equation}{section}

    \def \nn {{\mathbb N}}
    \def \ss {{\mathbb S}}
    \def \zz {{\mathbb Z}}
    \def \cc {{\mathbb C}}
    \def \rr {{\mathbb R}}
    
    \def \kk {{\mathbb K}}

    \stackMath
    \newcommand\tsup[2][2]{%
        \def\useanchorwidth{T}%
        \ifnum#1>1%
        \stackon[-1.3ex]{\tsup[\numexpr#1-1\relax]{#2}}{\mathchar"307E}%
        \else%
        \stackon[-1ex]{#2}{\mathchar"307E}%
        \fi%
    }

    \newcommand{\tp}{{\otimes}} %tensor product
    \newcommand{\tpp}{{\otimes'}} %tensor product prime
    \newcommand{\tpb}{{\Bar{\otimes}}} %barred tensor product
    \newcommand{\tpt}{{\Tilde{\otimes}}} %tilde tensor product
    \newcommand{\wpt}{{\Tilde{\wedge}}} %tilde tensor product
    \newcommand{\Hs}{{\mathcal{H}}} %Hilbert space H

    \newcommand{\tH}{{\mathcal{TH}}} %Tensor algebra on Hilbert
    \newcommand{\sH}{{\mathcal{SH}}} %Symmetrized Tensor algebra on Hilbert
    \newcommand{\cH}{{\mathcal{CH}}} %Cyclicized Tensor algebra on Hilbert
    \newcommand{\tHt}{{\mathcal{T}\tilde{\mathcal{H}}}} % tilded tensor space 
    \newcommand{\tHb}{{\mathcal{T}\bar{\mathcal{H}}}} % barred tensor space 
    \newcommand{\tHo}{{\tH_{\rm o}}} % Open strings tensor space 
    \newcommand{\tHc}{{\tH_{\rm c}}} % Closed strings tensor space 
    \newcommand{\Hst}{{\tilde{\mathcal{H}}}} % tilded Hilbert space
    \newcommand{\ttHt}{{\mathcal{T}\mathcal{T}\tilde{\mathcal{H}}}} % tilded tensor tensor space 
    \newcommand{\tHoc}{{\mathcal{T}\mathcal{H}_{\rm c}\tpt_{\rm oc}\mathcal{T}\mathcal{T}\mathcal{H}}_{\rm o}} % open closed tensor space
    \newcommand{\ttHoc}{{\mathcal{S}\mathcal{H}_{\rm c}\tpt_{\rm oc}\mathcal{S}\mathcal{C}\mathcal{H}}_{\rm o}} % open closed tensor space 
    \newcommand{\id}{{\mathbb{1}}} %Tensor identity
    \newcommand{\tid}{{\tilde{\mathbb{1}}}} %Tensor tilded identity
    \newcommand{\octah}{{\tH_{\rm c}\tpt\tH_{\rm o}}} %Open-closed tensor algebra
    \newcommand{\gl}{{\mathcal{G}}} %group like element
    \newcommand{\bgl}{{\bar{\mathcal{G}}}} % barred group like element
    \newcommand{\tgl}{{\tilde{\mathcal{G}}}} % tilded group like element
    \newcommand{\bDelta}{{\bar{\Delta}}} %barred delta
    \newcommand{\tDelta}{{\tilde{\Delta}}} %tilded delta
    \newcommand{\bnabla}{{\bar{\nabla}}} %barred nabla
    \newcommand\nabladot{%
        \mathrel{\ooalign{\hss$\nabla$\hss\cr%
        \kern0.8ex\raise0.5ex\hbox{\scalebox{0.7}{$\cdot$}}}}} % central dotted nabla

    \DeclareMathOperator{\Hom}{Hom}% Homomorphism
    \DeclareMathOperator{\Coder}{Coder}% Coderivation
    \newcommand{\HCoder}{{\mathfrak{coder}}}

    \theoremstyle{definition}
    
    \newtheorem{theorem}{Theorem}

\begin{document}
    
    \begingroup\allowdisplaybreaks
    \hspace*{\fill} UUITP--31/25

    \vspace*{1.1cm}
    
    \centerline{\Large \bf  Co-algebraic methods for }
    \vspace {.3cm}
    
    \centerline{\Large \bf String Field Theory and Quantum Field Theory  } 
    
    \vspace{.3cm}

    \begin{center}

    {\large Enrico Perron Cabus $^{(a,b)}$\footnote{Email: enrico.percabus@physics.uu.se}
    %, name 2$^{(a)}$\footnote{Email: name 2}  and  name 1$^{(a)}$\footnote{Email: name 1} }
    \vskip 1 cm
    $^{(a)}${\it Department of Physics and Astronomy, Uppsala University, Box 516, 75137 Uppsala, Sweden}
    \vskip 1 cm
    $^{(b)}${\it Dipartimento di Fisica, Università di Torino and
    INFN Sezione di Torino Via Pietro Giuria 1, I-10125 Torino, Italy}
    %\vskip .5 cm
    %\vskip .5 cm
    %$^{(b)}${\it institution 3 \\
    %	address
        }
    %\vskip .5 cm
    %$^{(c)}${\it Institute of Physics of the AS CR,\\   
    %Na Slovance 2,  Prague 8, Czech~Republic}
    
    \end{center}
    
    \vspace*{6.0ex}
    
    \centerline{\bf Abstract}
    \bigskip
    
    In this work we extend the notion of co-algebra, co-algebraic Wess-Zumino-Witten formulation of Lagrangian Field Theory and the Homotopy transfer theorem to many strings and particle systems. We discuss in detail the construction of higher dimensional co-algebras and the computational methods derived from them with a special interest regarding String Field Theory and Quantum Field Theory. As a result of this work we will be able to effortlessly extend some of the newly developed tools to study the algebraic structure, compute effective actions and compute scattering amplitudes of more complicated QFTs.

        \baselineskip=16pt
    \newpage
    \tableofcontents

    %%%%%%%%%%%%%%%%%%%%%%%%%
    % Beginning of the text %
    %%%%%%%%%%%%%%%%%%%%%%%%%

\section{Introduction and summary}\label{sec:1}

In the last two decades, many new tools based on homotopy algebras and co-algebras have been developed to facilitate the study of increasingly more complex classes of Quantum Field Theories (QFT)~\cite{gomis1995antibracket}.
\\
Homotopy algebras naturally enter in the interaction structure of all types of bosonic String Field Theories (SFT) 
\cite{zwiebach1993closed,Zwiebach_1998,Kajiura:2004xu,Kajiura:2005sn,Kajiura:2006mt,Munster:2011ij,Maccaferri:2022yzy,Maccaferri:2023gcg, Kunitomo:2022qqp}.
The homotopy algebraic structure of the interactions ensures that the action satisfies the Batalin-Vilkovisky equation \cite{gomis1995antibracket}, securing the existence of a space-time BRST charge at the classical level and an anomaly-free path-integral measure at the quantum level.
\\
While the language of homotopy algebra extends the possible types of QFTs studied \cite{gomis1995antibracket}, generalizing the notion of algebra, it also introduces a notable increase in notational and computational complexity. The use of co-algebras helps to tame the increase in complexity, introducing a $1:1$ map from the elements of the homotopy algebra into linear operators \cite{Erler_2014}. Thanks to the use of co-algebras, it is possible to reduce the entire interacting structure of a QFT to a single nilpotent linear operator acting on the Fock space, at least in absence of multi trace/non planar operators (see subsection \ref{N-DEGEN:SEC}).
\\
A valuable tool derived from the joint use of homotopy algebras and co-algebras is the homotopy transfer theorem. It allows us to integrate out fields in a QFT, providing the interacting structure of the full effective field theory in the process \cite{Maccaferri:2023gof,Maccaferri:2022yzy,Maccaferri:2024puc,koyama2020gaugeinvariantoperatorsopenbosonic}. The theorem also provides a way to compute amplitudes \cite{Macrelli:2019afx,Lopez-Arcos:2019hvg,Doubek:2017naz,Jurco:2019yfd,Jurco:2020yyu,Okawa:2022sjf,Konosu:2023pal,Konosu:2023rkm} and, in some specific cases, the result incorporates non-perturbative contributions \cite{Konosu:2024zrq}.
\\\\
The goal of this is paper is to provide the systematic generalization of the aforementioned tools and relate together different formulations of such tools. Our Results are three-fold:
\begin{itemize}
\item[I] We provide the explicit construction of co-algebras and, of specific interest, the construction of co-derivations on Fock spaces involving a finite and infinite number of particles/string types and boundaries on world-sheet topologies, i.e. including multi trace operators.\\
This allows for the formal extension of the notion of homotopy algebra and the homotopy transfer theorem, agreeing with the results already present in the literature \cite{Kajiura:2004xu,Kajiura:2005sn,Kajiura:2006mt,hoefel2009coalgebra,Erbin_2020,Maccaferri:2023gcg}.

\item[II] We prove that the co-derivation like objects from \cite{Maccaferri:2023gcg} are indeed fully fledged co-derivations. This enables the use of the homotopy transfer theorem without worrying about consistency issues in the context of bosonic oriented quantum open-closed SFT.
\\
This paper also provides the formal relations that link the proper definition of co-derivation to the more commonly used co-derivation like operators first introduced in \cite{Maccaferri:2023gcg, Kunitomo:2022qqp} and the definition given in \cite{hoefel2009coalgebra} for the specific case of the Open-Closed Homotopy Algebra SFT (OCHA) \cite{Kajiura:2004xu,Kajiura:2005sn,Kajiura:2006mt}.

\item[III] We generalize the method to compute amplitudes described in \cite{Okawa:2022sjf,Konosu:2023pal,Konosu:2023rkm,koyama2020gaugeinvariantoperatorsopenbosonic} to account for scalar QFTs with many different scalar fields. 
\end{itemize}
%\\\\
% The main result in this paper is the construction of co-algebras and, of specific interest, the construction of co-derivations on Fock spaces having a finite and infinite number of particles/string types and boundaries on world-sheet topologies. This allows for the formal extension of the notion of homotopy algebra and the homotopy transfer theorem, agreeing with the results already present in the literature \cite{Kajiura:2004xu,Kajiura:2005sn,Kajiura:2006mt,hoefel2009coalgebra,Erbin_2020,Maccaferri:2023gcg}.
%\\
% This paper also provides the formal relations that link the proper definition of co-derivation to the more commonly used co-derivation like operators first introduced in \cite{Maccaferri:2023gcg, Kunitomo:2022qqp} and the definition given in \cite{hoefel2009coalgebra} for the specific case of the Open-Closed Homotopy Algebra SFT (OCHA) \cite{Kajiura:2004xu,Kajiura:2005sn,Kajiura:2006mt}.
%\\\\
In order to simplify most of the computations present in this paper we provide an axiomatic definition of any Lagrangian Field Theory of particles and/or strings using only co-algebraic and homotopy algebraic ingredients regardless of any specific assumptions on the theory. We will refer to this axiomatic definition as Co-Algebraic Field Theory (CAFT). Assumptions on the Field Theory like the number of space-time dimensions or the spectrum can either be derived from the CAFT or used to link the specific CAFT to its specific Field Theory.
\\
The CAFT formulation provides a variety of shortcuts to otherwise time-consuming algebraic computations involving variations of the action of Lagrangian Field Theories. It also naturally reproduces the dual description of interaction vertices from the point of view of the open and closed string, i.e. open-closed channel duality, in the context of the open-closed sphere-disk SFT (SDHA) \cite{Maccaferri:2022yzy} and oriented bosonic quantum open-closed SFT \cite{Maccaferri:2023gcg}.
% As a consequence of the main result of this paper, it is possible to provide an axiomatic definition of any Lagrangian Field Theory using only co-algebraic and homotopy algebraic ingredients regardless of any specific assumptions on the theory. We will refer to this axiomatic definition as Co-Algebraic Field Theory (CAFT). Assumptions on the Field Theory like the number of space-time dimensions, locality of the interaction or the spectrum can either be derived from the CAFT or used to link the specific CAFT to its specific Field Theory.
% \\
% The CAFT formulation of the classical truncation of the open-closed sphere and disk SFT (SDHA) \cite{Maccaferri:2022yzy} naturally reproduces the dual description of interaction vertices from the point of view of the open string and closed string.
% \\\\
% The main result of the paper also provides a shortcut to many time-consuming algebraic computations involving variations of the action  of Lagrangian Field Theories.
% \\\\
% Lastly, the main result allows for the extension of the amplitude computation methods pioneered in \cite{Okawa:2022sjf,Konosu:2023pal,Konosu:2024zrq} to QFTs with a generic number of types of particles, including an infinite amount of particle species.
\\\\
In the section \ref{SEC:math} of this paper we provide a pedagogical introduction to the relevant mathematical structures we will be working with. In section \ref{SEC:APPQFTSFT} we show how the aforementioned structures are used in the study of SFTs and QFTs.\\\\
In section \ref{sec:caft} we introduce the concept of CAFT and review the computational benefits provided by the CAFT formulation of QFT and SFT .\\\\
In sections \ref{NCOALG:CH} and \ref{INFCOALG:CH} we extend the notion of co-algebra to finite and infinite dimensional tensor product spaces of co-algebras. We explicitly build co-derivations on these particular co-algebras and thoroughly explore the notion of cyclicity and the homotopy transfer theorem.\\\\
In sections \ref{SDHA:SEC} and \ref{QOCSFT}, we demonstrate how the CAFT formulation of open-closed SFT correctly reproduces results in the known literature \cite{Kajiura:2004xu,Kajiura:2005sn,Kajiura:2006mt,Maccaferri:2022yzy,Maccaferri:2023gcg} and simplifies the computations involved in the BV formulation of the theories.
\\\\
Lastly we show in section \ref{SEC:NSCATAMP} how the methods of computing correlators \cite{Okawa:2022sjf,Konosu:2023pal,Konosu:2023rkm} can be effortlessly extended to QFTs with more than one distinct particle family.

\section{Mathematical Preliminaries}\label{SEC:math}
In this section we provide a brief and self-contained introduction to the co-algebras, homotopy algebras and the homotopy transfer theorem 

\subsection{Co-algebras}\label{Sec_Co_algebras}
To introduce co-algebras it is necessary to start with understanding the Fock space of any QFT/SFT as a tensor product space $\tH$ spanning over the base Hilbert space $\Hs$ of the specific theory.
\\
Let $\Hs$ be a graded vector space over the field $\rr$ or $\cc$, $\tp$ an associative tensor product  and its identity $\boldsymbol{1}$ , then tensor product space $\tH$ is defined as 
    \begin{align}
        \tH := \bigoplus_{n=0}^{\infty}\Hs ^{\tp n},
    \end{align}
with $\Hs^{\tp 0}$ and the identity defined as
    \begin{align}
        \Hs ^{\tp 0}:=\boldsymbol{1},\,\,\,\,\,\boldsymbol{1}\tp \Hs =\Hs \tp \boldsymbol{1} = \Hs.
    \end{align}
Notice that the pair $\qty(\tH,\tp)$ forms an associative algebra 
    \begin{align}
        \tp:\tH\cross\tH\longrightarrow \tH.
    \end{align}

Some useful maps to define are the set of projectors acting on the tensor algebra
    \begin{align}
        \pi_n:\tH\longrightarrow \Hs ^{\tp n},\,\,\,\,\,\pi_n:=\pi_{n-i}\tp\pi_i,
    \end{align}
and the set of inclusion maps acting on a specific subspace of $\tH$
    \begin{align}
        \iota_n:\Hs\longrightarrow \Hs ^{\tp n+1},\,\,\,\,\,\iota_n A :=\sum_{j=0}^{n} \id^{\tp j}\tp A \tp \id^{\tp n-j}\, \forall A\in\Hs,
    \end{align}
where $\id$ is the identity map $\id:\Hs\longrightarrow\Hs$. The total inclusion map is defined as
    \begin{align}
        \iota:\Hs\longrightarrow\tH,\,\,\,\,\,\,\iota:=\sum_{n=0}^{\infty}\iota_n.
    \end{align}
    
We introduce the following notation which will considerably simplify most of the equations found in this paper, where elements of the tensor product space $\Hs^{\tp j-i+1}$ will be written in the following way:
    \begin{equation}\label{c_not}
        \begin{gathered}
            v_{i,j}=
            \begin{cases}
                \boldsymbol{1}\,\, \text{for}\,\, j=i-1\\
                v_i\tp v_{i+1}\tp...\tp v_j\,\, \text{for}\,\, 0\leq i\leq j\\
                \boldsymbol{1}\,\, \text{for}\,\, i=j+1\\
                0\,\, \text{else}
            \end{cases}
        \end{gathered}
    \end{equation}  
The tensor co-algebra over $\Hs$ is defined by the triple $\qty(\Hs,\tp,\Delta)$, where $\Delta$ is the co-product
    \begin{align}\label{co-prod-app}
        \Delta:\tH\longrightarrow \tH\tpp\tH,\,\,\,\,\,\,\,\,\Delta v_{1,n}:=\sum_{i=0}^{n}v_{1,i}\tpp v_{i+1,n}\,\,\,\,\,\forall\, v_{1,n}\in\tH,
    \end{align}
where the tensor product $\tpp$ is called external tensor product and a priori $\tp\neq\tpp$.\\
In the case of tensor co-algebras the co-product is said to be co-associative
    \begin{align}\label{COASS}
        \qty(\Delta\tpp\id)\Delta = \qty(\id\tpp\Delta)\Delta.
    \end{align}
It is then possible to define an object called concatenation product $\nabla$ that merges the split introduced by $\Delta$
    \begin{align}\label{conc_prod}
        \nabla:\tH\tp'\tH\longrightarrow\tH,\,\,\,\,\,\nabla(v_{1,i}\tp'v_{i+1,n}):=v_{1,i}\tp v_{i+1,n}=v_{1,n}.
    \end{align}
     This map basically turns $\tp'$ into $\tp$ and $\nabla$ is associative, satisfying
     \begin{align}
         \nabla(\nabla\tp'\id)=\nabla(\id\tp'\nabla).
     \end{align}
The concatenation product will be crucial when extending the definition of co-algebras to more complicated Fock spaces because it will provide a necessary ingredient to define the tensor algebra on those spaces.
\\
A special element in the co-algebra is what is called the group-like element $\gl$, which is a degree zero element of $\tH$ such that
    \begin{align}
        \Delta\gl := \gl\tpp\gl,
    \end{align}
    this can be built by choosing a degree zero element $\Psi\in\Hs$ in the following way
    \begin{align}
        \gl:=\sum_{n=0}^{\infty}\Psi^{\tp n} = \frac{1}{1-\tp\Psi}.
    \end{align}
Physically the group-like element is linked to the field present in the action functional of the theory and it will be used to provide a more compact formulation of the action \ref{WZWCO}.

\subsection{Co-derivations}
Co-derivations are linear operators $\boldsymbol{d}$ on $\tH$ that satisfy the co-algebraic equivalent of the Leibniz rules, the co-Leibniz rules
    \begin{align}\label{co-leib}
        \Delta\boldsymbol{d}=\qty(\boldsymbol{d}\tpp\id+\id\tpp\boldsymbol{d})\Delta.
    \end{align}
Co-derivations allow for the compact formulation of Lagrangian action functionals of QFTs/SFTs \cite{Erler_2014,vovsmera2020selected,Maccaferri:2023gcg} and use of the homotopy transfer theorem \cite{Okawa:2022sjf,Konosu:2023pal,Konosu:2024zrq}.\\
An alternative definition of the co-derivation, which is more prone to generalizations, is given by its action on the group-like element $\gl$
    \begin{align}\label{co-gl}
        \boldsymbol{d}\gl=\gl\tp(\pi_1\boldsymbol{d}\gl)\tp\gl.
    \end{align}
By applying the co-product on both sides of \eqref{co-gl} we recover \eqref{co-leib}. This second definition will help us to correctly define co-derivations for more complicated co-algebras later on in the paper.\\
The kinetic and interacting structure of QFTs/SFTs can be described by graded multilinear products $c_k$ acting on subspaces of $\tH$
    \begin{align}
        c_k:\Hs^{\tp k}\longrightarrow\Hs,\,\,\,\,\,,c_k\in\Hom(\Hs^{\tp k},\Hs):=\Hom_k,\,\,\,\,\,\,\forall k\in\nn,
    \end{align}
where $\Hom(\Hs^{\tp k},\Hs)$ is the space of multilinear products from $\Hs^{\tp k}$ to $\Hs$.\\
The key observations that makes the introduction of co-algebras useful is that it is possible to uniquely define a co-derivation $\boldsymbol{c}_k$ for each multilinear product $c_k$ in the following way
    \begin{align}\label{Hom_Coder}
        \boldsymbol{c}_k\pi_n=\sum_{i=0}^{n-k}\id^{\tp i}\tp c_k\tp \id^{\tp n-k-i} = \iota_{n-k+1}c_k\pi_n\,\,\,\,\,,c_k:=\pi_1\boldsymbol{c}_k,\,\,\,\,\,\,\boldsymbol{c}_k:\tH\rightarrow\tH,
    \end{align}
turning the kinetic and interacting structure of the given QFT/SFT into a linear operator acting on the Fock space $\tH$.\\
Note that the projector and inclusion maps define a 1:1 map between the spaces of multilinear products and co-derivations
    \begin{equation}\label{comm-diag1}
    \begin{tikzcd}
        \Coder_k(\tH) \arrow[r, bend left= 50,"\pi_1"] & \Hom_k(\tH) \arrow[l, bend left= 50,"\iota"]
    \end{tikzcd}
    \end{equation}
\subsection{Co-homomorphisms and cyclicity}\label{COHOM}
    There is a class of important objects called co-homomorphisms associated to mapping between theories, symmetries, field redefinitions and change of background.\\
    Let $(\tH_1,\Delta_1)$ and $(\tH_2,\Delta_2)$ be two different co-algebras, with co-products $\Delta_1$ and $\Delta_2$ obeying \eqref{co-prod-app}\footnote{Although it is not necessary that both co-algebras are tensor co-algebra.}. Co-homomorphisms are maps $\boldsymbol{F}$ from a co-algebra $(\tH_1,\Delta_1)$ to a potentially different co-algebra $(\tH_2,\Delta_2)$ which satisfy
    \begin{align}\label{co-hom}
        \Delta_2\boldsymbol{F}=(\boldsymbol{F}\tpp\boldsymbol{F})\Delta_1
    \end{align}
    If the co-homomorphism is graded zero then it maps group like elements into other group like elements. A special class of co-homomorphisms maps the co-algebra in itself and is associated to field transformations.\
    Co-homomorphisms are called invertible if there exists the inverse co-homomorphisms $\boldsymbol{F}^{-1}$ such that
    \begin{align}
        \id:=\boldsymbol{F}\boldsymbol{F}^{-1}=\boldsymbol{F}^{-1}\boldsymbol{F}.
    \end{align}
    Co-homomorphisms can be obtained via the exponentiation of co-derivations which can be accompanied by graded parameters $\varepsilon_i$
    \begin{align}\label{exp-cohom}
        \boldsymbol{F}_\varepsilon=\sum_{k=0}^{\infty}\frac{1}{k!}(\varepsilon_i\boldsymbol{d}^i)^{k}=e^{\varepsilon_i\boldsymbol{d}^i}.
    \end{align}
    By discarding higher orders in $\varepsilon_i$ the co-homomorphisms are associated to infinitesimal transformations.\\
    If the original vector space $\Hs$ is endowed with a symplectic form $\omega:\Hs\tp\Hs\rightarrow\cc$, which can be represented as
    \begin{align}\label{OMREP}
        v_1,v_2\in\Hs\,\,\,\omega(v_1,v_2)=\bra{w}\,\ket{v_1}\tp\ket{v_2}=\bra{w}\,\ket{v_1\tp v_2},
    \end{align}
    a cyclic co-homomorphisms (or co-symplectomorphisms) can be defined by requiring
    \begin{align}\label{cohom-cycl}
        \bra{\omega}\pi_2\boldsymbol{F}=\bra{\omega}\pi_2,\,\,\,\text{equivalently}\,\,\,\,\bra{\omega}\pi_1\boldsymbol{F}\tp\pi_1\boldsymbol{F}=\bra{\omega}\pi_1\tp\pi_1,
    \end{align}
    If the co-homomorphism is the exponentiation of a co-derivation, then at first order in $\varepsilon$ it will reproduce the usual definition of the cyclicity for co-derivations and multilinear-products \cite{Gaberdiel:1997ia,vovsmera2020selected}
      \begin{align}
        \bra{\omega}(\pi_1e^{\varepsilon \boldsymbol{d}})\tp(\pi_1e^{\varepsilon \boldsymbol{d}})=\bra{\omega}(\pi_1\tp\pi_1)\,\Longrightarrow\,
        \bra{\omega}(\pi_1\tp\pi_1 \boldsymbol{d})=-\bra{\omega}(\pi_1\boldsymbol{d}\tp\pi_1 )+\mathcal{O}(\varepsilon).
    \end{align}
    To get definition of a cyclic co-derivation in the most general setting we need to add two additional exponentiated co-derivations $(\boldsymbol{a},\boldsymbol{b})$
    \begin{align}\label{co-der-cycl}
        \bra{\omega}(\pi_1e^{\varepsilon \boldsymbol{d}})\tp(\pi_1e^{\varepsilon \boldsymbol{d}})(e^{\delta_1 \boldsymbol{a}}\tp e^{\delta_2 \boldsymbol{b}})=\bra{\omega}(\pi_1\tp\pi_1)(e^{\delta_1 \boldsymbol{a}}\tp e^{\delta_2 \boldsymbol{b}}),
    \end{align}
    and by unpacking order by order we get
    \begin{equation}\label{funccycl}
        \begin{split}
            &\mathcal{O}((\delta_1)^0,(\delta_2)^0)\Longrightarrow\,\,\,\bra{\omega}\pi_1\boldsymbol{F}_\varepsilon\tp\pi_1\boldsymbol{F}_\varepsilon=\bra{\omega}\pi_1\tp\pi_1\,\,\,\text{cyclicity of}\,\,\,\,\boldsymbol{F}_\varepsilon=e^{\varepsilon_i\boldsymbol{d}^i},\\
            &\mathcal{O}(\varepsilon^1,(\delta_1)^1,(\delta_2)^1)\Longrightarrow\,\,\,\omega(\pi_1\boldsymbol{d}\boldsymbol{a}\mathcal{G},\pi_1\boldsymbol{b}\mathcal{G})=-(-1)^{d(\boldsymbol{d})d(\boldsymbol{a})}\omega(\pi_1\boldsymbol{a}\mathcal{G},\pi_1\boldsymbol{d}\boldsymbol{b}\mathcal{G}),
        \end{split}
    \end{equation}
    where $d(\boldsymbol{x})$ is the grading of $\boldsymbol{x}$ and $\omega(\cdot,\cdot)$ is an alternative representation of the symplectic form $\bra{\omega}$.\\
    From the second equation in \eqref{funccycl} when specializing both $(\boldsymbol{a},\boldsymbol{b})$ to the identity co-derivation we fully recover the usual notion of cyclic co-derivation 
    \begin{align}
        (\boldsymbol{a},\boldsymbol{b})\rightarrow(\id,\id)\Longrightarrow \omega(\pi_1\boldsymbol{d}\mathcal{G},\pi_1\mathcal{G})=-\omega(\pi_1\mathcal{G},\pi_1\boldsymbol{d}\mathcal{G})
    \end{align}
    As we will later discuss in detail, the notion of cyclicity in more complicated tensor algebras will enforce dualities regarding the description of interaction vertices from the point of view of different particles/string taking part in the interaction. In bosonic open-closed SFT the duality enforced by cyclicity has been referred as open-closed channel duality \cite{Erbin_2020,Maccaferri:2023gcg}.

 \subsection{Types of tensor algebras}\label{co-alge-types}
    In the previous part we introduced the core informations about co-algebras. In the study of many QFTs and SFTs we will be working with symmetrized and cyclicized tensor algebras \cite{Maccaferri:2023gcg,vovsmera2020selected}. Because cyclicized and symmetrized tensor algebras are sub-algebras, we will only need to study the algebraic properties of our physical system in the normal tensor algebra. To relate the results discussed in this paper to the literature, we will introduce some operators that will allow us to project onto the relevant symmetrized and cyclicized tensor sub-algebras and co-algebras\\
    Let's introduce the symmetrization operator $\sigma$
    \begin{align}
        \sigma_{k}:\Hs^{\tp k}\longrightarrow\Hs^{\wedge k},
        \,\,\,\,\,v_{(1,k)}:=\sigma_k v_{1,k}=\sum_{\sigma\in\ss_k}(-1)^{\varepsilon(\sigma)}v_{\sigma(1)}\tp...\tp v_{\sigma(k)}=v_1\wedge...\wedge v_k,
    \end{align}
    where the wedge product $\wedge$ is the symmetrized tensor product and $v_{1,k}\in\Hs^{\tp k}$ and $\varepsilon(\sigma)$ is the Koszul sign of the permutation $\sigma$, which takes into account both the sign of the permutation and the signs relative to the grading of the objects involved.\\
    It is easy now to extend the action of $\sigma_k$ to the entire tensor algebra defining the symmetrized tensor algebra
    \begin{align}
        \sigma :=\sum_{k=1}^{\infty}\sigma_k\pi_k,\,\,\,\,\,\,\sigma\tH:=\sH.
    \end{align}
    Similarly we define the cyclic operator $\tau$
    \begin{align}
        \tau_{k}:\Hs^{\tp k}\longrightarrow\Hs^{\odot k},
        \,\,\,\,\,\tau_k v_{1,k}:=\sum_{\sigma\in\zz_k}(-1)^{\varepsilon(\sigma)}v_{\sigma(1)}\tp...\tp v_{\sigma(k)}:=v_1\odot...\odot v_k:=v_{\qty{1,k}},
    \end{align}
    where the product $\odot$ is the cyclicized tensor product.\\
    It is easy now to extend the action of $\tau_k$ to the entire tensor algebra defining the cyclicized tensor algebra
    \begin{align}
        \tau :=\sum_{k=1}^{\infty}\tau_k\pi_k,\,\,\,\,\,\,\tau\tH:=\cH.
    \end{align}
    All the definitions pertaining to the co-algebra, co-derivation, co-homomorphisms and group like element remain unaltered by the projection onto the sub-algebras $\sH$ and $\cH$. Interestingly enough the group like element on $\tH$ has an alternative expressions in terms of cyclicized and symmetrized tensor products
    \begin{align}\label{gl-elements}
        \gl = \sum_{n=0}^{\infty}\Psi^{\tp n}=\sum_{n=0}^{\infty}\frac{\Psi^{\wedge n}}{n!} := e^{\wedge\Psi} = \sum_{n=0}^{\infty}\frac{\Psi^{\odot n}}{n+\delta_{n,0}} := 1+\ln\qty(1-\odot \Psi).
    \end{align}
    All the representations of the group like element are equivalent.\\
    It will prove useful later on to explicitly write down the action of a co-derivation on a symmetrized group like element, which is
    \begin{align}
        \boldsymbol{d}e^{\wedge \Psi}=(\pi_1\boldsymbol{d}e^{\wedge \Psi})\wedge e^{\wedge \Psi}.
    \end{align}
    \subsection{Co-derivation algebra and homotopy algebras}\label{hom-alg}
    it is possible to define a product between elements of $\Hom(\tH)$. Given $c_k\in\Hom(\Hs^{\tp k},\Hs)$ and $d_l\in\Hom(\Hs^{\tp l},\Hs)$, the product is defined as 
    \begin{align}\label{FINPROD}
        c_kd_l=\sum_{j=0}^{k-1}c_k\qty(\qty(\id)^{\tp j}\tp d_l\qty(\qty(\id)^{\tp l})\tp \qty(\id)^{\tp k-1-j}),\,\,\,\,\,c_kd_l:\Hs^{\tp k+l-1}\longrightarrow\Hs,\,\,\,\,\,\forall k,l\in\nn.
    \end{align}
    From now on we will refer to the space of $k$ linear products graded $a$ as $\Hom^a_k:=\Hom^{a}(\Hs^{\tp k},\Hs)$. We will also refer to the entire space of graded multilinear products as $\Hom:=\sum_{k,a}\Hom^a_k$.\\
    From the above defined product it is possible to define the graded commutator
    \begin{align}\label{Hom_alg}
        \commutator{c_k}{d_l}:= c_kd_l-(-1)^{d(c_k)d(d_l)}d_lc_k,\,\,\,\,\,\,\commutator{\cdot}{\cdot}:\Hom^c_k\cross \Hom^d_l\longrightarrow\Hom^{c+l}_{k+l-1}.
    \end{align}
    The pair $(\Hom,\commutator{\cdot}{\cdot})$ forms a graded Lie algebra.\\
    The algebra structure can be extended to the space of graded co-derivations $\Coder$ because of the morphism \eqref{Hom_Coder}, therefore the pair $(\Coder,\commutator{\cdot}{\cdot})$ forms a graded Lie algebra with
    \begin{align}\label{Coder_alg}
        \commutator{\boldsymbol{c}_k}{\boldsymbol{d}_l}=\boldsymbol{c}_k\boldsymbol{d}_l-(-1)^{d(\boldsymbol{c}_k)d(\boldsymbol{d}_l)}\boldsymbol{d}_l\boldsymbol{c}_k=2\,\boldsymbol{c}_k\boldsymbol{d}_l,\,\,\,\,\,\,\commutator{\cdot}{\cdot}:\Coder^c_k\cross \Coder^d_l\longrightarrow\Coder^{c+l}_{k+l-1}.
    \end{align}
    The algebra $(\Coder,\commutator{\cdot}{\cdot})$, or equivalently $(\Hom,\commutator{\cdot}{\cdot})$, can become a differential graded Lie algebra if we can define a graded odd co-derivation $\boldsymbol{m}\in\Coder$ such that
    \begin{align}
        \commutator{\boldsymbol{m}}{\boldsymbol{m}}=0\Longleftrightarrow\boldsymbol{m}^{2}=0,
    \end{align}
    where the differential is defined as
    \begin{align}
        \boldsymbol{d}:=\commutator{\boldsymbol{m}}{\cdot}.
    \end{align}
    The condition $\boldsymbol{m}^{2}=0$ is the definition of an $A_\infty$ homotopy algebra, which is usually more appreciated in terms of multilinear products
    \begin{align}\label{co-der_hom_alg}
        \boldsymbol{m}^{2}=0\,\xlongrightarrow[]{\pi_1}\,\sum_{l=1}^{k}\frac{1}{2}\commutator{m_{l}}{m_{l-k}}=0.
    \end{align}
    When working on the subspace $\sH$ the $A_\infty$ algebra restricts to an $L_{\infty}$ algebra\cite{vovsmera2020selected,Erler_2014,zwiebach1993closed}.\\
    Although in bosonic SFT usually the interaction structures follows the full $A_\infty$ structure for the open SFT and the full $L_\infty$ structure for the closed SFT, there is a special instance in bosonic open SFT where, aside from the BRST charge $m_1$, there is only one interaction vertex $m_2$ called Witten star product \cite{witten1986non}. The Witten star product simplifies the $A_\infty$ algebra to a differential graded associative algebra,
    \begin{equation}
        \begin{split}
            &m_1(m_1(v_1))=0,\\
            &m_1(m_2(v_1,v_2))+m_2(m_1(v_1),v_2))+(-1)^{d(v_2)}m_2(v_1,m_1(v_2)))=0,\\
            &m_2(m_2(v_1,v_2),v_3)+(-1)^{d(v_1)}m_2(v_1,m_2(v_2,v_3))=0,
        \end{split}
    \end{equation}
    with $v_1,v_2,v_3\in\Hs$.
    \\
    In conclusion differential graded Lie algebras on the space of co-derivations, or equivalently multilinear products, define an homotopy algebras and vice versa. \\
    Homotopy algebras naturally appear when studying the classical field theories and SFTs as a result from the application of the classical BV master equation \cite{gomis1995antibracket}.
    \subsection{Higher order co-derivations}
     The algebraic structures of QFTs and quantum SFTs are given by  loop-algebras \cite{zwiebach1991quantum}, which is a generalization of the homotopy $L_\infty$ algebra by the addition of the multilinear map $U\in\Hom^{1}(\Hs^{\tp 0},\Hs^{\tp 2})$. The map $U$ in the co-algebraic formalism can't be mapped to a co-derivation because it doesn't satisfy the co-Leibniz rule \eqref{co-leib}, therefore the second order co-derivations are introduced, which are second order operators acting on $\tH$ and are in a 1:1 correspondence with $\Hom(\Hs^{\tp n},\Hs^{\tp 2})$. In general n-th order co-derivations satisfy the following relation
    \begin{align}
        \sum_{i=0}^{n}\sum_{\sigma\in\ss_n}(-1)^{i}\sigma\circ\qty(\Delta^{n-i+1}\circ \boldsymbol{D}_{;n}\tpp\id^{\tpp i})\circ\Delta^{i+1}=0,
    \end{align}
    where $\Delta^{j}$ is the repeated action of the co-product
    \begin{align}
        \Delta^{j}=(\Delta\tpp\id^{\tpp j})\circ\Delta^{j-1},\,\,\,\,\,\,\,\Delta^1=\Delta,
    \end{align}
    and the operator $\sigma$ permutes the $n$ elements splitted by the the action $\Delta^{n-1}$ according to the permutation group $\ss_n$.
    Lastly $\boldsymbol{D}_{;n}$ is the co-derivation of $n$-th order.
    \\
    For readability purposes we choose to represent the multilinear maps graded $a$, with $k$ inputs and $n$ outputs as $d_{k;n}\in\Hom^a_{k;n}$, we then choose to represent the associated n-th order co-derivation as $\boldsymbol{d}_{k;n}\in\Coder^a_{k;n}$.\\
    The 1:1 correspondence between multilinear maps $\Hom^a_{k;n}$ and $\Coder^a_{k;n}$ is similar to \eqref{Hom_Coder} with one difference in the usage of the projection map
    \begin{align}\label{N_Hom_Coder}
        \boldsymbol{d}_{k;n}\pi_m=\sum_{i=0}^{m-k}\id^{\tp i}\tp d_{k;n}\tp \id^{\tp m-k-i},\,\,\,\,\,d_{k;n}:=\pi_n\boldsymbol{d}_{k;n},\,\,\,\,\,\,\pi_{j<n}\boldsymbol{d}_{k;n}=0,
    \end{align}
    therefore the commutative diagram \eqref{comm-diag1} changes to
    \begin{equation}
    \begin{tikzcd}
        \Coder_{k;n}(\tH) \arrow[r, bend left= 50,"\pi_n"] & \Hom_{k;n}(\tH) \arrow[l, bend left= 50,"\iota"]
    \end{tikzcd}
    \end{equation}
    Furthermore, any $n$-th order co-derivation is contained in the space of $(n+1)$-th order co-derivations 
    \begin{align}
        \Coder_{;1}\subset\Coder_{;2}\subset...\subset\Coder_{;n}\subset\Coder_{;n+1}\subset...
    \end{align}
    \subsection{$IBL_\infty$ algebras}
    We remember that usual co-derivations together with the graded commutator $\commutator{\cdot}{\cdot}$ form an algebra \eqref{Coder_alg}. When adding higher order co-derivations the algebra opens up \cite{Markl_2001}
    \begin{align}
        \commutator{\cdot}{\cdot}:\Coder^c_{;n}\cross \Coder^d_{;m}\longrightarrow\Coder^{c+l}_{;n+m}.
    \end{align}
    To close\footnote{Recall that the product of co-derivations is proportional to the commutator of such co-derivations \eqref{Coder_alg}.} the co-derivation algebra it is necessary to enlarge the co-derivation space by including all higher order co-derivations
    \begin{align}
        \mathfrak{coder}:=\mathfrak{coder}(\tH,x):=\bigoplus_{n=1}^{\infty}x^{n-1}\Coder_{;n},
    \end{align}
    where $x\in\cc$ is some auxiliary parameter. An element $\mathfrak{D}\in\mathfrak{coder}$ can be expanded as
    \begin{align}
        \mathfrak{D}=\sum_{n=1}^{\infty}x^{n-1}\boldsymbol{d}_{;n}.
    \end{align}
    When working with $\HCoder$ the algebra of co-derivations closes again
    \begin{align}
        \commutator{\cdot}{\cdot}:\HCoder\cross\HCoder\longrightarrow\HCoder.
    \end{align}
    Similarly to the $A_{\infty}$ and $L_{\infty}$ algebras, on $\HCoder$ it is possible to search for a graded odd differential operator $\mathfrak{M}$ such that
    \begin{align}
        \frac{1}{2}\commutator{\mathfrak{M}}{\mathfrak{M}}=\qty(\mathfrak{M})^2=0.
    \end{align}
    The operator $\mathfrak{M}$ defines an $IBL_{\infty}$ algebra, which is the generalization of the $A_{\infty}$, $L_{\infty}$ and loop-algebra \cite{Munster:2011ij}.\\
    The $IBL_{\infty}$ contains homotopy algebras, loop-algebras and $IBL$ algebras.\\
    The loop-algebra is defined in terms of the $IBL_{\infty}$ algebra with the following differential
    \begin{align}
        \mathfrak{L}:=\sum_{g=0}^{\infty}x^{g}\boldsymbol{l}^{(g)}+x\boldsymbol{U},\,\,\,\,\boldsymbol{l}^{(g)}\in\Coder_{;1}^1,\,\,\,\,\boldsymbol{U}\in\Coder^1_{0;2}.
    \end{align}
    where $x\in\cc$ is a parameter, $g$ is the order of the expansion in $x$ and $\boldsymbol{U}$ is the second order co-derivation associated to the Poisson bi-vector $U$. The physical interpretation of $x,g,U$ in the case of the closed SFT is explained in \ref{WZWCO}.\\
    %In the context of the bosonic closed SFT $g$ will be attributed to the genus of the world-sheet topologies that defines that particular interaction vertex, while $\boldsymbol{U}$ is the second order co-derivation that will be associated to the Poisson bi-vector $U$ which joins together two closed string punctures together \cite{zwiebach1993closed}.\\
    The nilpotency of $\mathfrak{L}$, when expanded order by order in $x$, tells us that $\boldsymbol{l}^{(0)}$ form an $A_\infty$ or $L_\infty$ algebra and $\boldsymbol{U}$ is a nilpotent second order zero co-derivation
    \begin{align}
        \mathfrak{L}^{2}=0\,\longrightarrow\, (\boldsymbol{l}+x\boldsymbol{U})^2=0\Longrightarrow(\boldsymbol{l}^{(0)})^2=0,\,\,\,\,\,(\boldsymbol{U})^{2}=0.
    \end{align}
    We will call Q-$A_\infty$/Q-$L_\infty$ the quantum deformation of an $A_\infty$/$L_\infty$ algebra.\\
    Note that the nilpotency condition on $\boldsymbol{U}$ is trivially satisfied by it being an element of $\Coder^1_{0;2}$. 
     %
    %\begin{align}
    %    \pi_1\mathfrak{L}^{2}=0\,\longrightarrow\, (\boldsymbol{l}+x\boldsymbol{U})^2=0,
   %\end{align}
   %where $\boldsymbol{U}_{\rm c}$ is second order co-derivations associated to the closed Poisson bi-vector $U_{\rm c}$ and $\boldsymbol{l}$ is its interacting structure together with all quantum corrections
   %\begin{align}
  %     \boldsymbol{l}:=\sum_{g=0}^{\infty}x^{g}\boldsymbol{l}^{(g)}.
  % \end{align}

\subsection{Homotopy transfer theorem}\label{HTT}
If a QFT/SFT on $\Hs$ satisfies the BV master equation \cite{gomis1995antibracket}\eqref{BVC}\eqref{BVQ} then there exists a graded $1$ nilpotent operator $\partial$, i.e. the BRST charge, contained in the $A_\infty/ L_\infty$ interacting structure and its co-homology
\begin{align}
    Q=\partial:\Hs\longrightarrow\Hs,\,\,\,\,\,\,\,\partial^2=0,\,\,\,\,\,\,\,{\rm H}(\partial):=\eval{\frac{{\rm ker}(\partial)}{{\rm im}(\partial)}}_\Hs.
\end{align}
To understand how to extract effective theories or specific physical observables, using the homotopy transfer theorem, we firstly need to discuss what happens to the BRST co-homology when projections are involved.\\
The operation of integrating out fields in the path integral is algebraically equivalent to a projection from the full Hilbert space $\Hs$ to a subspace $\Hs_P$ \cite{Maccaferri:2024puc}
\begin{align}
    P:\Hs\longrightarrow\Hs_P\subseteq\Hs,\,\,\,\,\,\,P^2=P,\,\,\,\,\,\Hs=\Hs_P\oplus\Hs_\perp.
\end{align}
For a projection $\Hs_P$ to be physically relevant we require that $P$ is a \textit{co-chain map}\footnote{Note that $\partial$ forms a co-chain complex on $\Hs$ because $\partial$ is graded $1$. If $\partial$ properly suspended to a $-1$ differential then we are talking about chain complexes like in  \cite{Arvanitakis:2020rrk,Bonezzi:2023xhn}.\\When comparing this paper to \cite{Arvanitakis:2020rrk,Bonezzi:2023xhn} note that $P=\iota p$.} i.e. $P$ commutes with $\partial$
\begin{align}\label{CO-CH}
    \partial P = P \partial,\,\,\,\,\,{\rm Ker}(\partial)\subset {\rm Im}(P).
\end{align}
We can define a new differential $\partial'$ acting only on $\Hs_P$ with his co-homology
\begin{align}\label{cohom-map}
    \partial':=\partial P,\,\,\,\,\,{\rm H}(\partial'):=\eval{\frac{{\rm ker}(\partial')}{{\rm im}(\partial')}}_{\Hs_P}=\eval{\frac{{\rm ker}(\partial P)}{{\rm im}(\partial P)}}_\Hs.
\end{align}
The projector $P$ maps the co-homology of $\partial$ to the co-homology on the $\Hs_P$ restriction, where our effective field theories will live. In order not to lose any co-homological information by the projection $P$ we require that the two co-homologies are isomorphic to one another. The two co-homologies are isomorphic if $P$ is a quasi-isomorphism, i.e. it satisfies the Hodge-Kodaira decomposition 
\begin{align}\label{Hodge-Kodaira}
            \id = P+\alpha\qty{\partial h+h \partial}\,\Longrightarrow\,{\rm H}(\partial)\sim{\rm H}(\partial'),\,\,\,\,\alpha\in\cc,
\end{align}
where $h$ is called contracting homotopy and is a graded $-1$ map
\begin{align}
                h:\Hs\longrightarrow\Hs,
\end{align}
and $\alpha$ is complex number introduced to unify different notations of the decomposition\footnote{The choice of $\alpha=+1$ is notably used in \cite{Erbin_2020,Okawa:2022sjf} while $\alpha=-1$ in \cite{Arvanitakis:2020rrk,Bonezzi:2023xhn}.}.\\
When theories contain an $A_\infty/ L_\infty$ interacting structure instead of working on $\Hs$ and $\Hs_P$ we will work on the associated tensor algebras $\tH$ and $\tH_P$. To consistently extract effective theories or specific physical observables we rely upon the homotopy transfer theorem\cite{Macrelli:2019afx,Lopez-Arcos:2019hvg,Arvanitakis:2020rrk,Bonezzi:2023xhn}\footnote{The homotopy transfer theorem can be also given as a corollary of the homological perturbation lemma \cite{Erbin_2020,Maccaferri:2022yzy,Maccaferri:2024puc}}:
\begin{theorem}\label{HomTR}
    Given $(P,h)$, if $P$ is a quasi isomorphisms and $h$ satisfies the side conditions
    \begin{align}\label{side-cond}
        Ph=hP=h^2=0,
    \end{align}
    then there exists a suitable extension of $(P,h)$ on the tensor algebra $\tH$ such that the $A_\infty /L_\infty$ structure on $\tH$ can be transferred to a $A_\infty /L_\infty$ structure on $\tH_P$.
\end{theorem}
To uplift $(P,h)$ onto $\tH$ we rely on the co-algebraic framework where $P$ is naturally uplifted to a co-homomorphism
\begin{align}
    \boldsymbol{P}(v_{1,n}):= P(v_1)\tp...\tp P(v_n),\,\,\,\,\boldsymbol{P}(v_1)=P(v_1),\,\,\,\,\,v_{i}\in \Hs
\end{align}
while the $h$ has a more complicated extension with many equivalent recursive formulations
\begin{equation}\label{lr-h-def}
    \begin{split}
        &\boldsymbol{h}(v_{1,n}):= \boldsymbol{h}(v_{1,n-1})\tp v_n + \boldsymbol{P}(v_{1,n-1})\tp h(v_n) \\
        &\qquad\text{(or)}\,:= v_1\tp\boldsymbol{h}(v_{2,n}) +  h(v_1)\tp\boldsymbol{P}(v_{2,n}),\\
        &\boldsymbol{h}(v_{1}):=h(v_1),
    \end{split}
\end{equation}
with here reported only the two commonly used formulations\footnote{A detailed derivation of the recursive relations of $\boldsymbol{h}$ is given in \ref{h-action} or \cite{Arvanitakis:2020rrk}}.\\
The correct uplift of $(P,h)$ has the property of uplifting the equivalence relation \eqref{Hodge-Kodaira} onto the full tensor co-algebra $\tH$ 
\begin{align}
     \boldsymbol{\id} = \boldsymbol{P}+\alpha\qty{\boldsymbol{\partial h}+\boldsymbol{h \partial}},\,\,\,\boldsymbol{\partial}:=\iota\partial,\,\,\,\,\,\alpha\in\cc,
\end{align}
with the appropriate side conditions \eqref{side-cond}
\begin{align}\label{Co-side-cond}
    \boldsymbol{Ph}=\boldsymbol{h P}=\boldsymbol{h}^2=0,
\end{align}
where $\boldsymbol{\partial}$ is the co-derivation associated to $\partial$.\\
In the homotopy transfer theorem the $A_\infty/L_\infty$ structure on $\tH$ is identified by the nilpotent co-derivation $\boldsymbol{D}$ \eqref{co-der_hom_alg} where the differential $\partial$ on $\Hs$ is highlighted in the form of co-derivation
\begin{align}\label{D_rel}
    \boldsymbol{D}:= \boldsymbol{\partial}+\boldsymbol{B},\,\,\,\,\,\boldsymbol{D}^2=0\,\Rightarrow\, \boldsymbol{\partial}\boldsymbol{B}+\boldsymbol{B}\boldsymbol{\partial}+\boldsymbol{B}^2=0,
\end{align}
and $\boldsymbol{B}$ physically contains the interacting structure of the QFT/SFT we are studying.
The homotopy transported $A_\infty/L_\infty$ structure onto $\tH_P$ is encoded in the nilpotent co-derivation
\begin{align}
    \boldsymbol{D}':= \boldsymbol{\partial}'+\boldsymbol{B}',\,\,\,\,\,\,\,\,(\boldsymbol{D}')^2=0,
\end{align}
where the interacting structure $\boldsymbol{B}'$ is heavily modified by $\boldsymbol{h}$
\begin{align}\label{Morph0}
\boldsymbol{B}'=\boldsymbol{F}\boldsymbol{B}\boldsymbol{F}'=\boldsymbol{P}(\boldsymbol{B}+\alpha\boldsymbol{B}\boldsymbol{h}\boldsymbol{B}+\alpha^2\boldsymbol{B}\boldsymbol{h}\boldsymbol{B}\boldsymbol{h}\boldsymbol{B}+...)\boldsymbol{P}=\boldsymbol{P}\frac{1}{1-\alpha\boldsymbol{Bh}}\boldsymbol{BP}=\boldsymbol{PB}\frac{1}{1-\alpha\boldsymbol{hB}}\boldsymbol{P}  .
\end{align}
These modifications can be interpreted as the action of a $A_\infty/L_\infty$ morphism $\boldsymbol{F}$ and its right inverse $\boldsymbol{F}'$ which satisfy
\begin{align}\label{HTTMORPH}
  \boldsymbol{F}\boldsymbol{F}'=\id,\,\,\,\,\boldsymbol{F}\boldsymbol{\partial}\boldsymbol{F}'=\boldsymbol{\partial}'.
\end{align}
The choice of the $\boldsymbol{F}$ and $\tilde{\boldsymbol{F}}$ is not unique and the most commonly used are
\begin{align}\label{Morph1}
    \boldsymbol{F}:=\boldsymbol{P}\frac{1}{1-\alpha\boldsymbol{B}\boldsymbol{h}},\,\,\,\,\,\,\boldsymbol{F}'=(1-\alpha\boldsymbol{B}\boldsymbol{h})\boldsymbol{P},
\end{align}
or equivalently
\begin{align}\label{Morph2}
    \boldsymbol{F}:=\boldsymbol{P}(1-\alpha\boldsymbol{h}\boldsymbol{B}
    ),\,\,\,\,\,\,\boldsymbol{F}'=\frac{1}{1-\alpha\boldsymbol{h}\boldsymbol{B}}\boldsymbol{P},
\end{align}
with possible changes in the sign convention depending of the choice of $\alpha=\pm1$ \cite{Macrelli:2019afx,Lopez-Arcos:2019hvg,Doubek:2017naz,Jurco:2019yfd,Jurco:2020yyu,Arvanitakis:2020rrk,Bonezzi:2023xhn,Okawa:2022sjf,Erbin_2020,Maccaferri:2024puc}. A detailed derivation of $\boldsymbol{F}$ and $\boldsymbol{h}$ is present in \ref{Appendix_Hom}.

\section{Applications to QFT/SFT}\label{SEC:APPQFTSFT}
As mentioned before, the language of homotopy algebras and co-algebras together provides many benefits in studying QFTs/SFTs in a cost-effective way. The purpose of this section is to introduce the most relevant homotopy-co-algebraic tools while keeping the details in their respective sections.

\subsection{WZW Co-algebraic formulation of QFTs}\label{WZWCO}
Any Lagrangian QFT/SFT on a field $\Phi$ can formally be expressed using a non degenerate symplectic form $\omega$ together with a set of cyclic graded multilinear products $m_k$ containing the interacting structure of the theory \cite{Hohm:2017pnh}
\begin{align}
    S[\Phi]:=\sum_{k=0}^{\infty}\frac{g_k}{k+1}\,\omega\qty(\Phi,m_k(\Phi^{\tp k})),\,\,\,\,\,\Phi\in\Hs,
\end{align}
with the coupling constant $g_k$ associated to the interaction vertex $m_k$ and symmetry factor $k+1$. The field $\Phi$ can usually it taken as graded $0$ element of $\Hs$ in the following way
\begin{align}
    \Phi:=\sum_{a}\Phi^af_a,
\end{align}
where $f_a$ is a complete orthonormal basis of $\Hs$, $\Phi^a$ are the space-time fields which can be in different spin representations, and can be algebra valued\footnote{For usual QFTs like QED or QCD the algebra is a Lie algebra.}. The sum over $a$ hides the different contractions between spin representation indices, algebra representation indices and integration over momentum or space-time position.\\
This formulation becomes rather bulky when more fields are added to the QFT. By using notions from the co-algebras, it is possible to uplift $m_k$ to co-derivations $\boldsymbol{m}_k$ and $\Phi$ to the Group-like element $\gl$ in order to reduce the action to
\begin{align}
    S[\Phi]:=\sum_{k=0}^{\infty}\frac{g_k}{k+1}\,\omega\qty(\pi_1\gl,\pi_1\boldsymbol{m}_k\gl),\,\,\,\,\,\gl:=\frac{1}{1-\tp \Phi}.
\end{align}
We can get rid of the coupling constant if we redefine
\begin{align}\label{coupl-redef}
    g_k \boldsymbol{m}_k\,\longrightarrow\,\boldsymbol{m}_k,
\end{align}
We can also get rid of the $k+1$ factor if we interpolate the field $\Phi$ with a parameter $t\in\qty[0,1]$ as
\begin{align}
    \Phi\longrightarrow\Phi(t),\,\,\,\,\Phi(0)=0,\,\,\,\,\Phi(1)=\Phi,\,\,\,\,\,\gl\longrightarrow\gl(t),
\end{align}
and by introducing in the action the identity operator written as
\begin{align}
    \id=\int_{0}^{1}\dd{t}\pdv{t},
\end{align}
leading to the Wess-Zumino-Witten (WZW) co-algebraic formulation of the action \cite{vovsmera2020selected}
\begin{align}
    S[\Phi]:=\int_{0}^{1}\dd{t}\omega\qty(\pi_1\boldsymbol{\partial}_t\gl,\pi_1\boldsymbol{m}\gl),
\end{align}
where $\boldsymbol{\partial}_t$ is the co-derivation associated to $\pdv{t}$ and $\boldsymbol{m}$ is the linear operator defined as
\begin{align}
    \boldsymbol{m}:=\sum_{k=0}^{\infty}\boldsymbol{m}_k,
\end{align}
and $\gl$ is actually $\gl(t)$ to keep the notation contained.\\
In the WZW-co-algebraic formulation the equation of motions can be conveniently repackaged from
\begin{align}
    \sum_{k=0}^{\infty}m_k(\Phi^{\tp k})=0\,\,\,\,\,\,\,\text{to}\,\,\,\,\,\pi_1\boldsymbol{m}\gl=0.
\end{align} 
Similarly, the classical consistency relations imposed by the classical BV master equation are translated from 
\begin{align}
   \sum_{l=1}^{k}m_{l}m_{l-k}=\sum_{l=1}^{k}\frac{1}{2}\commutator{m_{l}}{m_{l-k}}=0,
\end{align}
to the nilpotent structure of $\boldsymbol{m}$
\begin{align}
     \boldsymbol{m}^2=0.
\end{align}
Gauge transformations can also be repackaged using co-algebraic quantities from
\begin{align}
    \delta_\Lambda\Phi=\sum_{k=1}^{\infty}\sum_{l=0}^{k-1}m_k(\Phi^{\tp l},\Lambda,\Phi^{\tp k-l-1})\,\,\,\,\text{to}\,\,\,\,\delta_{\Lambda}\Phi=\pi_1\commutator{\boldsymbol{m}}{\boldsymbol{\Lambda}}\gl,
\end{align}
where $\boldsymbol{\Lambda}$ is the zero co-derivation associated to $\Lambda$.\\
For systems that satisfy the quantum BV master equation, like the quantum closed SFT, the loop-algebra structure 
\begin{align}
    \sum_{g'=0}^{g}\sum_{k'=1}^{k}m_{k'}^{(g')}m_{k+1-k'}^{(g-g')}+m^{(g-1)}_{k+1}U=0,
\end{align}
can be repackaged as a nilpotent structure
\begin{align}
    \qty(\boldsymbol{m}+\kappa^2\boldsymbol{U})^2=0,\,\,\,\,\pi_1\boldsymbol{U}=0,\,\,\,\boldsymbol{U}^2=0,
\end{align}
where $\boldsymbol{m}$ and $\boldsymbol{U}$ are the co-derivation uplifts of the couplings $m^g_k$ and the Poisson bi-vector $U$, and $\kappa^2$ is the closed string coupling constant.\\ 
Thus the WZW co-algebraic formulation clearly provides a compact formulation of the action functional where many algebraically intensive computations are made simpler, and more compact and common algebraic structures shared between theories are highlighted by the formulation.

\subsection{EFTs from the homotopy transfer theorem}
Usually effective field theories (EFT) are a product of integrating out degrees of freedom/fields form the path integral, which restricts the Hilbert space $\Hs$ to a subspace $\bar{\Hs}$. Algebraically, the same process of integrating out fields can be done via the homotopy transfer theorem \cite{Arvanitakis:2020rrk,Erbin_2020}\ref{HTT}. Given an action functional rewritten in the WZW co-algebraic formulation that satisfies the quantum BV master equation
\begin{align}
    S[\Phi]:=\int_{0}^{1}\dd{t}\omega\qty(\pi_1\boldsymbol{\partial}_t\gl,\pi_1\boldsymbol{m}\gl),\,\,\,\,\,\,\,\boldsymbol{D}^2=(\boldsymbol{m}+g\boldsymbol{U})^2=0,
\end{align}
where $g$ is perturbative expansion parameter and $\boldsymbol{U}$ the co-derivation associated to the Poisson bi-vector $U$. If inside $\boldsymbol{m}$  exists an element $Q$, the BRST charge, such that
\begin{align}
    Q:\Hs\longrightarrow\Hs,\,\,\,\,Q^2=0,\,\,\,\,\pi_1\boldsymbol{m}= Q+\pi_1\tilde{\boldsymbol{m}},\,\,\,\partial=Q,
\end{align}
it is possible to use the homotopy transfer theorem to produce EFTs.\\ 
In this section we will use the BRST charge $Q$ instead of the differential $\partial$ because of the physical relevance of $Q$.\\
Let's define a projector $P$ and its co-algebraic equivalent
\begin{equation}
    \begin{split}
        &P:\Hs\longrightarrow\Hs_P\subseteq\Hs,\,\,\,\Phi'=P(\Phi),\\
        &\boldsymbol{P}:\tH\longrightarrow\tH_P\subseteq\tH,\,\,\,\gl'=\boldsymbol{P}\gl,
    \end{split}
\end{equation}
that satisfy the Hodge-Kodaira decomposition and its co-algebraic equivalent
\begin{equation}
    \begin{split}
        &\id =  P+\alpha\qty{Q h+h Q},\\
        &\boldsymbol{\id} = \boldsymbol{P}+\alpha\qty{\boldsymbol{Q h}+\boldsymbol{h Q}}.
    \end{split}
\end{equation}
The projector $P$ acts on the basis of $\Hs$ and not on the space-time field by selecting only the relevant elements we are interested in. As an example let us take $\Phi=\phi^1f_1+\phi^2f_2$. If we are interested only in the space-time field $\phi_1$ we define $P$ such that
\begin{align}
    P\circ f_1=f_1,\,\,\,P\circ f_2=0\Longrightarrow \Phi'=P\circ\Phi=\phi^1f_1.
\end{align}
The EFT interacting structure will be given by \eqref{Morph0}
\begin{align}
    \boldsymbol{m}':=\boldsymbol{F}\boldsymbol{m}\boldsymbol{F}':= \boldsymbol{P}\frac{1}{1+\alpha (\tilde{\boldsymbol{m}}+g\boldsymbol{U})\boldsymbol{h}}\boldsymbol{m}\boldsymbol{P}=\boldsymbol{P}\boldsymbol{m}\frac{1}{1+\alpha \boldsymbol{h}(\tilde{\boldsymbol{m}}+g\boldsymbol{U})}\boldsymbol{P}.
\end{align}
The EFT action is then defined as
\begin{align}
    S_{\rm eff}[\Phi]:=\int_{0}^{1}\dd{t}\omega\qty(\pi_1\boldsymbol{\partial}_t\gl,\pi_1\boldsymbol{F}\boldsymbol{m}\boldsymbol{F}'\gl),\,\,\,\,\,\,\,(\boldsymbol{m}')^2=(\boldsymbol{F}\boldsymbol{m}\boldsymbol{F}')^2=0.
\end{align}
The EFT just defined has in its interaction quantum corrections due to quantum effects of projected out fields. To recover the classical effective theory we just have to send the perturbative parameter and/or specific coupling constants entering in $\boldsymbol{F}$ and $\boldsymbol{F}'$ towards zero
\begin{align}
    \lim_{g\rightarrow0}\boldsymbol{F}:=\boldsymbol{F}_{\rm class},\,\,\,\lim_{g\rightarrow0}\boldsymbol{F}':=\boldsymbol{F}'_{\rm class},
\end{align}
remembering that specific coupling constants $g_k$ have been previously hidden in the definition of the couplings $m_k$ \eqref{coupl-redef}.

\subsection{Amplitudes from the homotopy transfer}\label{AHTT}
Recently, in the works \cite{Macrelli:2019afx,Lopez-Arcos:2019hvg,Doubek:2017naz,Jurco:2019yfd,Jurco:2020yyu,Bonezzi:2023xhn} and \cite{Okawa:2022sjf,Konosu:2023pal}, the homotopy transfer theorem together with the use of co-algebras provide a way to compute QFT amplitudes. In this paper we will mainly focus on the works of \cite{Okawa:2022sjf,Konosu:2023pal} where amplitudes are computed by completely integrating out all DOF of the path integral which, in terms of the homotopy transfer theorem, means setting
\begin{align}
    P=0\,\Longrightarrow\id=\alpha\qty{Q h+h Q}.
\end{align}
Given the action of a scalar bosonic QFT in $d$-dimensions with $N$ self interaction vertices and vanishing boundary terms
\begin{align}
    S[\Phi]:=\int\dd[d]x\qty[\frac{1}{2}\partial_\mu\Phi(x)\partial^\mu\Phi(x)+\frac{1}{2}m^2\Phi^2(x)+\sum_{n=3}^{N}g_n\frac{1}{n!}\mathcal{O}_n(\Phi^n(x))],\,\,\,\,\Phi\in\Hs_0
\end{align}
with $g_n$ the $n$-th coupling constant and $\mathcal{O}_n$ the $n$-th interaction vertex.\\
After rewriting the action using the symplectic form $\omega$ and multilinear products $m_k$ \footnote{The WZW formulation is not necessary in this context because we only need to identity the vertices with the appropriate multilinear products and their co-derivations. If we where to use the WZW formulation we will identify the action to a CAFT\ref{sec:caft}.} we have
\begin{align}
    S[\Phi]:=\frac{1}{2}\omega(\Phi,Q\Phi)+\sum_{n=3}^{N}\frac{1}{n}\omega(\Phi,m_{n-1}(\Phi^{\tp n-1})).
\end{align}
To identify the interaction vertices with $\omega$, $Q$ and $m_k$ we need to trivially extend $\Hs_0$, by adding ghosts, to a graded vector space
\begin{equation}
    \Hs:=\Hs_0\oplus\Hs_1,\,\,\,\,f_0(x)\in\Hs_0,\,\,d(f_0(x))=0,\,\,\,f_1(x)\in\Hs_1,\,\,d(f_1(x))=1,
\end{equation}
where $f_0$ is the basis of $\Hs_0$ and $f_1$ is the basis of $\Hs_1$. The trivial extension allows us to define maps
\begin{equation}
    \begin{split}
        &\omega:\Hs\cross\Hs\longrightarrow\cc,\,\,\,\,m_k:\Hs_0^{\tp k}\longrightarrow\Hs_1,\\
        &Q:\Hs_0\longrightarrow\Hs_1,\,\,\,Q:\Hs_1\longrightarrow0,
    \end{split}
\end{equation}
and it allows the following identifications \cite{Konosu:2023pal}
\begin{equation}
    \begin{split}
        &\Phi=\int\dd[d]{x}\Phi(x)f_0(x),\,\,\,\,\,\,\,\omega(f_0(x),f_1(y))=-\omega(f_1(x),f_0(y))=\delta^d(x-y),\\
        &Q f_0(x)=\qty(-\partial^2+m^2)f_1(x),\,\,\,\,Q f_1(x)=0\Longrightarrow Q^2=0.
    \end{split}
\end{equation}
In this paper we only look at polynomial type interactions vertices which can be identified as
\begin{equation}
    \begin{split}
        &S_{n}^{\rm int}=\int\dd[d]{x}\frac{g_n}{n!}\Phi^n(x):=\frac{1}{n}\omega(\Phi,m_{n-1}(\Phi^{\tp n-1})),\\
        &m_{n-1}(f_0(x_1)...f_0(x_{n-1})):=\frac{g_n}{n-1!} \int\dd[d]{x}\delta^d(x-x_1)...\delta^d(x-x_{n-1})f_1(x),\\
        &m_{n-1}(f_0(x_1)...f_1(x_j)...f_0(x_{n-1}))=0,\,\,\,\,\,\forall j\in[1,n-1].
    \end{split}
\end{equation}
it is clear that the interacting structure trivially satisfies the classical BV master equation and forms a $A_\infty$ algebra
\begin{equation}
    \qty(S,S)=0\Longrightarrow (\boldsymbol{Q}+\boldsymbol{m})^2=0,
\end{equation}
and if we introduce all renormalization vertices $\tilde{\boldsymbol{m}}$ then it satisfies the quantum BV master equation
%\footnote{For a physical QFT the $\alpha$ parameter in \eqref{BVQ} is usually identified with $\hbar$.}
, forming a loop-algebra
\begin{equation}
    \begin{gathered}
        \frac{1}{2}\qty(S,S)+\hbar\Delta S=0\Longrightarrow (\boldsymbol{Q}+\boldsymbol{m}+\hbar\tilde{\boldsymbol{m}}+\hbar\boldsymbol{U})^2=0,\\
        \pi_1\tilde{\boldsymbol{m}}_k=\tilde{m}_k:=\sum_{n=0}^\infty\hbar^ng_{k,n}\,m_k^{(n)},\,\,\,\,g_{k,n}\in\cc,
    \end{gathered}
\end{equation}
where to keep it contained we used co-derivations instead of the multilinear products \ref{WZWCO}. The Poisson bi-vector is expressed as
\begin{align}
        \boldsymbol{U}=\int\dd[d]{x}\boldsymbol{f}_0(x)\boldsymbol{f}_1(x),\,\,\,\pi_2\boldsymbol{U}=U,
\end{align}
where $\boldsymbol{f}_0$ and $\boldsymbol{f}_1$ are the zero co-derivations associated to the basis elements of $\Hs$
\begin{align}
    \pi_1\boldsymbol{f}_0(x)=f_0(x),\,\,\,\,\pi_1\boldsymbol{f}_1(x)=f_1(x).
\end{align}
The $n$-point functions are then computed via the homotopy transfer theorem by \cite{Okawa:2022sjf,Konosu:2023pal,Konosu:2023rkm}
\begin{equation}\label{npf}
    \begin{split}
        \expval{\Phi_1(x_1)...\Phi_n(x_n)}:=(-1)^{n}\omega_n\qty(\pi_n\boldsymbol{F}'\boldsymbol{1},f_1(x_1)\tp...\tp f_1(x_n)),\\
        \omega_n(a_1\tp...\tp a_n,b_1\tp...\tp b_n)=\prod_{i=1}^{n}\omega(a_i,b_i)(-1)^{d(b_i)d(\sum_{j=i+1}^na_j)},
    \end{split}
\end{equation}
where $\boldsymbol{F}'$ is the $A_\infty$ morphism \eqref{Morph1}
\begin{align}\label{BF-ref}
    \boldsymbol{F}'=\frac{1}{1-\alpha\boldsymbol{hB}},\,\,\,\boldsymbol{B}=\boldsymbol{m}+\hbar\tilde{\boldsymbol{m}}+\hbar\boldsymbol{U}.
\end{align}
The contracting homotopy map $h$ is a map
\begin{align}
    h:\Hs_1\longrightarrow\Hs_0,\,\,\,\,h:\Hs_0\longrightarrow0.
\end{align}
it is easy to see that, because $Q$ is the kinetic operator, $h$ has to be the propagator in order to satisfy the Hodge-Kodaria decomposition 
\begin{align}\label{h-prop}
    h f_0(x)=0,\,\,\,\,h f_1(x)=\int\dd[d]{y}\frac{1}{\alpha}\Delta(x-y)f_0(y),\,\,\,\,\Delta(x-y):=\int\frac{\dd[d]{k}}{(2\pi)^d}\frac{e^{ik\cdot(x-y)}}{k^2+m^2-\iota\varepsilon},
\end{align}
where
\begin{equation}
    Qh f_0(x)=0,\,\,\,\,Qh f_1(x)=\frac{1}{\alpha}f_1(x),\,\,\,\,hQ f_1(x)=0,\,\,\,\,hQ f_0(x)=\frac{1}{\alpha}f_0(x),
\end{equation}
such that
\begin{equation}
    \begin{split}
        &\id f_0(x)=\alpha\qty{Qh+hQ}f_0(x)=\alpha\qty{0+\frac{1}{\alpha}}f_0(x)=f_0(x),\\
        &\id f_1(x)=\alpha\qty{Qh+hQ}f_1(x)=\alpha\qty{\frac{1}{\alpha}+0}f_1(x)=f_1(x).
    \end{split}
\end{equation}
By fully unpacking \eqref{npf} we get that the $n$-point function is given by
\begin{align}
    \expval{\Phi_1(x_1)...\Phi_n(x_n)}:=(-1)^n\sum_{i=0}^{\infty}(\alpha\hbar)^i\omega_n\qty(\pi_n\qty{\boldsymbol{hB}}^i\boldsymbol{1},f_1(x_1)\tp...\tp f_1(x_n)).
\end{align}
By unpacking $\boldsymbol{B}$ we can distinguish between the different objects that enter the computation of the $n$-point function with specific non vanishing requirements
\begin{align}
        &\boldsymbol{hU}\Longrightarrow\pi_n\boldsymbol{hU}\pi_{n-2}
        \Longrightarrow \pi_n\qty{\boldsymbol{hU}}^j=\pi_n\qty{\boldsymbol{hU}}^j\pi_{n-2j},\,\,\,\,\pi_n\pi_m=\delta_{n,m}\pi_m\\
        &\boldsymbol{hm}_k\Longrightarrow\pi_n\boldsymbol{hm}_k\pi_{n+k-1}
        \Longrightarrow \pi_n\qty{\boldsymbol{hm}_k}^j=\pi_n\qty{\boldsymbol{hm}_k}^j\pi_{n+j(k-1)},\\
        &\boldsymbol{h\tilde{m}}_k\Longrightarrow\pi_n\boldsymbol{h\tilde{m}}_k\pi_{n+k-1}
        \Longrightarrow \pi_n\qty{\boldsymbol{h\tilde{m}}_k}^j=\pi_n\qty{\boldsymbol{h\tilde{m}}_k}^j\pi_{n+j(k-1)}.
\end{align}
The only difference between $\boldsymbol{m}$ and $\tilde{\boldsymbol{m}}$ is the additional $\hbar$ expansion present in $\tilde{\boldsymbol{m}}$. Since \eqref{npf} has in front of $\boldsymbol{F}$ only $\boldsymbol{1}=\boldsymbol{1}\pi_0$ the only non vanishing contributions a priori are given by
\begin{align}
    \pi_n\qty{\boldsymbol{hU}}^j\boldsymbol{1}=\pi_n\qty{\boldsymbol{hU}}^j\pi_{0}\delta_{2j,n},
\end{align}
which corresponds to the direct propagations of an even number of Bosons. Regarding the interaction vertices, their contributions are non vanishing only if there are enough powers of $\boldsymbol{hU}$ in order to saturate the all the entries of such vertices.

\subsection{$\Phi^3$ theory}
Let us take the $\Phi^3$ theory in $d=6$ dimensions in order to show a concrete computation of correlators via the homotopy transfer theorem \cite{Okawa:2022sjf}.
The classical $\Phi^3$ action without boundary contributions reads
\begin{align}
    S_{\rm cl}[\Phi]:=\int\dd[d]x\qty[-\frac{1}{2}\Phi(x)\partial_\mu\partial^\mu\Phi(x)+\frac{1}{2}m^2\Phi^2(x)-g_3\frac{1}{3!}\Phi^3(x)],\,\,\,\,\Phi\in\Hs_0.
\end{align}
The action can be repackaged using cyclic multilinear products as
\begin{align}
    S[\Phi]:=\frac{1}{2}\omega(\Phi,Q\Phi)+\frac{1}{3}\omega(\Phi,m_{2}(\Phi^{\tp 2})),
\end{align}
where we identify
\begin{equation}
    \begin{split}
        &Q f_0(x)=\qty(-\partial^2+m^2)f_1(x),\\
        &m_{2}(f_0(x_1),f_0(x_2)):=\frac{g_3}{2!} \int\dd[d]{x}\delta^d(x-x_1)\delta^d(x-x_2)f_1(x).
    \end{split}
\end{equation}
In order to compute correlators we need the quantum (UV) completion of the classical action 
\begin{align}
    S_{\rm ren}[\Phi]:=\sum_{k=0}^{\infty}\sum_{n=0}^\infty\hbar^ng_{k,n}\,\omega(\Phi,m_k^{(n)}(\Phi^{\tp k})),\,\,\,\,g_{k,n}\in\cc,
\end{align}
In the $\Phi^3$ model the quantum (UV) completion can be simplified using results from its renormalization, leading to a finite number of counter-terms, namely
\begin{equation}
    \begin{split}
        &{\rm Tadpole}\Rightarrow\,m^1_0\boldsymbol{1}:=-Yf_1=-Y\int\dd[d]{x}f_1(x),\\
        &{\rm Kinetic\,\,term}\Rightarrow\,m^1_1(f_0(x)):=\qty{(Z_M-1)-(Z_\Phi-1)\partial^2}f_1(x),\\
        &{\rm Vertex}\Rightarrow\,m^1_2(f_0(x_1),f_0(x_2)):=-\frac{(Z_{g_3}-1)}{2!}\int\dd[d]{x}\delta^d(x-x_1)\delta^d(x-x_2)f_1(x),
    \end{split}
\end{equation}
where all the other actions of $m^1_0,m^1_1,m^1_2$ on $\Hs$ are trivially zero, and where $Y$ and the $Z$ are the renormalization parameters and have to be expanded in powers of $g_3$
\begin{equation}
    \begin{split}
        &Y=g_3Y^{(1)}+\mathcal{O}(g_3^3),\,\,\,Z_\Phi=1+g_3^2Z_\Phi^{(1)}+\mathcal{O}(g_3^4),\,\,\,Z_{M}=1+g_3^2Z_M^{(1)}+\mathcal{O}(g_3^4)\\
        &Z_{g_3}=1+g_3^2Z_\lambda^{(1))}+\mathcal{O}(g_3^4).
    \end{split}
\end{equation}
The Quantum Homotopy structure necessary for the homotopy transfer theorem is defined by uplifting to co-derivation the multilinear products, resulting in
\begin{align}
    \boldsymbol{D}=\boldsymbol{Q}+\boldsymbol{B},\,\,\,\boldsymbol{B}=\boldsymbol{m}_2+\boldsymbol{m}^1_0+\boldsymbol{m}^1_1+\boldsymbol{m}^1_2+\hbar\boldsymbol{U},
\end{align}
where because we used information from the renormalization of $\Phi^3$, the $\hbar$ present in the definition  \eqref{BF-ref} are hidden in the definition of the $m^i_k$.\\
From \eqref{h-prop} we recall the form of the contracting homotopy map 
\begin{align}
    h f_1(x)=\int\dd[d]{y}\frac{1}{\alpha}\Delta(x-y)f_0(y),\,\,\,\,\Delta(x-y):=\int\frac{\dd[d]{k}}{(2\pi)^d}\frac{e^{ik\cdot(x-y)}}{k^2+m^2-\iota\varepsilon},
\end{align}
and the necessary elements needed for the computation of correlators
\begin{align}
        \expval{\Phi_1(x_1)...\Phi_n(x_n)}:=(-1)^n\omega_n\qty(\pi_n\boldsymbol{F}'\boldsymbol{1},f_1(x_1)\tp...\tp f_1(x_n)),\,\,\,\,\,\boldsymbol{F}'=\frac{1}{1-\alpha\boldsymbol{hB}}.
\end{align}
It's easy to see now that, in order to compute n-point correlators we need to evaluate the non vanishing contributions of
\begin{align}
    \pi_n\boldsymbol{F}'\boldsymbol{1}= \pi_n\qty{1+\alpha\boldsymbol{hB}+\alpha^2\boldsymbol{hB}\boldsymbol{hB}+...}\id.
\end{align}
for the $1$-point function at leading order in $g_3$ we encounter
\begin{align}
    \pi_1\boldsymbol{F}'\boldsymbol{1}\sim \alpha g_3\hbar^2 Y^{(1)}\pi_1\boldsymbol{h}\boldsymbol{m}_0+\alpha g_3\hbar^2\pi_1\boldsymbol{h}\boldsymbol{m}_2\boldsymbol{h}\boldsymbol{U}+\mathcal{O}(g_3^2),
\end{align}
which, after using the definition of $h$,$m^1_0$,$m_2$ and $U$, and the necessary regularization hypothesis \cite{Okawa:2022sjf}, leads to 
\begin{align}\label{Tadpole}
    \expval{\Phi(x)}=\frac{g_3\hbar^2}{m^2}\qty{\frac{1}{2}\int\frac{\dd[d]{k}}{(2\pi)^d}\frac{1}{k^2+m^2}+Y^{(1)}}+\mathcal{O}(g^2_3).
\end{align}
Note that whenever there are powers of $\alpha$ due to the expansion of $\boldsymbol{F}'$ they are allays cancelled by the $\alpha^{-1}$ present in the definition of $h$, therefore correlators are independent form the choice of Hodge-Kodaira decomposition.\\
A complete breakdown of the process for other correlators is provided in \cite{Okawa:2022sjf}, provided that definitions and normalization are slightly different from this paper.    
    
\section{Co-algebraic field theory}\label{sec:caft}
Since all local Lagrangian Field Theories can be rewritten in the WZW co-algebraic formulation \ref{WZWCO}, in this section we will define the concept of Co-Algebraic Field Theory (CAFT)\footnote{Any local QFT action can be rewritten using $\omega$,$m_k$ like we saw in \ref{AHTT}. }, its connection with generalized Lagrangian Field Theories and relevant mathematical properties. We will also review the computational benefits provided by the
CAFT formulation of QFT and SFT.\\ 
A co-algebraic field theory (CAFT) is defined given a tensor co-algebra $\tH$ together with a group like element $\gl$, a symplectic form $\omega$ and a co-derivation $\boldsymbol{m}$. The action of the co-algebraic field theory is
\begin{align}
S[\gl]:=\int_0^1\dd{t}\omega(\pi_1\boldsymbol{\partial}_t\gl,\pi_1\boldsymbol{m}\gl),\,\,\,\,\gl:=\gl(t).
\end{align}
If the co-derivation is cyclic then the CAFT is said to be cyclic.\\ 
The CAFT definition can be equivalently formulated in terms of a co-homomorphism\footnote{If we include more general formulations of co-homomorphisms the CAFT will define more general field theories.} generated by $\boldsymbol{m}$ together with a Grassman parameter $\varepsilon$
\begin{align}
    \boldsymbol{F}_\varepsilon:= e^{\varepsilon\boldsymbol{m}}\,\Longrightarrow\, S[\gl]:=\int\dd{\varepsilon}\int_0^1\dd{t}\omega(\pi_1\boldsymbol{\partial}_t\gl,\pi_1\boldsymbol{F}_\varepsilon\gl).
\end{align}
The two definitions are equal because of the following properties
\begin{align}
    \int\dd{\varepsilon}\varepsilon^n = -\delta_{1,n},\,\,\,\,\,\,\int\dd{\varepsilon}\omega(A,B(\varepsilon))=(-1)^{d(A)+1}\omega(A,\int\dd{\varepsilon}B(\varepsilon)).
\end{align}
By trivially parametrizing $\gl$ we recover the Wess-Zumino-Witten co-algebraic formulation of Lagrangian Field Theories \ref{WZWCO}. As we will later prove, given $n$ different field species with associated $\Hs_i$, any multilinear product 
\begin{align}
    m^j_{k_1,k_2,...,k_n}:\Hs_1^{\tp k_1}\tpt\Hs_2^{\tp k_2}\tpt...\tpt\Hs_n^{\tp k_n}\Longrightarrow\Hs_j,
\end{align}
 can be uniquely associated to a co-derivation $\boldsymbol{m}^j_{k_1,k_2,...,k_n}$ on $\tHt:=:\tH_1\tpt...\tpt\tH_n$, therefore any Lagrangian Field Theory can be derived from the appropriate CAFT and all properties due to the use of co-algebras and homotopy algebras at the CAFT level are universal between Lagrangian Field Theories.

\subsection{Field redefinitions, Variations and Symmetries}
As stated in \ref{COHOM} and \ref{WZWCO}, graded zero co-homomorphisms are responsible for mapping co-algebras into other co-algebras \footnote{A complete breakdown of the topics from a different perspective can be found in \cite{vovsmera2020selected,Gaberdiel:1997ia}.}. From a physical point of view given a co-homomorphism $\boldsymbol{F}$ such that
\begin{align}
    \boldsymbol{F}:\tH_1\longrightarrow\tH_2,\,\,\,\,\bra{\omega'}\pi_2\boldsymbol{F}=\bra{\omega}\pi_2,\,\,\,\,\gl':=\boldsymbol{F}\gl,
\end{align}
the co-homomorphism maps a CAFT with action $S$ to a new CAFT with action $S'$
\begin{equation}\label{DCAFT}
    \begin{split}
        S'[\gl']= \int_0^1\dd{t}\omega'(\pi_1\boldsymbol{\partial}_t\gl',\pi_1\boldsymbol{m}'\gl')=S[\gl]&+\int_0^1\dd{t}\omega'(\pi_1\commutator{\boldsymbol{\partial}_t}{\boldsymbol{F}}\gl,\pi_1\boldsymbol{m}'\boldsymbol{F}\gl)\\&+\int_0^1\dd{t}\omega'(\pi_1\boldsymbol{\partial}_t\boldsymbol{F}\gl,\pi_1(\boldsymbol{m}'\boldsymbol{F}-\boldsymbol{Fm})\gl)\\&+\int_0^1\dd{t}\omega'(\pi_1\commutator{\boldsymbol{\partial}_t}{\boldsymbol{F}}\gl,\pi_1(\boldsymbol{m}'\boldsymbol{F}-\boldsymbol{Fm})\gl).
    \end{split}
\end{equation}
If the co-homomorphism $\boldsymbol{F}$ satisfies
\begin{align}
    \commutator{\boldsymbol{\partial}_t}{\boldsymbol{F}}=0,\,\,\,\,\,\boldsymbol{m}'\boldsymbol{F}=\boldsymbol{Fm}\Longrightarrow S'[\gl']=S[\gl],
\end{align}
then the two actions are dual to each other and $\boldsymbol{F}$ referred to as the duality map, i.e. the two actions describe different theories with different $\tH_i$ but are in fact two distinct realizations of the same theory related by the mapping $\boldsymbol{F}$. \\
If the co-homomorphism maps the co-algebra to itself (co-endomorphism) then, instead of mapping between theories, it realized field redefinitions and changes of parametrization
\begin{equation}
    \begin{split}
        \boldsymbol{F}:\tH\rightarrow\tH,\,\,\,\,&\bra{\omega}\pi_2\boldsymbol{F}=\bra{\omega}\pi_2,\,\,\,\,\gl':=\boldsymbol{F}\gl,\\
        S'[\gl']=S[\gl]+&\int_0^1\dd{t}\omega(\pi_1\commutator{\boldsymbol{\partial}_t}{\boldsymbol{F}}\gl,\pi_1\boldsymbol{m}'\boldsymbol{F}\gl)\\
        +&\int_0^1\dd{t}\omega(\pi_1\boldsymbol{\partial}_t\boldsymbol{F}\gl,\pi_1(\boldsymbol{m}'\boldsymbol{F}-\boldsymbol{Fm})\gl)\\+&\int_0^1\dd{t}\omega(\pi_1\commutator{\boldsymbol{\partial}_t}{\boldsymbol{F}}\gl,\pi_1(\boldsymbol{m}'\boldsymbol{F}-\boldsymbol{Fm})\gl).
    \end{split}
\end{equation}
If the co-endomorphism satisfies
\begin{align}\label{GCON}
    \commutator{\boldsymbol{\partial}_t}{\boldsymbol{F}}=0,\,\,\,\,\,\boldsymbol{m}'\boldsymbol{F}=\boldsymbol{Fm}\Longrightarrow S'[\gl']=S[\gl],
\end{align}
then the field transformation $\boldsymbol{F}$ realizes a generalized Gauge transformation.\\
Infinitesimal field redefinitions are generated by the exponentiation of a set of co-derivations $\boldsymbol{T}_a$ of grading zero, together with a set of parameters $\varepsilon^a$
\begin{align}\label{0morph}
    \boldsymbol{F}:=e^{\varepsilon^a\boldsymbol{T}_a}:= \id + \varepsilon^a\boldsymbol{T}_a+\mathcal{O}(\varepsilon^2),
\end{align}
or if the generator is of grading differently then $0$ we uplift the graded parameter $\varepsilon$ to zero co-derivations $\boldsymbol{\varepsilon}^a$ and build the graded zero co-homomorphism as
\begin{align}\label{no0morph}
    \boldsymbol{F}:=e^{\commutator{\boldsymbol{\varepsilon}^a}{\boldsymbol{T}_a}},\,\,\,\,d(\boldsymbol{\varepsilon}^a)=-d(\boldsymbol{T}_a).
\end{align}
To generate infinitesimal Gauge transformations \eqref{0morph} and \eqref{no0morph} have to additionally satisfy \eqref{GCON}.

\subsection{Classical and Quantum algebraic structures}
To investigate the classical and quantum algebraic structure of a given CAFT with trivial parametrization, we ask that the CAFT satisfies the classical BV master equation 
\begin{align}\label{BVC}
    \qty(S,S)=0,
\end{align}
or the quantum BV master equation \footnote{More details on the BV master equation can be found in \cite{gomis1995antibracket}}.
\begin{align}\label{BVQ}
    \frac{1}{2}\qty(S,S) + \beta\Delta_{\rm BV} S=0,\,\,\,\,\beta\in\cc,
\end{align}
where $\beta$ of \eqref{BVQ} is a dimension-full constant like $\hbar$.\\
To explicitly write down the BV bracket $(\cdot,\cdot)$ and the the BV Laplacian $\Delta_{\rm BV}$ we need a suitable choice of basis of $f_a\in\Hs$ and its dual $f^a$ such that
\begin{equation}\label{omegaprop1}
    \begin{split}
        &\pi_1\gl=\Phi:=\phi^af_a=\phi_af^a,\,\,\,\,\omega(f^a,f_b)=-\omega(f_b,f^a)=\delta^a_{\,\,\,b}\\
        &\omega^{ab}=\omega(f^a,f^b)\,\,\,,\,\,\,\omega_{ab}=\omega(f_a,f_b)\\
        &d(f_a)=-d(\phi^a)\,\,\,,\,\,\,d(f^a)=-d(\phi_a)\,\,\,,\,\,\,d(\omega)=-1,
    \end{split}
\end{equation}
where the symplectic form is non trivial for
\begin{align}\label{omegaprop2}
    d(f^a)+d(f^b) = 1\Longleftrightarrow\omega(f^a,f^b)\neq0,\,\,\,\,\,&d(f_a)+d(f_b) = 1\Longleftrightarrow\omega(f_a,f_b)\neq0.
\end{align}
The BV bracket $(\cdot,\cdot)$ and the the BV Laplacian $\Delta_{\rm BV}$ can be formulated as
    \begin{align}\label{BVDIFFOP}
        \qty(X,Y):=X\overset{\leftarrow}{\pdv{}{\phi^a}}\omega^{ab}\overset{\rightarrow}{\pdv{}{\phi^b}}Y,\,\,\,\,\,\,\,\,\Delta_{\rm BV} X = \frac{(-1)^{d(\phi^a)}}{2}\omega^{ab}\overset{\rightarrow}{\pdv{}{\phi^a}}\overset{\rightarrow}{\pdv{}{\phi^b}}X,
    \end{align}
where $\Phi:=\sum_a\phi^af_a$, and $\phi^a$ are the space-time fields of the theory.\\
To compute the BV master equation we rely on a compact formula derived using \eqref{DCAFT} where repeated differentiation with respect to $\phi^a$ coincides to
\begin{align}\label{princ-result}
    \overset{\rightarrow}{\pdv{}{\phi^{a_n}}}...\overset{\rightarrow}{\pdv{}{\phi^{a_1}}}S[\gl]=\int_0^1\dd{t}\omega(\pi_1\boldsymbol{\partial}_t\gl,\pi_1\mathbf{n}\boldsymbol{f}_{a_n}...\boldsymbol{f}_{a_1}\gl)\,\,\,\forall\,n\geq0,\,\,\,\,\,\pi_1\boldsymbol{\partial}_t\boldsymbol{f}_{a_i}=0,
\end{align}
where $\boldsymbol{f}_a$ are the zero co-derivations associated to the basis $f_a$.\\
The case $n=1$ also has an alternative expression in terms independent of the $t$ parametrization of the field, which reads
\begin{align}\label{princ-result2}
    \overset{\rightarrow}{\pdv{}{\phi^a}}S=(-1)^{d(\phi^a)}\omega(\pi_1\boldsymbol{f}_a\gl,\pi_1\mathbf{n}\gl).
\end{align}
A complete and detailed proof for \eqref{princ-result} is given in appendix \ref{Appendix_CAFT}.\\
Thanks to \eqref{princ-result} it is fairly straight forward to derive that
\begin{align}
    \qty(S,S)= 2\int_0^1\dd{t}\omega(\pi_1\boldsymbol{\partial}_t\gl,\pi_1\mathbf{mm}\gl),
\end{align}
and that
\begin{align}
    \Delta_{\rm BV} S= \int_0^1\dd{t}\omega(\pi_1\boldsymbol{\partial}_t\gl,\pi_1\boldsymbol{m}\mathbf{U}\gl),\,\,\,\,\,\,\mathbf{U}=\frac{(-1)^{f^a}}{2}\omega^{ba}\boldsymbol{f}_a\boldsymbol{f}_b.
\end{align}
where properties \eqref{omegaprop1} and \eqref{omegaprop2} are used to complete the calculations.\\
The classical BV master equation tells us that a classically consistent CAFT has the algebraic structure of an Homotopy algebra
\begin{align}\label{CBV}
    \qty(S,S)= 0\,\Longleftrightarrow\,\boldsymbol{m}^2=0,
\end{align}
and a the quantum BV master action tells us that a quantum consistent CAFT has the algebraic structure of a loop-algebra
\begin{align}\label{QBV}
    \frac{1}{2}\qty(S,S)+\alpha\Delta S=0 \,\Longleftrightarrow\,\qty(\boldsymbol{m}+\alpha\mathbf{U})^2=0.
\end{align}
The results are consistent with what was observed in the different SFTs \cite{Erler_2014,Kajiura:2004xu,Kajiura:2005sn,Kajiura:2006mt,Maccaferri:2022yzy,Maccaferri:2023gcg}.\\
The results \eqref{princ-result},\eqref{CBV} and \eqref{QBV} have been derived using only co-algebraic informations and are therefore valid for any choice of Hilbert space and tensor algebra/Fock space. This includes also Fock spaces of many particles types/strings which are Fock spaces born form the tensor product of many Fock spaces together.    
    
\section{N components tensor co-algebras}\label{NCOALG:CH}
In this section we address the construction of the co-algebras required to describe systems with N different types of elements (e.g. particles/ fields/ strings).\\
Each different element lives in its base vector space $\Hs_i$ all over the field $\rr$ or $\cc$ and from it we can define the associated tensor algebra and co-algebra $(\Hs_i,\tp_i,\Delta_i)$ as we saw in \ref{Sec_Co_algebras}.\\
To build a co-algebra that encompasses all the $N$ potentially different objects we need to define the overall tensor algebra
\begin{equation}
    \begin{gathered}
        \tHt:=\tH_1\tpt\tH_2\tpt...\tpt\tH_N =\sum_{n_1,n_2,...,n_N=0}^{\infty}\Hs_1^{\tp_1 n_1}\tpt \Hs_2^{\tp_2 n_2}\tpt...\tpt\Hs_N^{\tp_N n_N} =\\
        = \sum_{n_1,n_2,...,n_N=0}^{\infty}\Hst^{(n_1,n_2,...,n_N)},
    \end{gathered}
\end{equation}
where $\qty{\Hs_i}_{0<i\leq N}$ are the base vector spaces of the specific particle/string/boundary\footnote{For world-sheet topologies with more then one boundary, to each boundary a open string field Fock space is required in order to effectively describe that specific surface, see section \ref{QOCSFT}.} described and $\tpt$ is, in principle, a tensor product that joins together the different $\tH_i$ together. Mathematically we have that $\tpt\simeq\tp_i=\tp$ because all $\Hs_j$ are defined on the same field $\kk$. We will keep the distinction between $\tpt$ and $\tp_i$ in order to help with the bookkeeping by clearly distinguishing between elements from the different spaces $\tH_i$.\\
To keep the notation as readable as possible elements of $\tHt$ are written as
\begin{align}
    v^1_{1,n_1}\tpt v^2_{1,n_2} \tpt ...\tpt v^N_{1,n_N} \in \Hst^{(n_1,n_2,...,n_N)} = \Hs_1^{\tp n_1}\tpt \Hs_2^{\tp n_2}\tpt...\tpt\Hs_N^{\tp n_N},
\end{align}
where the superscript $i$ in $v^i_{1,n}$ indicates from which base tensor algebra it originates, in this case $v^i_{1,n}\in\tH_i$.\\
An explicit example of $\tHt$ is the Fock space of QED where there are two base Hilbert spaces of states, namely the electron-positron Hilbert space $\Hs_{e\bar{e}}$ and the photon Hilbert space $\Hs_{\gamma}$. It's possible to build individual Fock spaces from the base Hilbert space $\tH_{e\bar{e}}$ and $\tH_\gamma$. A generic QED state is the tensor product between a state in $\tH_{e\bar{e}}$ and $\tH_{\gamma}$, therefore an element of $\tH_{e\bar{e}}\tpt\tH_{\gamma}$.\\
Like in \ref{Sec_Co_algebras} we define projectors $\pi$ and inclusions $\iota$ on $\tHt$. Projectors $\pi_{n_1,..,n_N}$ are defined using the base projectors $\pi^j_{n_j}$ on $\tH_j$ as
\begin{equation}\label{NPROJ}
    \begin{split}
        &\pi_{n_1,..,n_N}:\tHt\longrightarrow\Hst^{(n_1,..,n_N)}=\Hs_1^{\tp n_1}\tpt \Hs_2^{\tp n_2}\tpt...\tpt\Hs_N^{\tp n_N},\\
        &\pi_{n_1,..,n_N}:=\pi^1_{n_1}\tpt...\tpt\pi^N_{n_N},\,\,\,\,\pi^j_{n_j}:\tH_j\longrightarrow\Hs_j^{\tp n_j}.
    \end{split}
\end{equation}
To keep notation as readable as possible we define the following two special projectors
\begin{equation}
    \begin{split}
        &\pi^{(j)}:=\pi_{0_1,...,1_j,...0_N}:\tHt\longrightarrow\Hst^{(0_1,...,1_j,...,0_N)}=\boldsymbol{1}_1\tpt...\tpt\Hs_j\tpt...\tpt\boldsymbol{1}_N,\\
        &\pi_{(j)}:=\sum_{n_1+...+n_N=j}\pi_{n_1,...,n_N}:\tHt\longrightarrow\bigoplus_{n_1+...+n_N=j}\Hst^{(n_1,...,n_N)}:=\Hst^{(j)}.
    \end{split}
\end{equation}
Similarly to \eqref{NPROJ}, inclusions $\iota$ on $\tHt$ are defined using the base inclusions $\iota^j_{n_j}$ on $\tH_j$ as
\begin{equation}
    \begin{split}
        &\iota_{n_1,...,n_N}:=\iota^1_{n_1}\tpt...\tpt\iota^N_{n_N}:\Hst^{(0_1,...,1_j,...,0_N)}\longrightarrow\Hst^{(n_1,...,n_j+1_j,...,n_N)},\\
        &\iota:=\sum_{n_1,...,n_N=0}^{\infty}\iota_{n_1,...,n_N}.
    \end{split}
\end{equation}
Inclusions in the study of $N$ co-algebras do not feed into the remainder of this work and we will not discuss them further.

\subsection{Swapping map and co-product}
On $\tHt$ it's easier to define the tensor co-algebra instead of its tensor algebra due to the lack of a proper tensor product such that
\begin{align}
    \tpb:\tHt\cross\tHt\longrightarrow\tHt,
\end{align}
but $\tpb$ can be defined using the concatenation product \eqref{conc_prod}.\\
The co-product on $\tHt$ it is defined using the co-products of the base co-algebras $(\Hs_i,\tp_i)$ together with the addition of the swapping map $\Omega_N$\footnote{The map $\Omega_N$, and subsequently $\Omega_N^{-1}$, formally is defined by braiding maps \cite{article}.}
\begin{align}
    \Omega_N:(\tH_1\tp_1'\tH_1)\tpt...\tpt(\tH_N\tp_N'\tH_N) \longrightarrow (\tH_1\tpt...\tpt\tH_N)\tpb'(\tH_1\tpt...\tpt\tH_N),
\end{align}
where $\tpb'$ is the primed tensor product of the tensor algebra defined over $\tHt$ and $\tpb'\neq\tpb$ is the external tensor product.\\
To keep calculations more compact we chose $\Omega_N$ such that the swapping between elements of different $\tH_i$ does not generate phase contributions
\begin{align}\label{swap}
    \Omega_N\qty((v^1_{1,i_1}\tp_1'v^1_{i_1+1,n_1})\tpt...\tpt(v^N_{1,i_N}\tp_N'v^N_{i_N+1,n_N})):= (v^1_{1,i_1}\tpt...\tpt v^N_{1,i_N})\tpb'(v^1_{i_1+1,n_1}\tpt...\tpt v^N_{1,i_N}).
\end{align}
A different approach where the swapping map picks up phases but yields the same results is given in \cite{hoefel2009coalgebra}.\\
The co-product $\bDelta$ can be defined as
\begin{align}\label{NCOPROD}
    \bDelta = \Omega_N(\Delta_1\tpt\Delta_2\tpt...\tpt\Delta_N).
\end{align}
If all the base co-products $\Delta_i$ are co-associative \eqref{COASS} then $\bDelta$ is co-associative and vice-versa
\begin{align}\label{NCOASS}
    (\bDelta\tpb'\id)\bDelta = (\id\tpb'\bDelta)\bDelta\,\Longleftrightarrow\,(\Delta_i\tp_i'\id_i)\Delta_i = (\id_i\tp_i'\Delta_i)\Delta_i\,\,\,\,\,\forall i\in\qty[1,N].
\end{align}
The group like element $\bgl$ of $\tHt$ is defined as
\begin{align}
    \bDelta\bgl = \bgl\tpb'\bgl,
\end{align}
and it is entirely determined by the group like elements $\gl_i$ of the base co-algebras
\begin{equation}\label{NGL}
    \bgl=\gl_1\tpt\gl_2\tpt...\tpt\gl_N.
\end{equation}

\subsection{Concatenation product and natural tensor product}
To $\bDelta$ a concatenation product $\bnabla$ can be associated. Similarly to $\bDelta$, $\bnabla$ is built from the base co-algebras concatenation products $\nabla_j$ together with the inverse swapping map $\Omega_N^{-1}$ 
\begin{align}
    \bnabla:= (\nabla_1\tpt\nabla_2\tpt...\tpt\nabla_N)\Omega_N^{-1},\,\,\,\,\,\Omega_N\Omega_N^{-1}=\Omega_N^{-1}\Omega_N=\id,
\end{align}
where $\Omega_N^{-1}$ acts like
\begin{equation}
    \begin{split}
        &\Omega_N^{-1}: (\tH_1\tpt...\tpt\tH_N)\tpb'(\tH_1\tpt...\tpt\tH_N)\longrightarrow (\tH_1\tp'_1\tH_1)\tpt...\tpt(\tH_N\tp'_N\tH_N),\\
        &\Omega_N^{-1}\qty((v^1_{1,i_1}\tpt...\tpt v^N_{1,i_N})\tpb'(v^1_{i_1+1,n_1}\tpt...\tpt v^N_{1,i_N})):=(v^1_{1,i_1}\tp'_1v^1_{i_1+1,n_1})\tpt...\tpt(v^N_{1,i_N}\tp'_Nv^N_{i_N+1,n_N}).
    \end{split}
\end{equation}
Because in a co-algebra the concatenation product turns the external tensor product $\tpb'$ into the internal tensor product $\tpb$, $\bnabla$ provides the necessary tool to easily define the natural tensor product on $\tHt$
\begin{align}\label{NPROPPROD}
    (v^1_{1,i_1}\tpt...\tpt v^N_{1,i_N})\tpb(v^1_{i_1+1,n_1}\tpt...\tpt v^N_{1,i_N}):=\bnabla((v^1_{1,i_1}\tpt...\tpt v^N_{1,i_N})\tpb'(v^1_{i_1+1,n_1}\tpt...\tpt v^N_{1,i_N})).
\end{align}
Thanks to $\bnabla$ it is trivial to see that $\tpb$ turn $\tHt$ into a tensor algebra.

\subsection{Linear co-algebraic operators and co-Leibniz rules}
Note that $\bDelta$ and $\bnabla$ are $N+1$ linear operators. The fact that $\bDelta$ is of $N+1$ order complicates the definition of co-derivations $\boldsymbol{d}$ of $\tHt$ because the trivial generalization of the co-Leibniz rule on $\tHt$ implies
\begin{align}
    \bDelta\boldsymbol{d}=\qty(\boldsymbol{d}\tpb'\bar{\id}+\bar{\id}\tp'\boldsymbol{d})\bDelta\,\Longrightarrow\boldsymbol{d}=0.
\end{align}
To derive the properties $\boldsymbol{d}$ has to satisfy in order to be a co-derivation we start by trivially extending its action onto the group like element $\bgl$
\begin{align}\label{NCOGL}
    \boldsymbol{d}\bgl=\bgl\tpb(\pi_{(1)}\boldsymbol{d}\bgl)\tpb\bgl.
\end{align}
To recover the definition of the co-derivation we only need to apply \eqref{NCOPROD} to \eqref{NCOGL}. The results is a modified co-Leibniz rule that accounts for $\bDelta$ being a $N+1$ linear operator. 
\begin{align}\label{NCODERHOE}
    \bDelta\boldsymbol{d}\bgl=\bDelta(\bgl\tpb(\pi_{(1)}\boldsymbol{d}\bgl)\tpb\bgl).
\end{align}
The case for $N=2$ has been independently computed in \cite{hoefel2009coalgebra}.\\
A different path to define co-derivations $\boldsymbol{d}$ of $\tHt$ starts from noticing that $\bDelta$ can be rewritten as the consecutive action on $N$ mutually commuting linear operators $\bDelta_i$ together with the swapping map $\Omega_N$ 
\begin{align}\label{NLINCOP}
    \bDelta=\Omega_N\bDelta_1...\bDelta_N,\,\,\,\,\,\bDelta_i\bDelta_j=\bDelta_j\bDelta_i\,\,\,\,\,\,\,\,\forall_{i,j}\in\qty[1,N],
\end{align}
where $\bDelta_i$ are the trivial extensions of the base co-products $\Delta_i$
\begin{equation}
    \begin{gathered}
        \bDelta_i:=\id_1\tpt...\tpt\id_{i-1}\tpt\Delta_i\tpt\id_{i+1}\tpt...\tpt\id_{N},\\
        \bDelta_i:\tH_1\tpt...\tpt\tH_i\tpt...\tpt\tH_N\longrightarrow\tH_1\tpt...\tpt(\tH_i\tp_i'\tH_i)\tpt...\tpt\tH_N.
    \end{gathered}
\end{equation}
By applying \eqref{NLINCOP} to \eqref{NCOGL} and choosing to factor out $\Omega_N$ and $N-1$ instances of $\bDelta_j$ we recover 
\begin{align}
    \Omega_N\bDelta_1...\bDelta_{N-1}\qty(\bDelta_N\boldsymbol{d}\bgl-\bDelta_N\bgl\tpb(\pi_{(1)}\boldsymbol{d}\bgl)\tpb\bgl)=0,
\end{align}
which can be satisfied if $\boldsymbol{d}$ satisfies the linear co-Leibniz rule
\begin{align}
    \bDelta_N\boldsymbol{d}=(\boldsymbol{d}\tp'_N\id_N+\id_N\tp_N'\boldsymbol{d})\bDelta_N,
\end{align}
where on the RHS we use a short hand notation that tells us if we use the right or left split of $\tH_N$ previously broken by $\bDelta_N$.\\
By using the fact that the different $\bDelta_j$ mutually commute we can extract $N$ independent linear co-Leibniz rules, one for each $\bDelta_j$. Therefore, for $\boldsymbol{d}$ to be a co-derivation of $\tHt$ it has to satisfy the set of $N$ linearized co-Leibniz rules
\begin{align}\label{NCOLEIB}
    \qty{\bDelta_i \mathbf{d}=\qty(\id_i\tp'_i\mathbf{d}+\mathbf{d}\tp'_i\id_i)\bDelta_i}_{1\leq i \leq N},
\end{align}
where on the RHS we use a short hand notation that tells us if we use the right or left tensor algebra previously broken by $\bDelta_i$.\\
Because to derive the co-Leibniz rules we used in both cases only the properties of $\bDelta$ and \eqref{NCOGL}, definitions \eqref{NCOLEIB} and \eqref{NCODERHOE} are dual to each other. Furthermore, the difference between \eqref{NCODERHOE} and the case $N=2$ in \cite{hoefel2009coalgebra} differs only by the choice of swapping map, ensuring that co-derivations in \cite{hoefel2009coalgebra} are dual to \eqref{NCOLEIB}.

\subsection{Multilinear products and co-derivations}\label{NCODER:SEC}
In the literature, multilinear products are commonly defined as maps acting
\begin{align}\label{NIMPR}
    n^{j}_{i_1,...,i_n}:\Hs_1^{\tp_1 i_1}\tpt...\tpt \Hs_1^{\tp_1 i_1}\longrightarrow \Hs_j,\,\,\,\,\,n^{j}_{i_1,...,i_n}\in\Hom(\Hst^{(n_1,..,n_N)},\Hs_j)
\end{align}
where $j$ is the label of the output Hilbert space $\Hs_j$. Since we can trivially identify $\Hs_j$ with $\Hst^{(0_1,...,1_j,...,0_N)}$ we can identify
\begin{align}
    \Hom(\Hst^{(n_1,..,n_N)},\Hs_j)\simeq\Hom(\Hst^{(n_1,..,n_N)},\Hst^{(0_1,...,1_j,...,0_N)})\subset\Hom(\tHt,\tHt),
\end{align}
and co-derivations are a specific subset of elements in $\Hom(\tHt,\tHt)$ that satisfies \eqref{NCOLEIB}. A multilinear product $\boldsymbol{n}^j_{i_1,...,i_n}$ can be uplifted to a co-derivations $\bar{\boldsymbol{n}}^j_{i_1,...,i_n}$ in the following uplift procedure
\begin{align}\label{N-uplift}
    \mathbf{n}^{j}_{i_1,...,i_N}\pi_{n_1,...,n_N}=\sum_{j_1=0}^{n_1-i_1}...\sum_{j_N=0}^{n_N-i_N}\bar{\id}^{j_1,...,j_N}\tpb n^j_{i_1,...,i_N} \tpb\bar{\id}^{ n_1-j_1-i_1,..., n_N-j_N-i_N},
\end{align}
which is the generalization of \eqref{Hom_Coder} and $\bar{\id}^{i_1,...,i_N}$ is the identity operator
\begin{align}
    \bar{\id}^{i_1,...,i_N}:=\id_1^{\tp_1 i_1}\tpt...\tpt\id_N^{\tp_N i_N}.
\end{align}
\eqref{N-uplift} can also be rewritten using inclusion operators
\begin{align}
    \mathbf{n}^{j}_{i_1,...,i_N}\pi_{n_1,...,n_N}=\iota_{n_1-i_1,...,n_j-i_j+1,...,n_N-i_N}n^j_{i_1,...,i_N}\pi_{n_1,...,n_N}.
\end{align}
The operator $\mathbf{n}$, by construction, satisfies \eqref{NCOLEIB} and \eqref{NCOGL} and can be mapped back to the original multilinear product via the projections
\begin{align}
    n^{j}_{i_1,...,i_N} = \pi_{0_1,...,1_j,...,0_N}\,\mathbf{n}^{j}_{i_1,...,i_N},\,\,\,\,\,\mathbf{n}^{j}_{i_1,...,i_N}\pi_{i_1,...,i_j,...,i_N}=n^{j}_{i_1,...,i_N}.
\end{align}
Not only \eqref{N-uplift} provides the map from $\Coder(\tHt)$ to $\Hom(\tHt,\Hs_j)$ but it is also the unique way to uplift multilinear products to co-derivations. Our last statement and \eqref{N-uplift} can be proven starting by observing that $n^{j}_{i_1,...,i_N}$ implies that its associated co-derivation is a map
\begin{align}
    \mathbf{n}^{j}_{i_1,...,i_N}:\Hst^{(n_1,...,n_N)}\longrightarrow\Hst^{(n_1-i_1,...,n_j-i_j+1,...,n_N-i_N)}.
\end{align}
Then, working with $\mathbf{n}^{j}_{i_1,...,i_N}\pi_{n_1,...,n_N}$, we apply one of the co-Leibniz equations \eqref{NCOLEIB} resulting in 
\begin{align}\label{N_CO_UNIT1}
    \bDelta_l \mathbf{n}^{j}_{i_1,...,i_N}\pi_{n_1,..,n_N}=\sum_{p_l=0}^{n_l-i_l} & \left( \pi_{0_1,...,n_l-i_l-p_l,...,0_N}\tp'_l\mathbf{n}^{j}_{i_1,...,i_N}\pi_{n_1,...,p_l,...,n_N}+\right.\\
    &+\left.\mathbf{n}^{j}_{i_1,...,i_N}\pi_{n_1,...,p_l,...,n_N}\tp'_l\pi_{0_1,...,n_l-i_l-p_l,...,0_N}\right)\bDelta_l,
\end{align}
where we used the fact that
\begin{align}\label{NCOPROJ}
    \bDelta_l\pi_{i_1,...,i_N}=\sum_{p_l=0}^{i_l}\pi_{i_1,...,p_l,...,i_N}\tp'_l\pi_{0_1,...,i_l-p_l,...,0_N}=\sum_{p_l=0}^{i_l}\pi_{0_1,...,i_l-p_l,...,0_N}\tp'_l\pi_{i_1,...,p_l,...,i_N}.
\end{align}
We then iterate the process with different $\bDelta_l$ until there are only trivial splits possible. At this point all splitted co-derivations in \eqref{N_CO_UNIT1} are in the form
\begin{align}
    \mathbf{n}^{j}_{i_1,...,i_N}\pi_{i_1,...,i_j,...,i_N}=n^{j}_{i_1,...,i_N},
\end{align}
which are directly connected to defining multilinear products. In order to reproduce \eqref{N-uplift} we progressively remove all splits previously introduced using the necessary
set of concatenations products $\bnabla_p $ and we end up with \eqref{N-uplift}
\begin{equation}\label{N_CO_UNIT2}
    \begin{split}\bnabla_{l_1}\qty(\id_{l_1}\tp'_{l_1}\bnabla_{l_2})...\qty(\id_{l_1}\tp'_{l_1}\bDelta_{l_2})\bDelta_{l_1}&\mathbf{n}^{j}_{i_1,...,i_N}\pi_{n_1,..,n_N}=\\
    &=\sum_{j_1=0}^{n_1-i_1}...\sum_{j_N=0}^{n_N-i_N}\bar{\id}^{j_1,...,j_N}\tpb n^j_{i_1,...,i_N} \tpb\bar{\id}^{ n_1-j_1-i_1,..., n_N-j_N-i_N},
    \end{split}
\end{equation}
where in order to only find trivial splits we need to apply $\bDelta_j$ $n_j-1$ times for all $j\in\qty[1,N]$ \footnote{Explicit example are provided in appendix \ref{B1}}.\\
Equation \eqref{N_CO_UNIT2} proves that \eqref{N-uplift} not only is the right uplift procedure from $\Hom(\tHt,\Hs_j)$ to $\Coder(\tHt)$ but it is also unique, inducing the isomorphisms
\[
    \begin{tikzcd}
        \Coder^{j}_{i_1,...,i_N}(\tHt) \arrow[r, bend left= 50,"\pi^{(j)}"] & \Hom^{j}_{i_1,...,i_N}(\tHt) \arrow[l, bend left= 50,"\iota"]
    \end{tikzcd}
\]
\subsection{Co-homomorphisms and Cyclicity}
Co-homomorphisms of  $\tHt$ can be straightforwardly defined via the exponentiation of co-derivations, like in \ref{COHOM}
\begin{align}
    \mathbf{F}_{\varepsilon}:=\exp(\varepsilon\boldsymbol{n}),
\end{align}
where the co-derivation $\boldsymbol{n}$ is a generic element of $\Coder$
\begin{align}\label{generic-proper-co-der}
    \boldsymbol{n}:= \sum_{i_1,...,i_N=0}^{\infty}\sum_{j=1}^{N}\alpha_{i_1,...,i_N}^j\boldsymbol{n}^{j}_{i_1,...,i_N},\,\,\,\,\,\,\alpha_{i_1,...,i_N}^j\in\cc,\,\,\,\,\,\boldsymbol{n}^{j}_{i_1,...,i_N}\in\Coder.
\end{align}
If the base co-algebras $\tH_j$ are equipped with symplectic forms $\omega_j$ it's possible to define a general notion of cyclicity on the $N$ co-algebra $\tHt$.\\
Let $\omega_j$, or equivalently $\bra{\omega_j}$, be a non degenerate symplectic form such that
\begin{align}
    \omega_j:\Hs_j\tp_j\Hs_j\longrightarrow\cc,\,\,\,\omega_j\in\Hom(\Hs^{\tp_j 2}_j,\cc)\simeq\Hom(\Hst^{0_1,...,2_j,...,0_N},\cc).
\end{align}
The study of cyclicity on $\tHt$ is analogous to the study of cyclicity in normal co-algebras \ref{COHOM}.\\
Given a complete set of base symplectic forms $\qty{\omega_j}_{1\leq j\leq N}$ we proceed to define the symplectic form of $\tHt$ as linear combination of the base symplectic forms
\begin{align}
    \omega:=\sum_{j=1}^{N}c_j\omega_j,\,\,\,\,\,c_j\in\cc.
\end{align}
Remember that, just like for \eqref{OMREP}, the symplectic forms $\omega$ allows for more equivalent representations
\begin{align}
        v_1,v_2\in\Hst^{(1)}\,\,\,\omega(v_1,v_2)=\bra{\omega}\,\ket{v_1}\tpb\ket{v_2}=\bra{\omega}\,\ket{v_1\tpb v_2}.
\end{align}
The cyclicity of a co-homomorphism now is completely analogous to the cyclicity requirement in a normal co-algebra \eqref{cohom-cycl} and reads
\begin{align}\label{cycl-coder-N}
    \bra{\omega}\,\pi_{(2)}\mathbf{\mathbf{F}}=\bra{\omega}\,\pi_{(2)},\,\,\,\text{equivalently}\,\,\,\,\bra{\omega}\,\pi_{(1)}\mathbf{F}\tpb\pi_{(1)}\mathbf{F}=\bra{\omega}\,\pi_{(1)}\tpb\pi_{(1)},
\end{align}
If the co-homomorphisms is defined by exponentiating a co-derivation $\boldsymbol{n}$, together with two auxiliary co-derivations $(\boldsymbol{a},\boldsymbol{b})$ and their respective parameters $(\delta_1,\delta_2)$, we recover a similar definition to \eqref{co-der-cycl}
\begin{align}
    \bra{\omega}\,(\pi_{(1)}e^{\varepsilon \boldsymbol{n}})\tpb(\pi_{(1)}e^{\varepsilon \boldsymbol{n}})(e^{\delta_{1} {\mathbf{a}}}\tpb e^{\delta_2 {\mathbf{b}}})=\bra{\omega}\,(\pi_{(1)}\tpt\pi_{(1)})(e^{\delta_{1} {\mathbf{a}}}\tpb e^{\delta_2 {\mathbf{b}}}).
\end{align}
If we expand order by order is in the parameters $(\varepsilon,\delta_1,\delta_2)$, we will find the usual definition of cyclicity when the co-derivation output in the same base co-algebra $j$
\begin{equation}\label{NCYCLjj}
        \begin{split}
            &\mathcal{O}((\delta_1)^0,(\delta_2)^0)\Longrightarrow\,\,\,\omega_j(\pi_{(1)}\mathbf{d}^j\bgl,\pi_{(1)}\bgl)=-\omega_j(\pi_{(1)}\bgl,\pi_{(1)}\mathbf{d}^j\bgl),\\
            &\mathcal{O}(\varepsilon^1,(\delta_1)^1,(\delta_2)^1)\Longrightarrow\,\,\,\omega_j(\pi_{(1)}\mathbf{d}^j\mathbf{a}^j\bgl,\pi_{(1)}\mathbf{b}^j\bgl)=-(-1)^{d(\mathbf{d}^j)d(\mathbf{a}^j)}\omega_j(\pi_{(1)}\mathbf{a}^j\bgl,\pi_{(1)}\mathbf{d}^j\mathbf{b}^j\bgl).
        \end{split}
    \end{equation}
Additionally there are mixed relations for co-derivations outputting in different base co-algebras
\begin{align}\label{NCYCLij}
    \mathcal{O}(\varepsilon^1,(\delta_1)^1,(\delta_2)^1)\Longrightarrow\,\,\,c_j\omega_j(\pi_{(1)}\mathbf{d}^j\mathbf{a}^k\bgl,\pi_{(1)}\mathbf{b}^j\bgl)=-(-1)^{d(\mathbf{d}^j)d(\mathbf{a}^j)}c_k\omega_k(\pi_{(1)}\mathbf{a}^k\bgl,\pi_{(1)}\mathbf{d}^k\mathbf{b}^j\bgl).
\end{align}
The mixed cyclic relations between $\bar{\mathbf{d}}^j$ and ${\mathbf{d}}^k$ are naturally found in the study of open-closed SFT. The mixed relations account for the dual description of interaction vertices by open and closed strings \cite{Maccaferri:2022yzy,Maccaferri:2023gcg}. Furthermore, when dealing with world-sheet topologies with many boundaries, the mixed relations also account for the independent choice of boundary \cite{Maccaferri:2023gcg}. The co-algebraic formulation lets reinterpret many important dualities in SFT as a direct consequence of its cyclical algebraic structure.

\subsection{Co-derivation algebra and homotopy algebras}\label{NHOMALG:SUB}
Instead of following what has been done in \ref{hom-alg}, we can directly define a graded commutator over elements of $\Coder$
\begin{align}\label{co-der-alge}
    \commutator{\boldsymbol{c}}{\boldsymbol{d}}=\boldsymbol{c}\boldsymbol{d}-(-1)^{d(\boldsymbol{c})d({\boldsymbol{d})}}\boldsymbol{d}\boldsymbol{c},
\end{align}
which leads to the definition of the product between two multilinear products 
\begin{align}\label{NProdexpl}
    c^{j}_{i_1,...,i_N}d^q_{p_1,...,p_N}:=\sum_{l_1=0}^{i_1+p_1}...\sum_{l_q=0}^{i_q+p_q-1}...\sum_{l_N=0}^{i_N+p_N}c^{j}_{i_1,...,i_N}\qty(\bar\id_{l_1,...,l_N}\tpb d^q_{p_1,...,p_N}\tpb\bar\id_{i_1+p_1-l_1,...,i_N+p_N-l_N}),
\end{align}
by properly projecting and factoring out unnecessary elements of \eqref{co-der-alge} .\\
Thanks to the commutator we can define Homotopy algebras on a generalized $N$ tensor algebra $\tHt$ as a graded odd co-derivation $\boldsymbol{d}$ which obeys
\begin{align}\label{NHOM}
    \frac{1}{2}\commutator{\boldsymbol{d}}{\boldsymbol{d}}= \qty(\boldsymbol{d})^2 = 0,\,\,\,\,\boldsymbol{d}\in\Coder,
\end{align}
just like in \eqref{Coder_alg}.\\
Note that, when expressing 
\begin{align}
    \boldsymbol{d}:=\sum_{j=1}^{N}\sum_{n_1,...,n_N=0}^{\infty}\boldsymbol{d}^j_{n_1,...,n_N},
\end{align}
the homotopy algebraic structure factors into $N$ $A_\infty/L_\infty$ algebras when fed only from a single base tensor algebra $\tH_j$
\begin{align}\label{HAFAC}
    \boldsymbol{d}\boldsymbol{d}\pi_{0_1,...,n_j,...,0_N}=0\,\Longrightarrow \sum_{k_j=0}^{n_j}\boldsymbol{d}^j_{0_1,...,k_j,...,0_N}\boldsymbol{d}^j_{0_1,...,n_j-k_j,...,0_N}=0\,\,\,\,\forall j\in\qty[1,N].
\end{align}
The factored sub $A_\infty/L_\infty$ homotopy algebras when isolated give rise to physical\footnote{ They satisfy the Classical BV master equation.}  self interacting field theories.
\subsection{Homotopy transfer theorem}
Since on $\tHt$ it is possible to define homotopy algebras, in order to extract EFTs and observables from a CAFT we need to extend the validity of the homotopy transfer theorem to $N$-component tensor co-algebras.\\
Starting from the projector $P_i$ required to use the homotopy transfer theorem \ref{HTT} for the base Hilbert space $\Hs_i$ and its co-algebraic extensions $\boldsymbol{P}_i$
\begin{equation}
    \begin{split}
        P_i:\Hs_i\longrightarrow\Hs_{i,P_i},\,\,\,\,\,P_i:\tHt_i\longrightarrow\tHt_{i,P_i},
    \end{split}
\end{equation}
we can build a projector map on $\tHt$ as
\begin{equation}
    \begin{split}
        &\bar{\boldsymbol{P}}:=\boldsymbol{P}_1\tpt...\tpt\boldsymbol{P}_N,\,\,\,\,\,\,
        \bar{\boldsymbol{P}}:\tHt\longrightarrow\tH_{1,P}\tpt...\tpt\tH_{N,P}:=\tHt_P,\\
        &\bar{P}:=\sum_{j=1}^{N} P_j,\,\,\,\,\bar{P}:\tilde{\Hs}\longrightarrow\,\tilde{\Hs}_P.
    \end{split}
\end{equation}
Note that in most SFT literature the symbol $\bar{P}$ and $\bar{\boldsymbol{P}}$ are used in the homotopy transfer theorem context \ref{HTT} in order to indicate the the complementary part to the projection $P$ as in $\bar{P}=1-P$.\\
An alternative formulation of $\bar{\boldsymbol{P}}$, which will greatly simplify the following proofs and definitions, it is given by uplifting the individual $\boldsymbol{P}_i$ to operators $\bar{\boldsymbol{P}}_i$ acting on $\tHt$
\begin{align}
    \bar{\boldsymbol{P}}_i:=\boldsymbol{I}_1\tpt...\tpt\boldsymbol{P}_i\tpt...\tpt\boldsymbol{I}_N,
\end{align}
where $\boldsymbol{I}_i$ is the identity co-homomorphism of $\tH_i$
\begin{align}
    \boldsymbol{I}_i:=\sum_{n=0}^{\infty}\id_i^{\tp_i n}\pi^i_n,\,\,\,\,\boldsymbol{I}_i:\tHt_i\longrightarrow\,\tHt_i.
\end{align}
The uplift $\bar{\boldsymbol{P}}_i$ allows us to express $\bar{\boldsymbol{P}}$ as the successive action of $N$ mutually commuting operators on $\tHt$
\begin{align}\label{HTTOF}
    \bar{\boldsymbol{P}}:=\bar{\boldsymbol{P}_1}...\bar{\boldsymbol{P}_N},\,\,\,\,\commutator{\bar{\boldsymbol{P}}_i}{\bar{\boldsymbol{P}}_j}=0\,\,\,\forall\,i,j\in\qty[1,N].
\end{align}
It will prove useful to extend also the contracting homotopy maps $\boldsymbol{h}_i$ to graded $-1$ operators on $\tHt$
\begin{align}
    \bar{\boldsymbol{h}}_i:=\boldsymbol{I}_1\tpt...\tpt\boldsymbol{h}_i\tpt...\boldsymbol{I}_N,\,\,\,\,\,\commutator{\bar{\boldsymbol{h}}_i}{\bar{\boldsymbol{h}}_j}=0\,\,\forall\,i,j\in\qty[1,N].
\end{align}
Let us now turn our attention to the physical content of any QFT, i.e. the graded $1$ co-derivation $\boldsymbol{D}$ which usually contains the interaction terms $\boldsymbol{B}$ and a differential $\boldsymbol{\partial}$ such that
\begin{align}
    \boldsymbol{D}:=\boldsymbol{\partial}+\boldsymbol{B},\,\,\,\,\boldsymbol{D}\boldsymbol{D}=0\Rightarrow\,\boldsymbol{\partial}\boldsymbol{B}+\boldsymbol{B}\boldsymbol{\partial}+\boldsymbol{B}\boldsymbol{B}=0.
\end{align}
Because of \eqref{HAFAC} we also factor out $N$ $A_\infty/L_\infty$ algebras $\boldsymbol{D}_j$ from $\boldsymbol{D}$ 
\begin{align}
    \boldsymbol{D}:=\sum_{j=1}^N\boldsymbol{D}_j+{\rm mixed\,\,interactions},
\end{align}
which can be further factored in a differential $\boldsymbol{\partial}_j$ and a self interaction structure $\boldsymbol{B}_j$
\begin{align}
    \boldsymbol{D}_j\boldsymbol{D}_j=0,\,\,\,\,\boldsymbol{D}_j:=\boldsymbol{\partial}_j+\boldsymbol{B}_j\,\,\,\forall \,j\in\qty[1,N].
\end{align}
The factored $N$ homotopy algebras provide us with the link with homotopy transfer theorem \ref{HTT}.We recognise that the factored $A_\infty/L_\infty$ structures on $\tH_j$ are the usual structures transferred by the homotopy transfer theorem \ref{HTT}. Therefore, in order to transfer all $N$ $\boldsymbol{D}_j$ from $\tHt$ onto $\boldsymbol{D}_j'$ on $\tHt_P$, using \ref{HTT}, we require that
\begin{equation}\label{HTTNCOND}
    \begin{split}
        &P_i:\Hs_i\longrightarrow\Hs_{P,i},\,\,\,\,P_i\partial_i=\partial_iP_i,\\&\id_i=P_i+\alpha_i\qty{\partial_ih_i+h_i\partial_i},\,\,\,\,\,
        P_ih_i=h_iP_i=h_i^2=0\,\, \forall i\in\qty[1,N],
    \end{split}
\end{equation}
where the differentials $\partial_i$ are defined as
\begin{align}
    \partial_j:= \pi^{(j)} \boldsymbol{\partial}^{j},\,\,\,\,\,\,\partial_j:\Hs_j\longrightarrow\Hs_j.
\end{align}
The conditions \eqref{HTTNCOND} can be easily transferred on the operators $\bar{\boldsymbol{P}}_i$ as follows
\begin{equation}\label{CONHTTC}
    \begin{split}
        &\commutator{\bar{\boldsymbol{P}}_i}{\bar{\boldsymbol{h}}_j}=\bar{\boldsymbol{P}}_i\bar{\boldsymbol{h}}_j=\bar{\boldsymbol{h}}_i\bar{\boldsymbol{P}}_j=\bar{\boldsymbol{h}}_i\bar{\boldsymbol{h}}_j=0,\\
        &\bar{\id}=\bar{\boldsymbol{P}}_i+\alpha_i\qty{\boldsymbol{\partial}_i\bar{\boldsymbol{h}}_i+\bar{\boldsymbol{h}}_i\boldsymbol{\partial}_i},\,\,\,\,\forall\,i,j\in\qty[1,N].
    \end{split}
\end{equation}
 It is possible now to recognise the form of the differential $\partial$, and $\boldsymbol{\partial}$ thanks to \eqref{HTTNCOND}
\begin{align}
    \boldsymbol{\partial}:=\sum_{i=1}^N\boldsymbol{\partial}_i\,\,\,\,\,\,\,\,\partial:=\pi_{(1)}\boldsymbol{\partial}=\sum_{i=1}^N\partial_i.
\end{align}
Thanks to \eqref{CONHTTC} we recognize that $\bar{\boldsymbol{P}}$ is a chain map and maps the co-homology of $\boldsymbol{D}$ to the co-homology of $\boldsymbol{D}'$
\begin{align}\label{N-cohom-map}
    \boldsymbol{D}':=\boldsymbol{D}\bar{\boldsymbol{P}},\,\,\,\,\,{\rm H}(\boldsymbol{D}'):=\eval{\frac{{\rm ker}(\boldsymbol{D}')}{{\rm im}(\boldsymbol{D}')}}_{\tHt_P}=\eval{\frac{{\rm ker}(\boldsymbol{D} \bar{\boldsymbol{P}})}{{\rm im}(\boldsymbol{D} \bar{\boldsymbol{P}})}}_\tHt.
\end{align}
In order to consistently map co-homological data from $\tHt$ to $\tHt_P$ we require that the co-homologies of $\boldsymbol{D}$ and $\boldsymbol{D}'$ are isomorphic, therefore $\bar{\boldsymbol{P}}$ obeys the Hodge-Kodaira decomposition
\begin{align}
    \bar{\id}=\bar{\boldsymbol{P}}+\bar{\alpha}\qty{\boldsymbol{\partial}\bar{\boldsymbol{h}}+\bar{\boldsymbol{h}}\boldsymbol{\partial}}\,\Longrightarrow\,{\rm H}(\boldsymbol{D})\sim {\rm H}(\boldsymbol{D}'),\,\,\,\,\,\bar{\alpha}\in\cc,
\end{align}
hence the homotopy transfer has been established.
Therefore the homotopy transfer theorem for $N$ component tensor co-algebras, which is one of the main results of this work, can be stated as follows:
\begin{theorem}\label{NHTT}
    Let $\tHt$ be a $N$ component tensor co-algebra built from $N$ base co-algebras $\tH_i$, each equipped with a projector map $\boldsymbol{P}_i$ and a contracting homotopy map $\boldsymbol{h}_i$ satisfying the homotopy transfer theorem on $\tH_i$.
    Let $\bar{\boldsymbol{P}}$ be a projector and chain map that satisfies the Hodge-Kodaira decomposition
    \begin{align}
        \bar{\boldsymbol{P}}:=\boldsymbol{P}_1\tpt...\tpt\boldsymbol{P}_N,\,\,\,\,\,\,
        \bar{\boldsymbol{P}}:\tHt\longrightarrow\tHt_P,\,\,\,\,\commutator{\boldsymbol{\partial}}{\bar{\boldsymbol{P}}}=0,\,\,\,\,\bar{\id}=\bar{\boldsymbol{P}}+\bar{\alpha}\qty{\boldsymbol{\partial}\bar{\boldsymbol{h}}+\bar{\boldsymbol{h}}\boldsymbol{\partial}},\,\,\,\bar{\alpha}\in\cc,
    \end{align}
    together with a contracting homotopy map $\bar{\boldsymbol{h}}$ and differential $\boldsymbol{\partial}$  defined on $\tHt$.\\
    Then there exists a morphism that transfers the homotopy algebraic structure on $\tHt$ to an homotopy algebraic structure on the restriction $\tHt_P$, keeping the co-homologies isomorphic to one-another. 
\end{theorem}
In order to build the transferring morphism, let us explicitly write down the homotopy algebraic structure on $\tHt$ highlighting the differential $\boldsymbol{\partial}$
\begin{align}
    \boldsymbol{D}:=\boldsymbol{\partial}+\boldsymbol{B},\,\,\,\,\boldsymbol{D}^2=0,
\end{align}
and the transferred homotopy algebraic structure on $\tHt_P$
\begin{align}
    \boldsymbol{D}':=\boldsymbol{\partial}'+\boldsymbol{B}',\,\,\,\,(\boldsymbol{D}')^2=0,
\end{align}
By following the derivation of the morphisms maps of \ref{HTT} in \ref{Appendix_Hom}, working with $\bar{\boldsymbol{P}}$ and $\bar{\boldsymbol{h}}$ instead of $\boldsymbol{P}$ and $\boldsymbol{h}$ we recover that the morphism has properties
\begin{align}\label{NHTTMORPH}
  \bar{\boldsymbol{F}}\bar{\boldsymbol{F}}'=\bar{\id},\,\,\,\,\bar{\boldsymbol{F}}\boldsymbol{\partial}\bar{\boldsymbol{F}}'=\boldsymbol{\partial}'.
\end{align}
and the same structure of \eqref{HTTMORPH} where we recall the two most popular choices of the $\boldsymbol{F}$ and $\tilde{\boldsymbol{F}}$ 
\begin{align}\label{NMorph1}
    \bar{\boldsymbol{F}}:=\bar{\boldsymbol{P}}\frac{1}{1-\bar{\alpha}\boldsymbol{B}\bar{\boldsymbol{h}}},\,\,\,\,\,\,\bar{\boldsymbol{F}}'=(1-\bar{\alpha}\boldsymbol{B}\bar{\boldsymbol{h}})\bar{\boldsymbol{P}},
\end{align}
and
\begin{align}\label{NMorph2}
    \bar{\boldsymbol{F}}:=\bar{\boldsymbol{P}}(1-\bar{\alpha}\bar{\boldsymbol{h}}\boldsymbol{B}
    ),\,\,\,\,\,\,\bar{\boldsymbol{F}}'=\frac{1}{1-\bar{\alpha}\bar{\boldsymbol{h}}\boldsymbol{B}}\bar{\boldsymbol{P}}.
\end{align}
In order to derive the action of $\bar{\boldsymbol{h}}$ on elements of $\tHt$ we choose, for simplicity, to work with the operator formulation of $\bar{\boldsymbol{P}}$ \eqref{HTTOF}. Using \eqref{HTTOF} allows for a more compact derivation and result. Let us start by taking the Hodge-Kodaira decomposition of $\bar{\boldsymbol{P}}$ rewritten in terms of the operators $\bar{\boldsymbol{P}}_i$ and highlighting $\bar{\boldsymbol{h}}$
\begin{align}
    \bar{\alpha}\commutator{\boldsymbol{\partial}}{\bar{\boldsymbol{h}}}=\bar{\id}-\bar{\boldsymbol{P}}_1...\bar{\boldsymbol{P}}_N.
\end{align}
In order to isolate the single contacting homotopy map $\bar{\boldsymbol{h}}_j$ let us subtract the Hodge-Kodaira decomposition of a chosen $\bar{\boldsymbol{h}}_j$
\begin{align}
    \bar{\alpha}\commutator{\boldsymbol{\partial}}{\bar{\boldsymbol{h}}}=\bar{\id}-\bar{\boldsymbol{P}}_1...\bar{\boldsymbol{P}}_N-\bar{\id}+\bar{\boldsymbol{P}}_j+\alpha_j\commutator{\boldsymbol{\partial}_j}{\bar{\boldsymbol{h}}_j}\Rightarrow
    \bar{\alpha}\commutator{\boldsymbol{\partial}}{\bar{\boldsymbol{h}}}=\bar{\boldsymbol{P}}_j(1-\prod_{i\neq j}\bar{\boldsymbol{P}}_i)+\alpha_j\commutator{\boldsymbol{\partial}_j}{\bar{\boldsymbol{h}}_j}.
\end{align}
Because of \eqref{CONHTTC} we can upgrade $\commutator{\boldsymbol{\partial}_j}{\bar{\boldsymbol{h}}_j}$ to $\commutator{\boldsymbol{\partial}}{\bar{\boldsymbol{h}}_j}$ leading to
\begin{align}\label{NHPART}
    \bar{\alpha}\commutator{\boldsymbol{\partial}}{\bar{\boldsymbol{h}}}=\bar{\boldsymbol{P}}_j(1-\prod_{i\neq j}\bar{\boldsymbol{P}}_i)+\alpha_j\commutator{\boldsymbol{\partial}}{\bar{\boldsymbol{h}}_j}.
\end{align}
Let us now rewrite the product of $N-1$ $\bar{\boldsymbol{P}}_i$ operators using the Hodge-Kodaira decomposition in combination with \eqref{CONHTTC}
\begin{align}
    \prod_{\substack{i=1\\i\neq j}}^N\bar{\boldsymbol{P}}_i=\prod_{\substack{i=1\\i\neq j}}^N(\bar{\id}-\alpha_i\commutator{\boldsymbol{\partial}_i}{\bar{\boldsymbol{h}}_i})=\prod_{\substack{i=1\\i\neq j}}^N(\bar{\id}-\alpha_i\commutator{\boldsymbol{\partial}}{\bar{\boldsymbol{h}}_i}),
\end{align}
and by using the following identity
\begin{align}
    \prod_{\substack{i=1\\i\neq j}}^{N}(1-a_i)=1-\sum_{\substack{i=1\\ i\neq j}}^N a_i \prod_{\substack{l>i\\ l \neq j}}^N(1-a_l),
\end{align}
we rewrite the parenthesis of \eqref{NHPART} to
\begin{align}
    (1-\prod_{\substack{i=1\\i\neq j}}^N\bar{\boldsymbol{P}}_i)=\sum_{\substack{i=1\\ i\neq j}}^N \alpha_i\commutator{\boldsymbol{\partial}}{\bar{\boldsymbol{h}}_i}) \prod_{\substack{l>i\\ l \neq j}}^N(\bar{\id}-\alpha_l\commutator{\boldsymbol{\partial}}{\bar{\boldsymbol{h}}_l})=\sum_{\substack{i=1\\ i\neq j}}^N \alpha_i\commutator{\boldsymbol{\partial}}{\bar{\boldsymbol{h}}_i} \prod_{\substack{l>i\\ l \neq j}}^N\bar{\boldsymbol{P}}_l.
\end{align}
Thanks to the last manipulation we highlighted the graded commutator and differential on all elements of \eqref{NHPART} leading to
\begin{align}\label{NHPART2}
    \bar{\boldsymbol{h}}=\frac{\alpha_j}{\bar{\alpha}}\bar{\boldsymbol{h}}_j+\sum_{\substack{i=1\\ i\neq j}}^N \frac{\alpha_i}{\bar{\alpha}}\bar{\boldsymbol{h}}_i \bar{\boldsymbol{P}}_j\prod_{\substack{l>i\\ l \neq j}}^N\bar{\boldsymbol{P}}_l,
\end{align}
or the democratic version, averaging over all choices of $j$ in \eqref{NHPART2}
\begin{align}\label{NHPARTDEM}
    \bar{\boldsymbol{h}}=\frac{1}{N}\sum_{j=1}^N\qty{\frac{\alpha_j}{\bar{\alpha}}\bar{\boldsymbol{h}}_j+\sum_{\substack{i=1\\ i\neq j}}^N \frac{\alpha_i}{\bar{\alpha}}\bar{\boldsymbol{h}}_i \bar{\boldsymbol{P}}_j\prod_{\substack{l>i\\ l \neq j}}^N\bar{\boldsymbol{P}}_l}.
\end{align}
Note that this result works only if we apply $\bar{\boldsymbol{h}}$ on elements $\Hst^{(n_1,..,n_N)}$ with all $n_i\neq0$, namely 
\begin{align}
    \bar{\boldsymbol{h}}\pi_{n_1,..,n_N}=\qty{\frac{\alpha_j}{\bar{\alpha}}\bar{\boldsymbol{h}}_j+\sum_{\substack{i=1\\ i\neq j}}^N \frac{\alpha_i}{\bar{\alpha}}\bar{\boldsymbol{h}}_i \bar{\boldsymbol{P}}_j\prod_{\substack{l>i\\ l \neq j}}^N\bar{\boldsymbol{P}}_l}\pi_{n_1,..,n_N}\,\,\,\text{if}\,\,\, n_i\neq0\,\forall i\in\qty[1,N],
\end{align}
because when $n_i=0$ the Hodge-Kodaira for $\tH_i$ decomposition degenerates to
\begin{align}
    \id_i\pi^i_0=\qty{P_i+\alpha_i\qty{\partial_ih_i+h_i\partial_i}}\pi^i_0\,\Longrightarrow\,\id_i\pi^i_0=P_i\pi^i_0.
\end{align}
In order to extend the validity of \eqref{NHPART2} and \eqref{NHPARTDEM} when working on subspaces with some $n_i=0$ we define the set
\begin{align}
    \Sigma:=\qty{i\,|\,i\in\qty[1,N]\,\text{and }\, n_i\neq0},
\end{align}
where $\Sigma$ correctly identifies when it is possible to apply the Hodge-Kodaira decomposition.
Relations \eqref{NHPART2} and \eqref{NHPARTDEM} are then directly extended by taking $i,j,l$ from $\Sigma$
\begin{equation}\label{NHPARTFull}
    \begin{split}
        &\bar{\boldsymbol{h}}=\frac{\alpha_j}{\bar{\alpha}}\bar{\boldsymbol{h}}_j+\sum_{\substack{i\in\Sigma\\ i\neq j}} \frac{\alpha_i}{\bar{\alpha}}\bar{\boldsymbol{h}}_i \bar{\boldsymbol{P}}_j\prod_{\substack{l\in\Sigma\\l>i\,,\, l \neq j}}\bar{\boldsymbol{P}}_l\,\,\,\,\,\,\,\forall j\in\Sigma,\\
    &\bar{\boldsymbol{h}}=\frac{1}{{\rm Dim}(\Sigma)}\sum_{j\in\Sigma}\qty{\frac{\alpha_j}{\bar{\alpha}}\bar{\boldsymbol{h}}_j+\sum_{\substack{i\in\Sigma\\ i\neq j}} \frac{\alpha_i}{\bar{\alpha}}\bar{\boldsymbol{h}}_i \bar{\boldsymbol{P}}_j\prod_{\substack{l\in\Sigma\\l>i\,,\, l \neq j}}\bar{\boldsymbol{P}}_l},
    \end{split}
\end{equation}
with a special case when all $n_i=0$ except one $n_j\neq0$ where \eqref{NHPARTFull} reduces to 
\begin{align}
    \bar{\boldsymbol{h}}=\frac{\alpha_j}{\bar{\alpha}}\bar{\boldsymbol{h}}_j.
\end{align}

\subsection{Degeneration problem of $N$-component homotopy algebras}\label{N-DEGEN:SEC}
A curious but expected problem arises when dealing with $N$ component tensor co-algebras with two or more matching base vector spaces $\Hs_i=\Hs_j=\Hs$. When considering multi-linear products with image in $\Hs$ the space $\Hom(\tHt,\Hs)$ does not close under the product operation
\begin{align}
    \cdot:\Hom(\tHt,\Hs_j)\cross\Hom(\tHt,\Hs_i)\longrightarrow V\nsubseteq\Hom(\tHt,\tHt).
\end{align}
This is tied to the fact that $\tpt$ was introduce to order and separate different field contents tided to different base spaces $\tH_i$ \ref{NCOALG:CH}. Whenever two or more distinct base vector spaces match, their formally distinct tensor product of tensor algebras $\tH_i\tpt \tH_i$ degenerates into a single tensor algebra $\tH_i\tpt \tH_i\simeq\tH_i$ due to the fact $\Hs_i\tpt\Hs_i\simeq\Hs_i\tp\Hs_i$. Therefore, if we keep the formal distinction between spaces when applying multiple multilinear products we will find out that there will be elements that do not covered by the algebra structure of multilinear products. 
An example of this behaviour can be seen in QFT and CFT when adding to higher trace operators.
\\
In order to better understand the underlying problem let us discuss the following example: let us have a degenerated $2$ component co-algebra
\begin{align}
    \tHt:=\tH\tpt\tH,
\end{align}
and let us take two simple multilinear products $c^{1}_{1,1}$ and $d^{2}_{1,1}$
\begin{align}
    c^{1}_{1,1}:\Hs\tpt\Hs\longrightarrow\Hs,\,\,\,\,d^{2}_{1,1}:\Hs\tpt\Hs\longrightarrow\Hs.
\end{align}
The $c^{1}_{1,1}$ and $d^{2}_{1,1}$ can be composed in the following way
\begin{align}\label{N-degen-prod}
    c^{1}_{1,1}\cdot d^{2}_{1,1}=c^{1}_{1,1}d^{2}_{1,1}=c^{1}_{1,1}\qty(d^{2}_{1,1}(\id\tpt\id)\tpt\id)+c^{1}_{1,1}\qty(\id\tpt d^{2}_{1,1}(\id\tpt\id)).
\end{align}
The result of the composition is an element from a space outside of $\Hom(\tHt,\tHt)$, in fact it is an element of
\begin{align}
    c^{1}_{1,1}d^{2}_{1,1}\in\Hom(\Hs\tpt\Hs\tpt\Hs,\Hs),
\end{align}
which is a natural element of the set of homomorphisms of a $3$ component degenerate co-algebra.\\
As a consequence of \eqref{N-degen-prod} the algebra of $\Hom(\tH^{\tpt 2},\Hs)$, and also of $\Coder(\tH^{\tpt 2})$, opens up
\begin{align}
    &\commutator{\cdot}{\cdot}:\Hom(\tH^{\tpt 2},\Hs)\cross\Hom(\tH^{\tpt 2},\Hs)\longrightarrow \Hom(\tH^{\tpt 3},\Hs),\\
    &\commutator{\cdot}{\cdot}:\Coder(\tH^{\tpt 2})\cross\Coder(\tH^{\tpt 2})\longrightarrow \Coder(\tH^{\tpt 3}).
\end{align}
More generally, if a $N$ component co-algebra has degeneration $p$, we separate the non degenerate portion by defining $\tHt$ as
\begin{align}
    \tHt:=\tHb\tpt\tH^{\tpt p},\,\,\,\,\tHb:=\tH_1\tpt...\tpt\tH_{N-p},
\end{align}
and the homomorphism algebra opens in the following way
\begin{align}
    &\commutator{\cdot}{\cdot}:\Hom(\tHt,\Hs_j)\cross\Hom(\tHt,\Hs)\longrightarrow \Hom(\tHb\tpt\tH^{\tpt 2p-1},\Hs_j\oplus\Hs).
\end{align}
This behaviour is physically relevant when studying quantum open-closed SFT where, when composing different interaction vertices with more then one boundary, where we associate to each boundary a $\tH$ space of open strings, it naturally leads to an open algebra of interaction vertices. Therefore, in order to systematically study the algebraic properties of open string interaction vertices with more then one boundary we necessarily need to include all possible vertices with arbitrary number of boundaries.\\
In order to study and fully understand the degeneration phenomenon we need to study a different type of tensor co-algebra, namely the $\infty$ component tensor co-algebra.    
    
\section{$\infty$ components tensor co-algebras}\label{INFCOALG:CH}
When studying quantum open-closed SFT a new special structure appears. When dealing with open strings on surfaces with many boundaries, for each boundary there is a copy of $\tH_{\rm o}$, carrying matrix structures to accommodate stretched strings , the open string tensor algebra \cite{Maccaferri:2023gcg}. When constructing the action of quantum open-closed SFT we need to sum over all possible surfaces with fixed number of boundaries, creating the tensor product space $\tHt_{\rm oc}$ 
\begin{align}
    \tHt_{\rm oc}:=\tH_{\rm c}\tpt_{\rm oc}\qty[\bigoplus_{b=0}^\infty\tH_{\rm o}^{\tpt b}],
\end{align}
where $\tH_{\rm c}$ is the closed string tensor algebra.\\
In order to define co-derivations on $\tHt_{\rm oc}$, and prove the consistency of the definitions presented in \cite{Maccaferri:2023gcg}, we need to understand how to build a co-algebra on
\begin{align}
    \ttHt_{\rm o}:=\bigoplus_{b=0}^\infty\tH_{\rm o}^{\tpt b}.
\end{align}
\subsection{Preliminaries: projector and inclusion maps}
Let's take a moment to define some necessary elements. The tensor product space we will be working with is
\begin{align}
    \ttHt:=\bigoplus_{b=0}^\infty\tH^{\tpt b}=\cc\oplus\tH\oplus\tH\tpt\tH\oplus\cdot\cdot\cdot\,.
\end{align}
On $\ttHt$ projectors and inclusion are defined as
\begin{equation}
    \begin{split}
        &\Pi^b:\ttHt\longrightarrow\tH^{\tpt b}\,,\,I^b:\tH\longrightarrow\tH^{\tpt b+1},\\
        &I^bV:=\sum_{j=0}^b\id^{\tpt j}\tpt V\tpt\id^{\tpt b-j}\,\,\,\forall V\in\tH,
    \end{split}
\end{equation}
and the total inclusion map
\begin{align}
    I:=\sum_{b=1}^{\infty}I^b,\,\,\,I:\tH\longrightarrow\ttHt.
\end{align}
In order to project onto specific sub-spaces $\Hst^{(n_1,...,n_b)}$ we only need compose projectors
\begin{align}
    \Pi^b_{n_1,...,n_b}:=\pi_{n_1,...,n_b}\Pi^b:\ttHt\longrightarrow\Hst^{(n_1,...,n_b)}.
\end{align}

\subsection{Double co-algebra structure}
Looking closely at $\ttHt$ we can recognise that $\tpt$ turns $\ttHt$ into a tensor algebra
\begin{align}
    \tpt:\ttHt\cross\ttHt\longrightarrow\ttHt.
\end{align}
Working in analogy with \ref{Sec_Co_algebras} we can turn $\ttHt$ into a tensor co-algebra by defining a co-product $\tDelta$
\begin{equation}\label{inf-coprod}
    \begin{split}
        &\tDelta:\ttHt\longrightarrow\ttHt\tpt'\ttHt,\\
        & \tDelta A_{1,b}=\sum_{i=0}^{b}A_{1,i}\tpt A_{i+1,b},\,\,\,\,\,A_{1,b}\in\tH^{\tpt b},
    \end{split}
\end{equation}
where $\tpt'\neq\tpt$ is the external tensor product. The co-product $\tDelta$ is co-associative if
\begin{align}
    (\tDelta\tpt'\tid)\tDelta=(\tid\tpt'\tDelta)\tDelta.
\end{align}
The tensor co-algebra $\qty(\ttHt,\tpt,\tDelta)$ is built on tensor powers of $\tH$, like $\tH$ is built on tensor powers of $\Hs$, which hides many informations needed when defining co-derivations. In order to consistently define co-derivations on $\ttHt$ we need to realize that there is an additional co-algebra structure. Provided that $\qty(\tH,\tp,\Delta)$ is a co-algebra, thanks to the discussion in section \ref{NCOALG:CH}, any specific sub-space $\tH^{\tpt b}$ can be uplifted to a $b$ component tensor co-algebra where all the base vector spaces are the same ones
\begin{align}
    &\bDelta^{(b)}:=\Omega_b\underbrace{(\Delta\tpt...\tpt\Delta)}_{b},\,\,\,\,\,\bDelta^{(b)}:\tH^{\tpt b}\longrightarrow\tH^{\tpt b}\tpb'\tH^{\tpt b}.
\end{align}
Therefore $\ttHt$ can be uplifted to a co-algebra $\qty(\ttHt,\tpt,\tDelta)$ and independently every subspace $\tH^{\tpt b}$ can be uplifted to a $b$ component co-algebra $\qty(\tH^{\tpt b},\tpb,\bDelta^{(b)})$. This double co-algebra structure introduces a series of constraints when defining group like elements, co-derivations and other co-algebraic elements. As a notable example let us discuss the group like element $\tgl$ of $\ttHt$. From the point of view of $\qty(\ttHt,\tpt,\tDelta)$ for $\tgl$ to be a group like element it has to satisfy
\begin{align}
    \tDelta\tgl=\tgl\tpt'\tgl,\,\,\,\,\,\tgl:=\sum_{b=0}^{\infty}A^{\tpt b},\,A\in\tH,\,d(A)=0,
\end{align}
where $A$ is a generic graded $0$ element of $\tH$. The element $A$ is then fixed by requiring that, for any $b$ subspace $\tH^{\tpt b}$ of $\ttHt$, $A$ defines a group like element $\bgl^{(b)}$ of the associated $b$ component co-algebra. Therefore, due to \eqref{NCOGL}, $A$ has to be the group like element $\gl$ of the base co-algebra $\qty(\tH,\tp,\Delta)$
\begin{align}
    A=\gl=\sum_{n=0}^\infty \Psi^{\tp n},\,\,\,\,\,\Psi\in\Hs,\,d(\Psi)=0,
\end{align}
where $\Psi$ is a generic graded $0$ element of $\Hs$, which implies that 
\begin{align}\label{INFGL}
    \tgl := \sum_{b=0}^{\infty}\bgl^{(b)} = \sum_{b=0}^{\infty}\gl^{\tpt b}=\sum_{b=0}^\infty\qty[\sum_{p_1=0}^{\infty}...\sum_{p_b=0}^{\infty}\Psi^{\tp p_1}\tpt...\tpt \Psi^{\tp p_b}]= \frac{1}{1-\tpt\frac{1}{1-\tp\Psi}}.
\end{align}

\subsection{Multilinear products}
Let us start by defining multilinear products on $\ttHt$ as maps
\begin{align}
    c^b:\tH^{\tpt b}\longrightarrow\tH.
\end{align}
This definition is consistent if we treat $\tH^{\tpt b}$ like we treated $\Hs^{\tp n}$ in \ref{Sec_Co_algebras}. This partial definition will help us to identify the right co-Leibniz rules later in the paper.\\
The definition can be further refined by including more details 
\begin{align}
    c^{j,b}_{i_1,...,i_b;n}:\Hs^{\tp i_1}\tpt...\tpt\Hs^{\tp i_b}\longrightarrow\Hs^{\tp n}_j=\Hs^{\tp n},
\end{align}
where $j$ indicates the output to be on the $j$-th boundary.\\
Note that multilinear products on $\ttHt$ comprehend multilinear products and maps of $\tH^{\tpt b}$. This crucial detail makes it possible to define co-derivations on the $b$ component co-algebra \eqref{NCOPROD} and then transfer such properties to the double co-algebra on $\ttHt$.

\subsection{Co-derivations in the double co-algebra structure}\label{INFCODER:SEC}
A co-derivation of $\ttHt$, with respect to $\tDelta$, is a linear map $\boldsymbol{D}^b$
\begin{align}
    \boldsymbol{D}^b:\ttHt\longrightarrow\ttHt,
\end{align}
which satisfies the co-Leibniz rule
\begin{align}\label{infty-coleib}
    \tDelta \boldsymbol{D}^b = \qty(\tid\tpt'\boldsymbol{D}^b+\boldsymbol{D}^b\tpt'\tid)\tDelta,
\end{align}
where $\tid$ is the identity operator on $\ttHt$.\\
In a SFT, and more generally in a CAFT, only specific linear operators $\boldsymbol{D}^{j,b}_{i_1,...,i_b;1}$ will appear, which will be directly linked to multilinear products $c^{j,b}_{i_1,...,i_b;1}$. In order for $\boldsymbol{D}^{j,b}_{i_1,...,i_b;1}$ to be a co-derivation of $\ttHt$ it has to satisfy \eqref{infty-coleib} and simultaneously the modified $b$ co-Leibniz rule \eqref{NCOLEIB}
\begin{equation}\label{BCODERMOD}
    \begin{split}
    &\bDelta_j \boldsymbol{D}^{j,b}_{i_1,...,i_b;1}\Pi^b=\qty(\id_j\tp'_j\boldsymbol{D}^{j,b}_{i_1,...,i_b;1}+\boldsymbol{D}^{j,b}_{i_1,...,i_b;1}\tp'_j\id_j)\bDelta_j\Pi^b.
    \end{split}
\end{equation}
The modification to the co-Leibniz rule is introduced because 
\begin{align}
    \boldsymbol{D}^{j,b}_{i_1,...,i_b;1}:\tH^{\tpt b}\longrightarrow\tH,
\end{align}
therefore if it had to satisfy the $b$ co-Leibniz rules it would have to violate \eqref{infty-coleib} because to $\boldsymbol{D}^{j,b}_{i_1,...,i_b;1}\Pi^b$ we could directly associate a $b$ component co-derivation $\boldsymbol{d}^{j,b}_{i_1,...,i_b;1}$ and then we would find that
\begin{equation}
    \begin{split}
        \tDelta\boldsymbol{D}^{j,b}_{i_1,...,i_b;1}\Pi^b&=\sum_{i=0}^{b}\qty(\Pi^{i}\tpt'\boldsymbol{D}^{j,b}_{i_1,...,i_b;1}\Pi^{i}+\boldsymbol{D}^{j,b}_{i_1,...,i_b;1}\Pi^{i}\tpt'\Pi^{b-i})\tDelta\\
        &=\qty(\Pi^{0}\tpt'\boldsymbol{D}^{j,b}_{i_1,...,i_b;1}\Pi^{b}+\boldsymbol{D}^{j,b}_{i_1,...,i_b;1}\Pi^{b}\tpt'\Pi^{0})\tDelta\\
        &=\qty(\Pi^{0}\tpt'\boldsymbol{d}^{j,b}_{i_1,...,i_b;1}\Pi^{b}+\boldsymbol{d}^{j,b}_{i_1,...,i_b;1}\Pi^{b}\tpt'\Pi^{0})\tDelta\neq\tDelta\boldsymbol{d}^{j,b}_{i_1,...,i_b;1}\Pi^{b}=\boldsymbol{D}^{j,b}_{i_1,...,i_b;1}\Pi^{b},\\
        &\Longrightarrow \tDelta\boldsymbol{D}^{j,b}_{i_1,...,i_b;1}\Pi^b\neq \tDelta\boldsymbol{D}^{j,b}_{i_1,...,i_b;1}\Pi^b.
    \end{split}
\end{equation}
which is a contradiction. Although we can associate a $b$ component co-derivation to $\boldsymbol{D}^{j,b}_{i_1,...,i_b;1}$ in the following way
\begin{align}
    \boldsymbol{D}^{j,b}_{i_1,...,i_b;1}\Pi^b_{k_1,..,k_b}=\pi_{0_1,...,k_1-i_j+1,...,0_b}\boldsymbol{d}^{j,b}_{i_1,...,i_b;1}\Pi^b_{k_1,..,k_b},
\end{align}
which does satisfy both \eqref{infty-coleib} and \eqref{BCODERMOD}.\\
Following a similar reasoning of \ref{NCODER:SEC} a multilinear product $d^{j,b}_{k_1,..,k_b;1}$ can be uplifted to a co-derivation in the following way
\begin{equation}\label{INFTY_UPLIFT}
    \begin{split}
        \boldsymbol{D}^{j,b}_{k_1,..,k_b;1}\Pi^{b'}_{p_1,...,p_{b'}}&:=\sum_{i=0}^{b'-b}\tid^{\tpt i}\tpt\qty[\Pi^1
        \boldsymbol{d}^{j,b}_{k_1,..,k_b;1}
        ]\tpt\tid^{\tpt b'-i}\\
        &:=\sum_{i=0}^{b'-b}\tid^{\tpt i}\tpt\qty[\sum_{q_j = 0}^{p_{j+i}-k_j}\id_j^{\tp_j q_j}\tp_j
        d^{j,b}_{k_1,..,k_b;1}\tp_j\id_j^{\tp_j p_{j+i} - q_j}\circ\pi_{k_1,...,p_{j+1},...,k_b}
        ]\tpt\tid^{\tpt b'-i},
    \end{split}
\end{equation}
where $\boldsymbol{d}^{j,b}_{k_1,..,k_b;1}$ is the $b$ component co-derivation defined in \eqref{N-uplift} and the last formula is the explicit definition starting from the multilinear product $d^{j,b}_{k_1,..,k_b;1}$.\\
Definition \eqref{INFTY_UPLIFT} matches the definition of the co-derivation like objects in \cite{Maccaferri:2023gcg} proving that they are in fact fully fledged co-derivations.\\
Co-derivations \eqref{INFTY_UPLIFT} are also well behaved when acting on the group like element $\tgl$
\begin{align}
    \boldsymbol{D}^{j,b}_{k_1,..,k_b;1}\tgl = \tgl\tpt\qty[\Pi^1_1\boldsymbol{D}^{j,b}_{k_1,..,k_b;1}\tgl]\tpt\tgl.
\end{align}
From now onwards, if not specified otherwise, we omit the index $\qty{;1}$ when referring to co-derivations and multilinear products on $\ttHt$
\begin{align}
    &c^{j,b}_{k_1,..,k_b;1}\longrightarrow c^{j,b}_{k_1,..,k_b},
    \\
    &\boldsymbol{D}^{j,b}_{k_1,..,k_b;1}\longrightarrow\boldsymbol{D}^{j,b}_{k_1,..,k_b}.
\end{align}

\subsection{Co-homomorphisms and cyclicity}
Thanks to \ref{INFCODER:SEC} co-homomorphisms of $\ttHt$ are  defined by exponentiating a co-derivation $\boldsymbol{D}$ together with a graded parameter $\varepsilon$
\begin{align}
    \tilde{\boldsymbol{F}}_{\varepsilon}=\exp(\varepsilon \boldsymbol{D}).
\end{align}
Generally $\boldsymbol{D}$ is a generic element of the space of co-derivations of $\Coder(\ttHt)$ 
\begin{align}
    \boldsymbol{D}:=\sum_{b=0}^{\infty}\sum_{j=1}^{b}\sum_{i_1,...,i_b=0}^{\infty} \alpha^{j,b}_{i_1,...,i_b}\boldsymbol{D}^{j,b}_{i_1,...,i_b},\,\,\,\,\,\alpha^{j,b}_{i_1,...,i_b}\in\cc.
\end{align}
By introducing a non-degenerate symplectic form $\bra{\omega}$ on $\tH$ we endow all the $b$ co-algebras with a symplectic form $\bra{\omega}^b$, and we endow $\ttHt$ with a symplectic form as well. The notion of cyclicity for a co-homomorphisms on $\ttHt$ is defined as
\begin{align}
    \bra{\omega}\,\Pi^1_2\tilde{\boldsymbol{F}} = \bra{\omega}\,\Pi^1_2,
\end{align}
in direct analogy with a normal co-algebra. This definition of cyclicity could be generalised by introducing multi-symplectic forms that act on higher projections $\Pi^b\ttHt$, but this topic is outside the scope of this paper.\\
Following the definition of cyclicity for the co-derivations provided in \eqref{co-der-cycl} and \eqref{cycl-coder-N} we provide the definition of a cyclical co-derivation of $\ttHt$. Given three co-derivations $\boldsymbol{D},\boldsymbol{A}$ and $\boldsymbol{B}$ together with graded parameters $\varepsilon,\delta_1$ and $\delta_2$, $\boldsymbol{D}$ is cyclical if and only if
\begin{align}
    \bra{\omega}(\Pi^1_1e^{\varepsilon \boldsymbol{D}})\tp(\Pi^1_1e^{\varepsilon \boldsymbol{D}})(e^{\delta_1 \boldsymbol{A}}\tp e^{\delta_2 \boldsymbol{B}})=\bra{\omega}(\Pi^1_1\tp\Pi^1_1)(e^{\delta_1 \boldsymbol{A}}\tp e^{\delta_2 \boldsymbol{B}}),
\end{align}
If we expand order by order in $\varepsilon,\delta_1$ and $\delta_2$, and remember the possible representations of $\bra{\omega}$ \eqref{OMREP}, we find the generalization of the cyclicity relations found for the $b$ co-algebra \eqref{cycl-coder-N}
\begin{equation}
        \begin{split}
            &\mathcal{O}((\delta_1)^0,(\delta_2)^0)\Longrightarrow\,\,\,\omega(\Pi^1_1\boldsymbol{D}\tgl,\Pi^1_1\tgl)=-\omega(\Pi^1_1\tgl,\Pi^1_1\boldsymbol{D}^j\tgl),\\
            &\mathcal{O}(\varepsilon^1,(\delta_1)^1,(\delta_2)^1)\Longrightarrow\,\,\,\omega(\Pi^1_1\boldsymbol{D}\boldsymbol{A}\tgl,\Pi^1_1\boldsymbol{B}\tgl)=-(-1)^{d(\boldsymbol{D})d(\boldsymbol{A})}\omega(\Pi^1_1\boldsymbol{A}\tgl,\Pi^1_1\boldsymbol{D}\boldsymbol{B}\bgl).
        \end{split}
    \end{equation}
If we now expand the co-derivation $\boldsymbol{D}$ into components $\boldsymbol{D}^{j,b}_{i_1,...,i_b}$ we will observe a generalization of the usual cyclicity and the mixing highlighted in \eqref{NCYCLij}
\begin{align}
    \alpha^{j,b}\omega(\Pi^1_1\boldsymbol{D}^{j,b}&\boldsymbol{A}^{j_1,b_1}\bgl,\Pi^1_1\boldsymbol{B}^{j,b_2}\bgl)\\
    &=-(-1)^{d(\boldsymbol{D}^{j,b})d(\boldsymbol{A}^{j_1,b_1})}\alpha^{j_1,b}\omega(\Pi^1_1\boldsymbol{A}^{j_1,b_1}\bgl,\Pi^1_1\boldsymbol{D}^{j_1,b}\boldsymbol{B}^{j,b_2}\bgl),
\end{align}
based on the choice of $\qty{j,j_1,b,b_1,b_2}$. If $b_1=b_2=b$ we recover \eqref{NCYCLij}.\\
The dualities induced by cyclicity are naturally found in open-closed SFT \cite{Maccaferri:2023gcg} and account for the equivalent description of interaction vertices with respect to the closed or the open string and the equivalent choice of reference boundary when using the open string description.

\subsection{Co-derivation algebra and homotopy algebras}
As previously observed in \ref{N-DEGEN:SEC}, whenever there is degeneration in a $N$ component co-algebra the co-derivation algebra opens. This can be seen if we work on the various subspaces $\Pi^b\ttHt$ where the commutator and product between co-derivations is defined by \eqref{co-der-alge} and \eqref{NProdexpl}
\begin{align}
    \commutator{\cdot}{\cdot}:\Coder(\tH^{\tpt b})\cross\Coder(\tH^{\tpt b'})\longrightarrow \Coder(\tH^{\tpt b+b'-1}).
\end{align}
Because the space of $\Coder(\tH^{\tpt b})$ is a subset of $\Coder(\ttHt)$ the algebra of co-derivations closes for
\begin{align}\label{Inf-co-der-alge}
    \commutator{\cdot}{\cdot}:&\Coder(\ttHt)\cross\Coder(\ttHt)\longrightarrow \Coder(\ttHt),\\
    &\Coder(\tH^{\tpt b})\subseteq\Coder(\ttHt)\,\forall\, b\in \nn.
\end{align}
A product between elements of $\Hom(\tH^{\tpt b},\Hs)$ and $\Hom(\tH^{\tpt b'},\Hs)$, potentially with $b\neq b'$, is then defined by projecting out the excess from \eqref{Inf-co-der-alge} and performing the right identifications
\begin{align}\label{INFIMPPROD}
    c^{j,b}_{i_1,...,i_b}d^{p,b'}_{p_1,...,p_{b'}}:=\sum_{j=0}^{b-1}c^{j,b}_{i_1,...,i_b}\qty(\id^{\tpt j}\tpt d^{p,b'}_{p_1,...,p_{b'}}\tpt\id^{\tpt b-j-1}).
\end{align}
Note the similarities between \eqref{INFIMPPROD} and \eqref{FINPROD} where this time the product cycles between powers $\tH^{\tpt j}$ instead of powers of $\Hs^{\tp j}$.\\
Thanks to the commutator we can define Homotopy algebras on $\infty$ tensor algebras $\ttHt$ as a graded odd co-derivation $\boldsymbol{D}$ which obeys
\begin{align}\label{INFHOM}
    \frac{1}{2}\commutator{\boldsymbol{D}}{\boldsymbol{D}}= \qty(\boldsymbol{D})^2 = 0,\,\,\,\,\boldsymbol{D}\in\Coder(\ttHt),
\end{align}
just like in \eqref{Coder_alg}.\\
Note that, when expressing 
\begin{align}
    \boldsymbol{D}:=\sum_{b=1}^{\infty}\sum_{j=1}^{b}\sum_{n_1,...,n_N=0}^{\infty}\boldsymbol{D}^{j,b}_{n_1,...,n_N},
\end{align}
the homotopy algebraic structure factors into an $A_\infty/L_\infty$ algebra when $\boldsymbol{D}$ acts on the subspace $\tH$
\begin{align}\label{HAFAC2}
    \boldsymbol{D}\boldsymbol{D}\Pi^1=0\,\Longrightarrow \sum_{k=0}^{n}\boldsymbol{D}^{1,1}_{k}\boldsymbol{D}^{1,1}_{n-k}=0.
\end{align}
The factored sub $A_\infty/L_\infty$ homotopy algebra when isolated gives rise to physical\footnote{ They satisfy the Classical BV master equation.} self interacting theories. Note that only one $A_\infty/L_\infty$ sub algebra has been factored due to the fact that the co-derivation algebra is open for every $b>1$.

\subsection{Homotopy transfer theorem}
The homotopy transfer theorem on $\ttHt$ is completely fixed by the homotopy transfer theorem on $\tH$. Because $\ttHt$ can be decomposed in $b$ component co-algebras $\tH^{\tpt b}$, from theorem \ref{NHTT}, we know how to build the maps $\bar{\boldsymbol{h}}^b$ for each $\tH^{\tpt b}$. The projector $\tilde{\boldsymbol{P}}$ from $\ttHt$ to a restriction $\ttHt_P$ is defined using theorem \ref{NHTT} to be
\begin{align}
    \tilde{\boldsymbol{P}}:=\sum_{b=1}^{\infty}\bar{\boldsymbol{P}}^b\Pi^b,\,\,\,\,\bar{\boldsymbol{P}}^b:=\underbrace{(\boldsymbol{P}\tpt...\tpt\boldsymbol{P})}_\text{b},\,\,\,\boldsymbol{P:}\tH\longrightarrow\tH_P.
\end{align}
We can then use the contracting homotopy maps $\bar{\boldsymbol{h}}^{b}$ to define the contracting homotopy map $\tilde{\boldsymbol{h}}$ on $\ttHt$ 
\begin{align}
        \tilde{\boldsymbol{h}}:=\sum_{b=1}^{\infty} \bar{\boldsymbol{h}}^{b}\Pi^b,
\end{align}
where the maps $\bar{\boldsymbol{h}}$ were constructed in \eqref{NHPARTFull}.\\
This map satisfies the Hodge-Kodaira decomposition
\begin{align}\label{infHKdec}
    \tilde{\id}=\tilde{\boldsymbol{P}}+\tilde{\alpha}\qty{\tilde{\boldsymbol{\partial}}\tilde{\boldsymbol{h}}+\tilde{\boldsymbol{h}}\tilde{\boldsymbol{\partial}}},\,\,\,\tilde{\alpha}\in\cc,
\end{align}
where $\tilde{\id}$ is the identity operator on $\ttHt$ and the differential $\tilde{\boldsymbol{\partial}}$ is the trivial uplift of $\partial$ to be a co-derivation of $\ttHt$.\\
The Hodge-Kodaira decomposition can be proven by rewriting $\tilde{\boldsymbol{h}}\tilde{\boldsymbol{\partial}}$ in the following way
\begin{align}
    \tilde{\boldsymbol{h}}\tilde{\boldsymbol{\partial}}=\qty[\sum_{b=1}^{\infty} \bar{\boldsymbol{h}}^{b}\Pi^b]\qty[\sum_{b=1}^{\infty} \bar{\boldsymbol{\partial}}^{b}\Pi^b]=\sum_{b=1}^{\infty} \bar{\boldsymbol{h}}^{b}\bar{\boldsymbol{\partial}}^{b}\Pi^b,
\end{align}
which implies that \eqref{infHKdec}
\begin{align}
    \tilde{\alpha}\qty{\tilde{\boldsymbol{\partial}}\tilde{\boldsymbol{h}}+\tilde{\boldsymbol{h}}\tilde{\boldsymbol{\partial}}}=\sum_{b=1}^{\infty}\tilde{\alpha}\qty{\bar{\boldsymbol{h}}^{b}\bar{\boldsymbol{\partial}}^{b}+\bar{\boldsymbol{\partial}}^{b}\bar{\boldsymbol{h}}^{b}}\Pi^b.
\end{align}
If the conditions of the homotopy transfer theorem are satisfied on $\tH$ the $\bar{\boldsymbol{h}}^{b}$ and $\bar{\boldsymbol{\partial}}^{b}$ satisfy the Hodge-Kodaira decomposition for every sub co-algebra $\Pi^b\ttHt$ implying that
\begin{align}
    \tilde{\id}-\tilde{\boldsymbol{P}}=\sum_{b=1}^{\infty}\tilde{\alpha}\qty{\bar{\boldsymbol{h}}^{b}\bar{\boldsymbol{\partial}}^{b}+\bar{\boldsymbol{\partial}}^{b}\bar{\boldsymbol{h}}^{b}}\Pi^b=\sum_{b=0}^{\infty}\frac{\tilde{\alpha}}{\bar{\alpha}^b}\qty[\tid^b-\bar{\boldsymbol{P}}^b]\Pi^b
\end{align}
which fixes the choice $\tilde{\alpha}=\bar{\alpha}^b=\alpha$ for all $b\in\nn$ in order to satisfy \eqref{infHKdec}.\\
Therefore the Homotopy transfer theorem for $\infty$ component co-algebras, which is one of the main results of this work, can be stated as:
\begin{theorem}\label{INFHTT}
    Given a $\infty$ component co-algebra built from the base co-algebra $\tH$. Let it be  equipped with a projector map $\boldsymbol{P}$ and a contracting homotopy map $\boldsymbol{h}$. If $\boldsymbol{h}$ and $\boldsymbol{P}$ satisfy the homotopy transfer conditions for $\tH$ the map $\tilde{\boldsymbol{P}}$ is a projector, a chain map and satisfies the Hodge-Kodaira decomposition together with the contracting homotopy $\tilde{\boldsymbol{h}}$ 
    \begin{align}
        \tilde{\boldsymbol{P}}:=\sum_{b=0}^{\infty}\boldsymbol{P}^{\tpt b}\Pi^b,\tilde{\boldsymbol{h}}:=\sum_{b=0}^{\infty}\bar{\boldsymbol{h}}^{b}\Pi^b,\,\,\,\,\,\,
        {\tilde{\boldsymbol{P}}}=0,\,\,\,\,\bar{\id}=\tilde{\boldsymbol{P}}+\alpha\qty{\tilde{\boldsymbol{\partial}}\tilde{\boldsymbol{h}}+\tilde{\boldsymbol{h}}\tilde{\boldsymbol{\partial}}},\,\,\,\alpha\in\cc,
    \end{align}
    where $\bar{\boldsymbol{h}}$ is the $b$ component contracting homotopy map, $\tilde{\boldsymbol{\partial}}$ the uplift of the differential $\partial$ to element of $\Coder(\ttHt)$ and $\alpha$ is fully determined by the Hodge-Kodaira decomposition on $\tH$. Furthermore there exists a morphism that transfers the homotopy algebraic structure on $\ttHt$ to an homotopy algebraic structure on the restriction $\ttHt_P:=\tilde{\boldsymbol{P}}\ttHt$, keeping the co-homologies isomorphic to one-another.
\end{theorem}
The morphism is then defined as
\begin{align}
    \tilde{\boldsymbol{F}}:=\sum_{b=0}^{\infty}\bar{\boldsymbol{F}}^b\Pi^b,\,\,\,\,\tilde{\boldsymbol{F}}':=\sum_{b=0}^{\infty}\bar{\boldsymbol{F}}'^b\Pi^b,
\end{align}
where all $\bar{\boldsymbol{F}}^b$ and $\bar{\boldsymbol{F}}'^b$ are the morphism maps from the $b$ component homotopy transfer theorem \ref{NHTT} and implies that
\begin{align}
  \tilde{\boldsymbol{F}}\tilde{\boldsymbol{F}}'=\tilde{\id},\,\,\,\,\tilde{\boldsymbol{F}}\tilde{\boldsymbol{\partial}}\tilde{\boldsymbol{F}}'=\tilde{\boldsymbol{\partial}}'.
\end{align}    
    
\section{Classical open-closed SFT}\label{SDHA:SEC}
In this section we discuss how the classical truncation of SFT to spheres and disks (SDHA) \cite{Maccaferri:2022yzy} naturally arises from a $2$ component cyclic CAFT. Furthermore, we highlight how open-closed channel duality is a direct consequence of cyclicity. Lastly we recover the OCHA SFT \cite{Kajiura:2004xu,Kajiura:2005sn,Kajiura:2006mt} by breaking cyclicity in the CAFT.
\subsection{$2$ component CAFT}\label{SDHACAFT}
Although the open-closed tensor algebra $\octah$ is cyclicized in the open sector and symmetrized in the closed sector, we will work without cyclicization and symmetrization due to the fact that algebraic properties on $\octah$ can be directly transferred to cyclicized and symmetrized sub-tensor algebras.\\
The SDHA CAFT starts from the $2$ component tensor algebra
\begin{align}
    \tHt:=\tHc\tpt\tHo=\bigoplus_{k=0}^\infty\bigoplus_{n=0}^{\infty}\Hs_{\rm c}^{\tp_{\rm c} k}\tpt\Hs_{\rm o}^{\tp_{\rm o} n},
\end{align}
built from the Hilbert space of the first quantized open string $\Hs_{\rm o}$ and closed string $\Hs_{\rm c}$ after a suitable choice of background, i.e. D-brane system and space-time metric.\\
Let us identify on $\Hs_{\rm o}$ and $\Hs_{\rm c}$ the open string field $\Psi\in\Hs_{\rm o}$ and closed string field $\Phi\in\Hs_{\rm c}$ to be
\begin{align}
    \Psi:=\sum_{a}\psi^ao_a,\,\,\,o_a\in\Hs_{\rm o},\,\,\,d(\Psi)=0,\\
    \Phi:=\sum_{a}\phi^ac_a,\,\,\,c_a\in\Hs_{\rm c},\,\,\,d(\Phi)=0,
\end{align}
where $d(\cdot)$ is the grading, $c_a$ and $o_a$ are the base elements of their respective Hilbert spaces and $\phi^a,\psi^a$ graded parameters.\\
The group like element $\bgl\in\tHt$ is then defined as
\begin{align}
    \bgl:=\sum_{k,n=0}^{\infty}\Phi^{\tp_{\rm c} k}\tpt\Psi^{\tp_{\rm o} n},\,\,\,\chi:=\pi_{(1)}\bgl,
\end{align}
with $\chi$ the open-closed string field. A common base vector $f_a$ is defined as
\begin{align}
    f_a:=c_a+o_a.
\end{align}
Let us now endow $\Hs_{\rm c}$ and $\Hs_{\rm o}$ with two respective symplectic forms
\begin{align}
    \omega_{\rm c}:\Hs_{\rm c}^{\tp_{\rm c} 2}\longrightarrow \cc\,,\,\omega_{\rm o}:\Hs_{\rm o}^{\tp_{\rm o} 2}\longrightarrow \cc,
\end{align}
which endow $\tHt$ with a symplectic form 
\begin{align}\label{SDHASYMPL}
    \omega:=\alpha_{\rm c}\omega_{\rm c}+\alpha_{\rm o}\omega_{\rm o}, \,\,\,\,\alpha_{\rm c},\alpha_{\rm o}\in\cc.
\end{align}
Constants $\alpha_{\rm c},\alpha_{\rm o}$ will later on be directly linked to string coupling constants of the topological expansion when concerning spheres and disks.\\ 
The symplectic form $\omega$ allows for the definition of the dual basis $f^a$ such that
\begin{align}
    f^a:=\alpha_{\rm c} c^a+\alpha_{\rm o} o^a,\,\,\,\omega(f^a,f_b)=-\omega(f_b,f^a)=\delta^a_{\,\,b},
\end{align}
with the evaluations of the symplectic form
\begin{align}
    \omega^{ab}:=\omega(f^a,f^b),\,\,\,\omega_{ab}:=\omega(f_a,f_b).
\end{align}
Topologically $\octah$ only describes disks due to the explicit presence of $\tH_{\rm o}$ associated to a boundary. In order to also describe the sphere contributions we need to add the closed tensor algebra to $\tHt$ in the following way
\begin{align}
    \tHt_{\rm SD}:=\tH_{\rm c}\oplus\tHt.
\end{align}
In order to work in the simplest way we trivially extend $\tH_{\rm c}\simeq\tHt$, but in order to distinguish sphere contributions from the disk contributions we divide co-derivations $\boldsymbol{n}$ on $\tHt$ in the following way
\begin{align}
    \boldsymbol{n}:=\boldsymbol{l}^0+\kappa\boldsymbol{l}^1+\boldsymbol{m}^1,
\end{align}
where  $\boldsymbol{l}^0$ are closed co-derivations associated to the sphere, $\boldsymbol{l}^1$ are closed co-derivation associated to the disk $(b=1)$ and $\boldsymbol{m}^1$ are the open co-derivations associated to the disk. The constant $\kappa$ has been added because, according to the topological expansion, closed strings on the disk are the first quantum correction of weight $\kappa$. The co-derivation can be subsequently expanded in terms of multilinear products
\begin{align}
    \pi_{(1)}\boldsymbol{l}^0=\sum_{k=0}^{\infty}l^0_k,\,\,\pi_{(1)}\boldsymbol{l}^1=\sum_{k,n=0}^{\infty}l^1_{k,n},\,\,\pi_{(1)}\boldsymbol{m}^1=\sum_{k,n=0}^{\infty}m^1_{k,n},
\end{align}
where, in order to have a consistent description of the sphere contributions, we need to set that
\begin{align}
    \boldsymbol{l}^0_{k,0}\neq0,\,\,\,\,\boldsymbol{l}^0_{k,n\geq1}=0\,\Longleftrightarrow\,l^0_{k,n\geq1}=0,l^0_{k,0}= l^0_k\neq0,
\end{align}
because the sphere does not allow for open string insertions.\\
At last we can define the cyclical SDHA CAFT according to section \ref{sec:caft}, which gives the action 
\begin{align}
S[\bgl]_{\rm SD}:=\int_0^1\dd{t}\bar{\omega}(\pi_{(1)}\boldsymbol{\partial}_t\tgl(t),\pi_{(1)}\boldsymbol{n}\tgl(t)),\,\,\,\,\bgl(0):=0,\bgl(1):=\bgl,
\end{align}
and asking that $\boldsymbol{n}$ is cyclical with respect to $\bar{\omega}$.\\
By expanding the action functional into sphere and disk contributions we find that, in order to have consistency with the topological expansion we have to fix $\alpha_{\rm c}=\frac{1}{\kappa^2}$ and $\alpha_{\rm o}=\frac{1}{\kappa}$ leading to
\begin{align}
    S[\bgl]_{\rm SD}:&=\frac{1}{\kappa^2}\int_0^1\dd{t}\omega_{\rm c}(\pi_{(1)}\boldsymbol{\partial}_t\tgl(t),\pi_{(1)}\boldsymbol{l}^0\tgl(t))+\frac{1}{\kappa}\int_0^1\dd{t}\omega_{\rm o}(\pi_{(1)}\boldsymbol{\partial}_t\tgl(t),\pi_{(1)}\boldsymbol{m}^1\tgl(t))\\&+\frac{1}{\kappa}\int_0^1\dd{t}\omega_{\rm c}(\pi_{(1)}\boldsymbol{\partial}_t\tgl(t),\pi_{(1)}\boldsymbol{l}^1\tgl(t)).
\end{align}
\subsection{The classical BV master equation and SDHA}
Let us now ask the SDHA CAFT to obey the classical BV master equation up to $\mathcal{O}(\kappa)$ for the open strings and  up to $\mathcal{O}(\kappa^2)$ for the closed strings. Using \eqref{CBV} we quickly compute that 
\begin{align}\label{SDHABV}
    \qty(S,S)_{\rm oc}= 0\,\Longleftrightarrow\,\boldsymbol{n}^2=\qty(\boldsymbol{l}^0+\kappa\boldsymbol{l}^1+\boldsymbol{m}^1)^2=0,
\end{align}
where $\qty(\cdot,\cdot)_{\rm oc}$ is the open-closed BV bracket defined according to \eqref{BVDIFFOP}
\begin{align}
    \qty(X,Y):=X\overset{\leftarrow}{\pdv{}{\chi^a}}\omega^{ab}\overset{\rightarrow}{\pdv{}{\chi^b}}Y.
\end{align}
By expanding $\boldsymbol{n}$ in the nilpotent relation truncating at the powers of $\kappa$ we recover the SDHA relations \cite{Maccaferri:2022yzy}, namely at order $\mathcal{O(\kappa^0)}$ we recover the classical closed $L_\infty$
\begin{align}\label{SDHAREL}
    &\kappa^{1}\pi_{0,1}:\,\text{Nothing},\\
    &\kappa^{0}\pi_{1,0}:\,\,\boldsymbol{l}^0\boldsymbol{l}^0=0,
\end{align}
at order $\mathcal{O}(\kappa^1)$ we recover the SDHA relation, which includes the OCHA relations, 
\begin{align}\label{SDHAREL2} 
    &\kappa^{1}\pi_{0,1}:\,\boldsymbol{m}^1\boldsymbol{m}^1+\boldsymbol{m}^1\boldsymbol{l}^0=0,\\
    &\kappa^{0}\pi_{1,0}:\boldsymbol{l}^1\boldsymbol{l}^0+\boldsymbol{l}^0\boldsymbol{l}^1+\boldsymbol{l}^1\boldsymbol{m}^1=0,
\end{align}
and discard the higher order terms $\mathcal{O(\kappa^2)}$ which do not satisfy \eqref{SDHABV} (higher loop and non planar corrections).\\
In the specific case no closed string are fed in the first line of \eqref{SDHAREL2} we recover the $A_\infty$ relations for the classical open SFT.\\
By expanding the co-derivations in \eqref{SDHAREL2} in terms of multilinear products we can explicitly recover the SDHA relations of \cite{Maccaferri:2022yzy}

\subsection{Cyclicity and dualities}
In \ref{SDHACAFT} we required the $2$ component CAFT to be cyclic. The physical relevance of cyclicity can be explained by expanding $\boldsymbol{n}$ in terms of $\boldsymbol{l}^0,\boldsymbol{l}^1$ and $\boldsymbol{m}^1$ and recalling \eqref{NCYCLjj} and \eqref{NCYCLij} together with two place holder co-derivations $\boldsymbol{a}^j,\boldsymbol{b}^j$
\begin{equation}\label{CAFTCYCL}
    \begin{split}
        &\omega_{\rm c}(\pi_{1,0}\boldsymbol{l}\boldsymbol{a}^{\rm c}\tgl,\pi_{1,0}\boldsymbol{b}^{\rm c}\tgl) = -(-1)^{d(\boldsymbol{l})d(\boldsymbol{a}^{\rm c})}\omega_{\rm c}(\pi_{1,0}\boldsymbol{a}^{\rm c}\tgl,\pi_{1,0}\boldsymbol{l}\boldsymbol{b}^{\rm c}\tgl),\\
        &\omega_{\rm o}(\pi_{0,1}\boldsymbol{m}^1\boldsymbol{a}^{\rm o}\tgl,\pi_{0,1}\boldsymbol{b}^{\rm o}\tgl) = -(-1)^{d(\boldsymbol{m}^1)d(\boldsymbol{a}^{\rm o})}\omega_{\rm o}(\pi_{0,1}\boldsymbol{a}^{\rm o}\tgl,\pi_{0,1}\boldsymbol{m}^1\boldsymbol{b}^{\rm o}\tgl),\\
        &\omega_{\rm c}(\pi_{1,0}\boldsymbol{l}^1\boldsymbol{a}^{\rm o}\tgl,\pi_{1,0}\boldsymbol{b}^{\rm c}\tgl) = -(-1)^{d(\boldsymbol{l}^1)d(\boldsymbol{a}^{\rm c})}\omega_{\rm o}(\pi_{0,1}\boldsymbol{a}^{\rm o}\tgl,\pi_{0,1}\boldsymbol{m}^1\boldsymbol{b}^{\rm c}\tgl),
    \end{split}
\end{equation}
where, for simplicity, we have defined $\boldsymbol{l}$ as
\begin{align}
    \boldsymbol{l}:=\boldsymbol{l}^0+\kappa\boldsymbol{l^1}.
\end{align}
In the first two rows of \eqref{CAFTCYCL} we recognise the usual cyclicity relation of open SFT and closed SFT that describes that, in order to describe the interaction vertex, all string punctures on the surface/boundary are equivalent.\\
The last row of \eqref{CAFTCYCL} tells us that an interaction vertex on the disk can be equivalently described using an open or a closed string puncture as reference puncture (open-closed duality).\\
The cyclicity conditions successfully reproduce important aspects of open-closed SFT and allow for an alternative understanding of classical open-closed SFT as a cyclical $2$ component CAFT generated by the nilpotent operator $\boldsymbol{n}$. Mathematically, the cyclicity of $\boldsymbol{n}$ implies that its exponentiation generates a co-homomorphisms that preserves the symplectic structure of $\tHt_{\rm SD}$.

\subsection{OCHA: breaking cyclicity of the SDHA}\label{OCHA:sec}
The OCHA, thoroughly studied in \cite{Kajiura:2004xu,Kajiura:2005sn,Kajiura:2006mt},
has similar relations to \eqref{SDHAREL} but without the $\kappa^1$ corrections. Therefore, in order to consistently reproduce the OCHA relations we need to truncate the $\boldsymbol{n}^2=0$ at order $\kappa^0$ leaving 
\begin{align}\label{OCHAREL}
    \kappa^{0}\Longrightarrow&\,\,\boldsymbol{l}^0\boldsymbol{l}^0=0,\,\,\,\,\,\,\boldsymbol{m}^1\boldsymbol{m}^1+\boldsymbol{m}^1\boldsymbol{l}^0=0.
\end{align}
Note that now \eqref{OCHAREL} are not invariant under the cyclical transformations of $\boldsymbol{n}$ that swap between $\boldsymbol{m}^1\leftrightarrow\boldsymbol{l}^1$.\\
The OCHA truncation of $\boldsymbol{n}^2=0$ can, in principle, be done by sending $\kappa\rightarrow 0$, which results in
\begin{align}
    \boldsymbol{n}=\boldsymbol{l}^0+\kappa\boldsymbol{l}^1+\boldsymbol{m}^1\,\overset{\kappa\rightarrow0}{\longrightarrow}\,\boldsymbol{n}=\boldsymbol{l}^0+\boldsymbol{m}^1,
\end{align}
where open-closed channel duality is effectively broken because the cyclicity factors into the separate cyclicity of $\boldsymbol{l}^0$ with respect to $\omega_{\rm c}$ and $\boldsymbol{m}^1$ with respect to $\omega_{\rm o}$.\\
Physically, the theory with broken cyclicity results inconsistent \cite{Maccaferri:2022yzy} due to the lack of closed string verities on the disk.\\
From a pure mathematical point of view\footnote{If we forget that we are studying SFT.}, the breaking of cyclicity implies that we need to change the WZW parametrization \ref{WZWCO} of the action to the following parametrization in order to keep using co-algebraic methods to study its properties
\begin{align}\label{OCHAACT}
    S_{\rm OCHA}[\Phi,\Psi]:=\frac{1}{\kappa^2}\int_0^1\dd{t}\omega_{\rm c}(\pi_1\boldsymbol{\partial}_t\gl_{\rm c}(t),\pi_1\mathbf{l}^{0}\gl_{\rm c}(t))+\frac{1}{\kappa}\int_0^1\dd{t}\bar{\omega}_{\rm o}(\dot{\Psi}(t),\pi_{0,1}\mathbf{m}^1(e^{\wedge \Phi}\tpt \gl_{\rm o}(t))),
\end{align}
where in the disk portion of the action only the open string field  $\Psi$ is trivially parametrized according to the WZW prescription \ref{WZWCO}.\\ If we apply the classical BV master equation to \eqref{OCHAACT} we will directly get exactly the OCHA relations of \eqref{OCHAREL} without the need for truncations.    
    
\section{Quantum open-closed SFT}\label{QOCSFT}
In this section we discuss how the quantum open-closed SFT naturally arises from a $2$ component with degeneration cyclic CAFT. Furthermore, we highlight how open-closed duality is a direct consequence of cyclicity.

\subsection{$N=2$ and $\infty$ tensor co-algebra}
A notable difference of quantum open-closed SFT with its classical truncation, as anticipated in \ref{N-DEGEN:SEC}, is the fact that the algebra governing the gluing of surfaces with more than one boundary opens up. This fact is evident in the quantum open-closed SFT tensor algebra
\begin{align}
    \ttHoc:=\bigoplus_{k=0}^{\infty}\Hs_{\rm c}^{\wedge_{\rm c} k}\tpt \bigoplus_{b=0}^{\infty}\qty[\bigoplus_{p_1,...,p_b=0}^{\infty}\Hs_{\rm o}^{\odot_{\rm o} p_1}\wpt...\wpt \Hs_{\rm o}^{\odot_{\rm o} p_b}],
\end{align}
where $b$ indicates the number of boundaries present on the surface and $\wedge,\odot$ have been introduced in section \ref{co-alge-types} as the symmetrized and cyclicized tensor products.\\
In order to study the properties of quantum open-closed SFT with the tools defined in this paper we will work on the tensor algebra
\begin{align}
    \ttHt:=\tHoc=\bigoplus_{k=0}^{\infty}\Hs_{\rm c}^{\tp_{\rm c} k}\tpt \bigoplus_{b=0}^{\infty}\qty[\bigoplus_{p_1,...,p_b=0}^{\infty}\Hs_{\rm o}^{\tp_{\rm o} p_1}\tpt...\tpt \Hs_{\rm o}^{\tp_{\rm o} p_b}],
\end{align}
because $\ttHoc\subseteq\ttHt$ and properties of $\ttHt$ can be directly mapped onto $\ttHoc$.\\
The tensor algebra $\ttHt$ is a $\infty$ component tensor algebra with respect to the open sector, and a $2$ component co-algebra with degeneration $b$ with respect to the subspace $\tH_{\rm c}\tpt\tH_{\rm o}^{\tpt b}$. From sections \ref{NCOALG:CH} and \ref{INFCOALG:CH} we can introduce the $b+1$ component co-product $\bDelta^b$ and the co-product on the boundary $\tDelta$ starting from the closed co-product $\Delta_{\rm c}$ and the open co-product $\Delta_{\rm o}$
\begin{align}
    &\bDelta^b:=\Omega_b\qty(\Delta_{\rm c}\tpt\Delta_{\rm o}^{\tpt b}),\,\,\,\bDelta^b:\tH_{\rm c}\tpt\tH^{\tpt b}_{\rm o}\longrightarrow\tH_{\rm c}\tpt\tH^{\tpt b}_{\rm o}\tpb'\tH_{\rm c}\tpt\tH^{\tpt b}_{\rm o},\\
    &\tDelta:=\Omega_{\rm oc}\qty(\Delta_{\rm c}\tpt\tDelta_{\rm o}),\,\,\,\,\tDelta:\tH_{\rm c}\tpt\ttHt_{\rm o}\longrightarrow\tH_{\rm c}\tpt\ttHt_{\rm o}\tpt'_{\rm oc}\tH_{\rm c}\tpt\ttHt_{\rm o},\\
    &\tDelta_{\rm o}:\ttHt_{\rm o}\longrightarrow \ttHt_{\rm o}\tpt'\ttHt_{\rm o},
\end{align}
which covers the dual co-algebra structure of $\ttHt$.
$\ttHt$ can become a $2$ component tensor co-algebra according to section \ref{NCOALG:CH} by defining the tensor product $\tpt_{\rm oc}\neq\tpt_{\rm oc}'$ using \eqref{NPROPPROD},  
\begin{align}
    \tpt_{\rm oc}:\ttHt\cross\ttHt\longrightarrow\ttHt.
\end{align}
In order to build the quantum open-closed CAFT we need to build the open and closed string field from the Hilbert space of the first quantized open string $\Hs_{\rm o}$ and closed string $\Hs_{\rm c}$. After a suitable choice of background, i.e. D-brane system and space-time metric, we identify the open string field $\Psi\in\Hs_{\rm o}$ and closed string field $\Phi\in\Hs_{\rm c}$ to be
\begin{align}
    \Psi:=\sum_{a}\psi^ao_a,\,\,\,o_a\in\Hs_{\rm o},\,\,\,d(\Psi)=0,\\
    \Phi:=\sum_{a}\phi^ac_a,\,\,\,c_a\in\Hs_{\rm c},\,\,\,d(\Phi)=0,
\end{align}
where $c_a$ and $o_a$ are the base elements of their respective Hilbert spaces and $\phi^a,\psi^a$ graded parameters.\\
From sections \ref{NCOALG:CH} and \ref{INFCOALG:CH} we know that the group like element $\tgl$ is completely fixed by $\Phi$ and $\Psi$ and reads
\begin{align}
    \tgl :=\gl_{\rm c} \tpt \frac{1}{1-\tpt \gl_{\rm o}} =\frac{1}{1-\tp_{\rm c}\Phi}\tpt\frac{1}{1-\tpt \frac{1}{1-\tp_{\rm o} \Psi}}.
\end{align}

\subsection{Open-closed co-derivation}
The literature \cite{Maccaferri:2023gcg} features multilinear products of the form
\begin{align}
    &n^{(g,b,j\neq 0)}_{k,p_1,...,p_b}:\Hs_{\rm c}^{\tp_{\rm c} k}\tpt\Hs_{\rm o}^{\tp_{\rm o} p_1}\tpt...\tpt \Hs_{\rm o}^{\tp_{\rm o} p_b}\longrightarrow \Hs_{\rm o},\\
    &n^{(g,b,j = 0)}_{k,p_1,...,p_b}:\Hs_{\rm c}^{\tp_{\rm c} k}\tpt\Hs_{\rm o}^{\tp_{\rm o} p_1}\tpt...\tpt \Hs_{\rm o}^{\tp_{\rm o} p_b}\longrightarrow \Hs_{\rm c}.
\end{align}
Such multilinear products can be uplifted to co-derivations using the following uplift procedure.
In this formulation multilinear products can be directly uplifted to co-derivations
\begin{equation}
    \begin{split}
        \boldsymbol{N}^{(g,b,j\neq0)}_{k;p_1,...,p_b}\Pi^{b'}_{k',q_1,...,q_{b'}}:=&\sum_{i=0}^{k'-k}\sum_{l=0}^{b'-b}\qty(\id_{\rm c}^{\tp_{\rm c} i}\tpt\id_{\rm o}^{\tpt l})\tpt_{\rm oc}\\
        &\tpt_{\rm oc}\qty[\sum_{p=0}^{q_j-p_j}\qty(\id_{\rm o}^{\tp_{\rm o} p}\tpb n^{(g,b,j\neq0)}_{k;p_1,...,p_b}\tpb\id_{\rm o}^{\tp_{\rm o} q_j-p_j-p})\pi_{p_1,...,q_j,...,p_b}]\tpt_{\rm oc}\\
        &\tpt_{\rm oc}\qty( \id_{\rm c}^{\tp_{\rm c} k'-k-i}\tpt\id_{\rm o}^{\tpt b'-b-l}),
    \end{split}
\end{equation}
for the $j$ boundary and for the closed sector $j=0$ 
\begin{equation}
    \begin{split}
        \boldsymbol{N}^{(g,b,j=0)}_{k;p_1,...,p_b}\Pi^{b'}_{k',q_1,...,q_{b'}}:=&\sum_{i=0}^{k'-k}\sum_{l=0}^{b'-b}\qty(\id_{\rm c}^{\tp_{\rm c} i}\tpt\id_{\rm o}^{\tpt l})\tpt_{\rm oc}n^{(g,b,j=0)}_{k;p_1,...,p_b}\tpt_{\rm oc}\\
        &\tpt_{\rm oc}\qty( \id_{\rm c}^{\tp_{\rm c} k'-k-i}\tpt\id_{\rm o}^{\tpt b'-b-l}).
    \end{split}
\end{equation}
The co-derivations $\boldsymbol{N}$ satisfy the co-Leibniz rules according to sections \ref{NCOALG:CH} and \ref{INFCOALG:CH} and match the co-derivation like elements firstly introduced in \cite{Maccaferri:2023gcg} proving them to be fully fledged co-derivations of $\ttHt$.
\subsection{Quantum open-closed cyclical CAFT}
Knowing now what co-derivations and group like element are on $\ttHt$ we can recall the symplectic form $\bar{\omega}$ from \eqref{SDHASYMPL}
\begin{align}
    \omega:=\frac{1}{\kappa^2}\omega_{\rm c}+\frac{1}{\kappa}\omega_{\rm o},
\end{align}
and define the co-derivation $\boldsymbol{N}$ as
\begin{align}
    \boldsymbol{N}:=\sum_{g,b=0}^{\infty}\kappa^{2g+b}\frac{1}{b}\sum_{j=0}^{b}\sum_{k,p_1,...,p_b=0}^{\infty}\boldsymbol{N}^{(g,b,j=0)}_{k;p_1,...,p_b},
\end{align}
where now the different contributions are weighted according to the topological expansion of the string with $g$ being the genus and $b$ the boundaries of the associated surface.\\
Thanks to all the previously defined elements we can define the Quantum open-closed cyclical CAFT with action
\begin{align}\label{SMRV}
    S_{\rm MRV}[\tgl]:=\int_0^1\dd{t}\bar{\omega}(\Pi^1_{(1)}\boldsymbol{\partial}_t\tgl(t),\Pi^1_{(1)}\boldsymbol{N}\tgl(t)),\,\,\,\tgl(0):=0,\,\tgl(1):=\tgl,
\end{align}
where MRV stands for Maccaferri-Ruffino-Vo\v smera, the autors of \cite{Maccaferri:2023gcg}.
\subsection{The nilpotent structure of open-closed SFT}
Like we did in section \ref{SDHA:SEC}, we ask for the action \eqref{SMRV} to satisfy the quantum BV master equation which, thanks to the co-algebraic tools established in section \ref{sec:caft}, can be quickly computed using \eqref{QBV} which leads to 
\begin{align}\label{MRVNILL}
    \frac{1}{2}\qty(S,S)_{\rm oc}+\Delta_{\rm oc} S=0 \,\Longleftrightarrow\,\qty(\boldsymbol{N}+\boldsymbol{U})^2=0,
\end{align}
where $\qty(\cdot,\cdot)_{\rm oc}$ is the open-closed BV bracket and $\Delta_{\rm oc}$ the BV Laplacian, and the dependency of $\kappa$ is hidden in the definition of $\bar{\omega}$.
The Poisson bi-vector $\boldsymbol{U}$ is defined as
\begin{align}
    \boldsymbol{U}:= \kappa^2\boldsymbol{U}_{\rm c}+\kappa \boldsymbol{U}_{\rm o},\,\,\,\,\boldsymbol{U}^2=0,
\end{align}
with individual open and closed bi-vectors
\begin{align}
    \boldsymbol{U}_{\rm c}:= \frac{(-1)^{c^b}}{2}\omega_{\rm c}^{ab}\boldsymbol{C}_b\boldsymbol{C}_a,\,\,\,\,\boldsymbol{U}_{\rm o}:= \frac{(-1)^{o^b}}{2}\omega_{\rm o}^{ab}\boldsymbol{O}_b\boldsymbol{O}_a,
\end{align}
where $\bar{\boldsymbol{C}}_a$ and $\bar{\boldsymbol{O}}_a$ are the co-derivations associated to the base vectors $c_a\in\Hs_{\rm c}$ and $o_a\in\Hs_{\rm o}$.\\
The result \eqref{MRVNILL} is precisely the nilpotent structure of open-closed SFT  \cite{Maccaferri:2023gcg} derived in a more cost effective way compared to \cite{Maccaferri:2023gcg}.
\subsection{Cyclical structure of quantum open-closed SFT}
Like with the SDHA CAFT, we asked for the quantum open-closed CAFT to be cyclical, i.e. the co-derivation $\boldsymbol{N}$ is cyclical with respect to $\bar{\omega}$. The implication of the cyclical structure can be better appreciated by rearranging $\boldsymbol{N}$ in open and closed contributions at fixed $(g,b)$ in the following way
\begin{align}
    \boldsymbol{N}:=\sum_{g=0}^{\infty}\sum_{b=0}^{\infty}\kappa^{2g+b}\boldsymbol{L}^{(g,b)}+\sum_{g=0}^{\infty}\sum_{b=1}^{\infty}\kappa^{2g+b}\boldsymbol{M}^{(g,b)},
\end{align}
where $\boldsymbol{L}^{(g,b)}$ is linked to $j=0$ and $\boldsymbol{M}^{(g,b)}$ contains all the other $j\neq0$. The cyclicity of $\boldsymbol{N}$ together with placeholder co-derivations $\boldsymbol{A}^i$ and $\boldsymbol{B}^i$ implies that, other than the usual cyclicity conditions \ref{COHOM}, for $b\geq1$ we have open-closed duality in the description of interaction vertices
\begin{align}\label{QOCSFTD1}
    \omega_{\rm c}(\Pi^1_{1,0}\boldsymbol{L}^{(g,b)}\boldsymbol{A}^{\rm o}\tgl,\Pi^1_{1,0}\boldsymbol{B}^{\rm c}\tgl) = -(-1)^{d(\boldsymbol{L})d(\boldsymbol{A}^{\rm c})}\kappa\omega_{\rm o}(\Pi^1_{0,1}\boldsymbol{A}^{\rm o}\tgl,\Pi^1_{0,1}\boldsymbol{M}^{(g,b)}\boldsymbol{B}^{\rm c}\tgl),
\end{align}
 and, when explicitly writing out the reference boundary on $\boldsymbol{M}^{(g,b,j\neq0)}$, cyclicity provides duality between the choice of special boundary
\begin{align}\label{QOCSFTD2}
    \omega_{\rm o}(\Pi^1_{0,1}\boldsymbol{M}^{(g,b,j)}\boldsymbol{A}^{\rm o}\tgl,\Pi^1_{0,1}\boldsymbol{B}^{\rm o}\tgl) = -(-1)^{d(\boldsymbol{L})d(\boldsymbol{A}^{\rm c})}\omega_{\rm o}(\Pi^1_{0,1}\boldsymbol{A}^{\rm o}\tgl,\Pi^1_{0,1}\boldsymbol{M}^{(g,b,j')}\boldsymbol{B}^{\rm o}\tgl),\,\,\,\,\forall j\neq j'\in\qty[1,b],
\end{align}
where the index $i={\rm c,o}$ of $\boldsymbol{A}^i$ and $\boldsymbol{B}^i$ refers to closed or open co-derivation.\\
The results reconstructed in this section are in accordance with known literature and provide an axiomatic definition of quantum open-closed SFT. It also provides a way to make use of co-algebraic manipulation techniques in order to facilitate many algebraic computations, like the computation of the BV master equations. Lastly, being the quantum open-closed SFT a CAFT, it allows for the direct use of the homotopy transfer theorem without worrying about the consistency of the co-algebraic object entering the theorem.    
    
\section{N Bosonic field scattering amplitudes via homotopy transfer}\label{SEC:NSCATAMP}
Thanks to the extension to $N$ component co-algebras we were able to extend the validity of the homotopy transfer theorem to more complex co-algebras. Thanks to theorem \ref{NHTT} it is possible to directly extend the method of computing correlation functions reviewed in \ref{AHTT} to QFT with $N$ different fields.\\
In order to illustrate why it is possible to extend the validity of \ref{AHTT} to $N$ component co-algebras we need to remember that the QFT studied in \ref{AHTT} was firstly uplifted to CAFT, then the application of the homotopy transfer theorem was performed at the CAFT level. Because all CAFTs share the same overall formulation and algebraic properties, including the homotopy transfer theorem, the validity of \ref{AHTT} can be extended to more complicated QFT.
\subsection{Amplitudes for $N$ bosonic field}\label{NAHTT}
Given $N$ different fields $\phi_j$, with their respective Hilbert spaces $\Hs_{j,0}$, and an action functional
\begin{align}
    S[\phi_1,...,\phi_n]:=\sum_{j=1}^{N}\int\dd[d]x\qty[-\frac{1}{2}\phi_j(x)\partial_\mu\partial^\mu\phi_j(x)+\frac{1}{2}m_j^2\phi_j^2(x)]+V(\phi_1,...,\phi_N),\,\,\phi_j\in\Hs_{j,0},
\end{align}
with interactions $V(\phi_1,...,\phi_N)$ of the form 
\begin{align}
    V(\phi_1,...,\phi_N)=\sum_{n_1,...,n_N=0}^{m_1,...,m_N}\frac{g_{n_1,...,n_N}}{n_1!...n_N!}\mathcal{O}_{n_1,...,n_N}(\phi_1^{n_1},...,\phi_N^{n_N}),
\end{align}
where $\mathcal{O}$ is the interaction vertex. To each element we can associate a cyclical multilinear product in order to rewrite $S[\phi_1,...,\phi_n]$ in a way that can be easily uplifted to CAFT
\begin{equation}\label{NAHTTREWRITE}
    \begin{split}
        S[\phi_1,...,\phi_n]:&=\sum_{j=1}^N\frac{1}{2}\omega_j(\phi_j,Q_j\phi_j)+\\&+\frac{1}{N}\sum_{j=1}^N\sum_{n_1,...,n_N=0}^{m_1,...,m_N}\frac{1}{n_j}\omega_j(\phi_j,m^j_{n_1,...,n_j-1,...,n_N}(\phi_1^{\tp n_1}\tpt...\phi_j^{\tp n_j-1}\tpt...\phi_N^{\tp n_N})).
    \end{split}
\end{equation}
Just like with \ref{AHTT}, in order to correctly identify the interaction vertices with $\omega_j,Q_j$ and $m^j_{n_1,...,n_N}$ we need to trivially extend $\Hs_j$, by adding ghosts, to a graded vector space 
\begin{equation}
    \Hs_j:=\Hs_{j,0}\oplus\Hs_{0,1},\,\,\,\,f_{j,0}(x)\in\Hs_{j,0},\,\,d(f_{j,0}(x))=0,\,\,\,f_{j,1}(x)\in\Hs_{j,1},\,\,d(f_{j,1}(x))=1,
\end{equation}
where $f_{j,0}$ is the basis of $\Hs_{j,0}$ and $f_{j,1}$ is the basis of $\Hs_{j,1}$. The trivial extension allows us to define maps
\begin{equation}
    \begin{split}
        &\omega_j:\Hs_j\cross\Hs_j\longrightarrow\cc,\,\,\,\,m^j_{n_1,...,n_N}:\Hs_{1,0}^{\tp n_1}\tpt...\tpt\Hs_{N,0}^{\tp n_N}\longrightarrow\Hs_{j,1},\\
        &Q_j:\Hs_{j,0}\longrightarrow\Hs_{j,1},\,\,\,Q_j:\Hs_{i\neq j,0}\longrightarrow0,\,\,\,Q_j:\Hs_{i,1}\longrightarrow0,
    \end{split}
\end{equation}
and it allows the following identifications 
\begin{equation}
    \begin{split}
        &\phi_j=\int\dd[d]{x}\phi_j(x)f_{j,0}(x),\,\,\,\,\,\,\,\omega_j(f_{j,0}(x),f_{j,1}(y))=-\omega_j(f_{j,1}(x),f_{j,0}(y))=\delta^d(x-y),\\
        &Q_j f_{j,0}(x)=\qty(-\partial^2+m_j^2)f_{j,1}(x),\,\,\,\,Q_j f_{j,1}(x)=0\Longrightarrow Q_j^2=0.
    \end{split}
\end{equation}
In this paper we only look at polynomial type interactions vertices which can be identified as
\begin{equation}
    \begin{split}
        S_{n_1,...,n_N}^{\rm int}&=\frac{g_{n_1,...,n_N}}{n_1!...n_N!}\phi_1^{n_1}\phi_2^{n_2}...\phi_N^{n_N}\\&:=\frac{1}{N}\sum_{j=1}^N\frac{1}{n_j}\omega_j(\phi_j,m^j_{n_1,...,n_j-1,...,n_N}(\phi_1^{\tp n_1}\tpt...\phi_j^{\tp n_j-1}\tpt...\phi_N^{\tp n_N})),
    \end{split}
\end{equation}
where there are $N$ equivalent ways to rewrite the interaction vertex using multilinear products. The equivalence between all the $N$ possible rewriting is ensured by the cyclicity condition on $m^j_{n_1,..,n_N}$ due to \eqref{NCYCLij}. The multilinear products can then be written as follows
\begin{equation}
    \begin{split}
        m^j_{n_1,...,n_N}&\qty(\bigotimes_{l_1=1}^{n_1} f_{1,0}(x^1_{l_1})\tpt...\bigotimes_{l_N=1}^{n_N} f_{N,0}(x^N_{l_N}))\\&:=\frac{g_{n_1,...,n_N}}{\prod_{i=1}^N(n_i-\delta_{i,j})!} \int\dd[d]{x}\prod_{l_1=1}^{n_1} \delta^d(x-x^1_{l_1})...\prod_{l_N=1}^{n_N} \delta^d(x^N_{l_N})f_{j,1}(x),
    \end{split}
\end{equation}
and if any $f_{i,1}$ enters $m^j_{n_1,...,n_N}$  it is evaluated to be zero.\\
It is clear that, by construction, the interacting structure trivially satisfies the classical BV master equation and forms an $N$ component homotopy algebra \ref{NHOMALG:SUB}
\begin{equation}
    \qty(S,S)=0\Longrightarrow (\boldsymbol{Q}+\boldsymbol{m})^2=0,
\end{equation}
with
\begin{align}\label{baseDEF}
    \boldsymbol{Q}:=\sum_{j=1}^N\boldsymbol{Q}_j,\,\,\,\boldsymbol{m}:=\sum_{j=1}^N\sum_{n_1,...,n_N=0}^{m_1,...,m_N}\boldsymbol{m}^j_{n_1,...,n_N},\,\,\,\bar{\omega}:=\sum_{j=1}^N\omega_j.
\end{align}
If we formally introduce all renormalization vertices $\tilde{\boldsymbol{m}}$ then it satisfies the quantum BV master equation, forming a generalization of the loop-algebra
\begin{equation}
    \begin{gathered}
        \frac{1}{2}\qty(S,S)+\hbar\Delta S=0\Longrightarrow (\boldsymbol{Q}+\boldsymbol{m}+\hbar\tilde{\boldsymbol{m}}+\hbar\boldsymbol{U})^2=0,\\
        \pi_{(1)}\tilde{\boldsymbol{m}}^j_{n_1,...,n_N}=\tilde{m}^j_{n_1,...,n_N}:=\sum_{k=0}^\infty\hbar^kg^j_{k;n_1,...,n_N}\,m_{n_1,...,n_N}^{j;k},\,\,\,\,g_{k;n_1,...,n_N}\in\cc,
    \end{gathered}
\end{equation}
where to keep it contained we use co-derivations instead of the multilinear products \ref{NCODER:SEC}. The Poisson bi-vector is expressed as
\begin{align}
        \boldsymbol{U}=\sum_{j=1}^N\boldsymbol{U}_j=\sum_{j=1}^N\int\dd[d]{x}\boldsymbol{f}_{j,0}(x)\boldsymbol{f}_{j,1}(x),\,\,\,\pi_2\boldsymbol{U}=U,
\end{align}
where $\boldsymbol{f}_{j,0}$ and $\boldsymbol{f}_{j,1}$ are the zero co-derivations associated to the basis elements of $\Hs_j$
\begin{align}
    \pi_{(1)}\boldsymbol{f}_{j,0}(x)=f_{j,0}(x),\,\,\,\,\pi_{(1)}\boldsymbol{f}_{j,1}(x)=f_{j,1}(x).
\end{align}
The $\qty{n_1,...,n_N}$-point functions, in total analogy with \eqref{npf}, can then be computed via the homotopy transfer theorem by
\begin{equation}\label{Nnpf}
    \begin{split}
        &\expval{\prod_{l_1=1}^{n_1} \phi_1(x^1_{l_1})...\prod_{l_N=1}^{n_N} \phi_N(x^N_{l_N})}\\
        &\qquad\qquad:=(-1)^{n_1+...+n_N}{\omega}_{n_1,...,n_N}\qty(\pi_{n_1,...,n_N}\boldsymbol{F}'\boldsymbol{1},\bigotimes_{l_1=1}^{n_1} f_{1,1}(x^1_{l_1})\tpt...\bigotimes_{l_N=1}^{n_N} f_{N,1}(x^N_{l_N})),\\
        &{\omega}_{n_1,...,n_N}\qty(\bigotimes_{l_1=1}^{n_1} a^1_{l_1}(x^1_{l_1})\tpt...\bigotimes_{l_N=1}^{n_N} a^N_{l_N}(x^N_{l_N}),\bigotimes_{l_1=1}^{n_1} b^1_{l_1}(y^1_{l_1})\tpt...\bigotimes_{l_N=1}^{n_N} b^N_{l_N}(y^N_{l_N}))\\
        &\qquad\qquad=\prod_{j=1}^{N}\prod_{l_j=1}^{n_j}\omega_j(a^j_{l_j}(x^j_{l_j}),b^j_{l_j}(y^j_{l_j}))(-1)^{d(b^j_{l_j})d(\sum_{p=l_j+1}^{n_j}a^j_{p_j})},
    \end{split}
\end{equation}
where $\bar{\boldsymbol{F}}'$ is the morphism \eqref{NMorph2}
\begin{align}
    \bar{\boldsymbol{F}}'=\frac{1}{1-\bar{\alpha}\bar{\boldsymbol{h}}\boldsymbol{B}},\,\,\,\boldsymbol{B}=\boldsymbol{m}+\hbar\tilde{\boldsymbol{m}}+\hbar\boldsymbol{U}.
\end{align}
Like with \ref{AHTT} the contracting homotopy map $h_j$ is a map
\begin{align}
    h_j:\Hs_{j,1}\longrightarrow\Hs_{j,0},\,\,\,\,h_j:\Hs_{i,0}\longrightarrow0,\,\,\,h_j:\Hs_{i\neq j,1}\longrightarrow 0.
\end{align}
and is the propagator of the $j$-th field $h_j$, according to \ref{AHTT} has to be the propagator in order to satisfy the Hodge-Kodaria decomposition with $\bar{\boldsymbol{P}}=0$
\begin{align}\label{jh-prop}
    h_j f_{j,1}(x)=\int\dd[d]{y}\frac{1}{\alpha_j}\Delta_j(x-y)f_{j,0}(y),\,\,\,\,\Delta_j(x-y):=\int\frac{\dd[d]{k}}{(2\pi)^d}\frac{e^{ik\cdot(x-y)}}{k^2+m_j^2-\iota\varepsilon}.
\end{align}
By fully unpacking \eqref{Nnpf} we get that the $\qty{n_1,...,n_N}$-point function is given by
\begin{equation}\label{Nnpf2}
    \begin{split}
        &\expval{\prod_{l_1=1}^{n_1} \phi_1(x^1_{l_1})...\prod_{l_N=1}^{n_N} \phi_N(x^N_{l_N})}\\
    &=(-1)^{n_1+...n_N}\sum_{i=0}^{\infty}(\alpha\hbar)^i{\omega}_{n_1,...,n_N}\qty(\pi_{n_1,...,n_N}\qty{\boldsymbol{\bar{h}B}}^i\boldsymbol{1},\bigotimes_{l_1=1}^{n_1} f_{1,1}(x^1_{l_1})\tpt...\bigotimes_{l_N=1}^{n_N} f_{N,1}(x^N_{l_N})).
    \end{split}
\end{equation}
In order to simplify future computations we repackage \eqref{Nnpf} using $\bar{\omega}$ and the basis element $\bar{f}_a\in\bar{\Hs}$ defined as
\begin{align}\label{comBASE}
    \bar{f}^i_a:=\sum_{j=1}^Nf_{j,a}.
\end{align}
We also define the field element of $\Phi\in\bar{\Hs}$ in following way
\begin{align}\label{comField}
    \Phi:=\sum_{j=1}^{N}\phi_j=\int\dd[d]{x}\sum_{j=1}^N\phi_j(x)f_{j,0}(x).
\end{align}
Thanks to \eqref{baseDEF},\eqref{comBASE} and  \eqref{comField} we rewrite \eqref{Nnpf} from the correlator of $\qty{n_1,...,n_N}$ particles together to the correlator of $m=n_1+...+n_N$ fields $\Phi$
\begin{equation}\label{Nmpf}
    \begin{split}
        &\expval{\Phi(x_1)...\Phi(x_m)}:=(-1)^m\bar{\omega}_{m}\qty(\pi_{(m)}\boldsymbol{F}'\boldsymbol{1}, \bar{f}_{1}(x_1)\tpb...\bar{f}_{1}(x_m)),\\
        &\bar{\omega}_{m}\qty(\bar{a}_1(x_1)\tpb...\tpb\bar{a}_m(x_m),\bar{b}_1(y_1)\tpb...\tpb\bar{b}_m(y_m))=   \prod_{i=1}^{m}\bar{\omega}(\bar{a}_{i}(x_i),b_i(y_i))(-1)^{d(b_i)d(\sum_{p=i+1}^{n_j}a_{p_j})}.
    \end{split}
\end{equation}
Relation \eqref{Nmpf} ties the $N$ field amplitude formula \eqref{Nnpf} in form with the $1$ field amplitude formula from \ref{AHTT}\cite{Okawa:2022sjf,Konosu:2023pal,Konosu:2023rkm}. \eqref{Nmpf} will prove useful in order to prove that \eqref{Nnpf} satisfies the Schwinger-Dyson equation because we directly follow the proof provided in \cite{Okawa:2022sjf,Konosu:2023rkm}.\\
Note that \eqref{Nmpf} computes all correlators involving $m=n_1+...+n_N$ external fields, which are more correlators than what \eqref{Nnpf} computes but contains the result computed by \eqref{Nnpf}.

\subsection{Free theory and self interacting theory}
Correlators of free theories and self interacting theories with $N$ different fields are particularly easy to compute. Let us recall from \ref{NHOMALG:SUB} that in every $N$ component homotopy algebra $N$ $A_\infty/L_\infty$ algebras factor. In the case of free and self interacting theories, all multilinear products corresponding to interactions between different fields are set to zero, therefore the only homotopy algebraic structures entering the homotopy transfer theorem are the $N$ factored $A_\infty/L_\infty$ algebras. The implication of the factoring can be immediately seen in \eqref{Nnpf} because
\begin{align}\label{FACNnpf} 
    &\expval{\prod_{l_1=1}^{n_1} \phi_1(x^1_{l_1})...\prod_{l_N=1}^{n_N} \phi_N(x^N_{l_N})}:=\prod_{j=1}^N\expval{\prod_{l_j=1}^{n_j} \phi_j(x^j_{l_j})}_j,
\end{align}
where $\expval{\cdot}_j$ is the $j$-th correlator computed using the $1$ field method \ref{AHTT}. Therefore for free theories and self interacting theories only \eqref{Nnpf} simplifies to \eqref{FACNnpf} and in order to build correlators we only need to rely upon \ref{AHTT}. Furthermore, the consistency of \eqref{Nnpf} in the free/self interacting case, i.e. the Schwinger-Dyson equations, reduces to proving the Schwinger-Dyson equations for each field $\phi_j$ because the path-integral factors
\begin{equation}
    \begin{split}
        Z:&=\int\dd[d]{\phi_1}...\dd[d]{\phi_N}e^{\frac{\iota}{\hbar}S_{\rm free/self}[\phi_1,..,\phi_N]}=\int\dd[d]{\phi_1}...\dd[d]{\phi_N}e^{\sum_{j=1}^N\frac{\iota}{\hbar}S_{\rm free/self}[\phi_j]}\\&=\prod_{j=1}^N\int\dd[d]{\phi_j}e^{\frac{\iota}{\hbar}S_{\rm free/self}[\phi_j]}=\prod_{j=1}^NZ_j,
    \end{split}
\end{equation}
where $Z_j$ are the path integrals of the single field $\phi_j$.

\subsection{$\phi\phi\Phi$ Toy model}
In order to explicitly show the validity of \eqref{Nnpf} beyond the self interacting case, let us consider the following action functional
\begin{align}
    S[\phi,\Phi]:=\int\dd[d]x\qty[\frac{1}{2}\phi(x)\qty(m^2-\partial_\mu\partial^\mu)\phi(x)+\frac{1}{2}\Phi(x)\qty(M^2-\partial_\mu\partial^\mu)\Phi(x)-\lambda\frac{1}{2}\phi(x)^2\Phi(x)],
\end{align}
where $\phi\in\Hs_{1,0}$ and $\Phi\in\Hs_{2,0}$. According to \eqref{NAHTTREWRITE} we rewrite the action as
\begin{equation}
    \begin{split}
        S[\phi,\Phi]=\frac{1}{2}\omega_1(\phi,Q_1\phi)+\frac{1}{2}\omega_2(\Phi,Q_2\Phi)+\frac{1}{2}\frac{1}{2}\omega_1(\phi,m^1_{1,1}(\phi\tpt\Phi))+\frac{1}{2}\omega_2(\Phi,m^2_{2,0}(\phi^{\tp 2})),
    \end{split}
\end{equation}
where $m^2_{2,0}$ and $m^1_{1,1}$ are related by cyclicity \eqref{NCYCLij} in the following way
\begin{equation}
    \omega_2(\Phi,m^2_{2,0}(\phi^{\tp 2}))=\omega_1(\phi,m^1_{1,1}(\phi\tpt\Phi)).
\end{equation}
According to \ref{NAHTT}, after we trivially extend $\Hs_{j,0}$ to the graded $\Hs_j$ we can identify
\begin{equation}
    \begin{split}
        &Q_1f_{1,0}(x)=\qty(m^2-\partial_\mu\partial^\mu)f_{1,1}(x),\,\,Q_2f_{2,0}(x)=\qty(M^2-\partial_\mu\partial^\mu)f_{2,1}(x),\\
        &m^1_{1,1}(f_{1,0}(x_1)\tpt f_{2,0}(x_2))=-\lambda\int\dd[d]x\delta^d(x-x_1)\delta^d(x-x_2)f_{1,1}(x),\\
        &m^2_{2,0}(f_{1,0}(x_1)\tp f_{1,0}(x_2))=-\frac{\lambda}{2}\int\dd[d]x\delta^d(x-x_1)\delta^d(x-x_2)f_{2,1}(x),
    \end{split}
\end{equation}
and the associated contracting homotopy maps $h_1,h_2$ necessary to use the $N$ component homotopy transfer theorem \ref{NHTT} are the propagators of the free fields $\phi$ and $\Phi$
\begin{equation}
    \begin{split}
        &h_1 f_{1,1}(x)=\int\dd[d]{y}\frac{1}{\alpha_1}\Delta_1(x-y)f_{1,0}(y),\,\,\,\,\Delta_1(x-y):=\int\frac{\dd[d]{k}}{(2\pi)^d}\frac{e^{ik\cdot(x-y)}}{k^2+m^2-\iota\varepsilon},\\
        &h_2 f_{2,1}(x)=\int\dd[d]{y}\frac{1}{\alpha_2}\Delta_2(x-y)f_{2,0}(y),\,\,\,\,\Delta_2(x-y):=\int\frac{\dd[d]{k}}{(2\pi)^d}\frac{e^{ik\cdot(x-y)}}{k^2+M^2-\iota\varepsilon},
    \end{split}
\end{equation}
where $\alpha_1$ and $\alpha_2$ are the sign choice of the Hodge-Kodaira decomposition for each contracting homotopy map \ref{NHTT}.\\
In order to compute correlators we need the quantum (UV) completion of the classical action 
\begin{align}
    S_{\rm ren}[\phi,\Phi]:=\sum_{k,l=0}^{\infty}\sum_{n=0}^\infty\hbar^n\qty[g^1_{k,n,m}\,\omega_1(\phi,m_{k,l}^{1,n}(\phi^{\tp k}\tpt\Phi^{\tp l}))+g^2_{k,l,n}\,\omega_2(\Phi,m_{k,l}^{2,n}(\phi^{\tp k}\tpt\Phi^{\tp l}))],
\end{align}
with $g^j_{k,l,n}\in\cc$. Let us assume that we work in $d$ dimensions such that $\phi^2\Phi$ is renormalizable and we simplify the quantum (UV) completion by introducing counter-terms, namely
\begin{equation}
    \begin{split}
        &\Phi\,\,{\rm Tadpole}\Rightarrow\,m^{2,1}_{0,0}\boldsymbol{1}:=-Yf_{2,1}=-Y\int\dd[d]{x}f_{2,1}(x),\\
        &\phi\,\,{\rm Kinetic\,\,term}\Rightarrow\,m^{1,1}_{1,0}(f_{1,0}(x)):=\qty{(Z_m-1)-(Z_\phi-1)\partial^2}f_{1,1}(x),\\
        &\Phi\,\,{\rm Kinetic\,\,term}\Rightarrow\,m^{2,1}_{0,1}(f_{2,0}(x)):=\qty{(Z_M-1)-(Z_\Phi-1)\partial^2}f_{2,1}(x),\\
        &\phi\,\,{\rm Vertex}\Rightarrow\,m^{1,1}_{1,1}(f_{1,0}(x_1)\tpt f_{2,0}(x_2)):=-\frac{(Z_{\lambda}-1)}{2}\int\dd[d]{x}\delta^d(x-x_1)\delta^d(x-x_2)f_{1,1}(x),\\
        &\Phi\,\,{\rm Vertex}\Rightarrow\,m^{2,1}_{2,0}(f_{1,0}(x_1)\tp f_{1,0}(x_2)):=-\frac{(Z_{\lambda}-1)}{2}\int\dd[d]{x}\delta^d(x-x_1)\delta^d(x-x_2)f_{2,1}(x),
    \end{split}
\end{equation}
 where $Y$ and the $Z_j$ have to be expanded in terms of $\lambda$
\begin{equation}
    \begin{split}
        &Y=\lambda Y^{(1)}+\mathcal{O}(\lambda^3),\\
        &Z_\phi=1+\lambda^2Z_\phi^{(1)}+\mathcal{O}(\lambda^4),\,\,\,Z_{m}=1+\lambda^2Z_m^{(1)}+\mathcal{O}(\lambda^4)\\
        &Z_\Phi=1+\lambda^2Z_\Phi^{(1)}+\mathcal{O}(\lambda^4),\,\,\,Z_{M}=1+\lambda^2Z_M^{(1)}+\mathcal{O}(\lambda^4)\\
        &Z_{\lambda}=1^2Z_{\lambda}^{(1)}+\mathcal{O}(\lambda^4).
    \end{split}
\end{equation}
In order to compute ${n,m}$-point functions we need to evaluate the non vanishing contributions of
\begin{align}\label{NAHTTEXP}
    \pi_{n,m}\bar{\boldsymbol{F}}'\boldsymbol{1}= \pi_n\qty{1+\bar{\alpha}\bar{\boldsymbol{h}}\boldsymbol{B}+\bar{\alpha}^2\bar{\boldsymbol{h}}\boldsymbol{B}\bar{\boldsymbol{h}}\boldsymbol{B}+...}\id =\pi_n\qty{\bar{\alpha}\bar{\boldsymbol{h}}\boldsymbol{B}+\bar{\alpha}^2\bar{\boldsymbol{h}}\boldsymbol{B}\bar{\boldsymbol{h}}\boldsymbol{B}+...}\id,
\end{align}
which are linked to 
\begin{equation}
    \begin{split}
        &\pi_{n,m}\bar{\boldsymbol{h}}_1\boldsymbol{U}_1\pi_{n-2,m}\Longrightarrow\pi_{n,m}\qty{\bar{\boldsymbol{h}}_1\boldsymbol{U}_1}^j\pi_{n-2j,m},\\
        &\pi_{n,m}\bar{\boldsymbol{h}}_2\boldsymbol{U}_2\pi_{n,m-2}\Longrightarrow\pi_{n,m}\qty{\bar{\boldsymbol{h}}_2\boldsymbol{U}_1}^j\pi_{n,m-2j},\\
        &\pi_{n,m}\bar{\boldsymbol{h}}_1\boldsymbol{m}^1_{k,l}\pi_{n+k-1,m+l}\Longrightarrow \pi_{n,m}\qty{\bar{\boldsymbol{h}}_1\boldsymbol{m}^1_{k,l}}^j\pi_{n+j(k-1),m+jl},\\
        &\pi_{n,m}\bar{\boldsymbol{h}}_2\boldsymbol{m}^2_{k,l}\pi_{n+k,m+l-1}\Longrightarrow \pi_{n,m}\qty{\bar{\boldsymbol{h}}_2\boldsymbol{m}^2_{k,l}}^j\pi_{n+jk,m-j(l-1)},
    \end{split}
\end{equation}
plus other mixed relations we will not be using in this paper.
\subsubsection{One-point function example}
The only non vanishing contribution from \eqref{NAHTTEXP} is for the one point function $\expval{\Phi(x)}$, which reads  
\begin{align}
    \pi_{0,1}\bar{\boldsymbol{F}}'\boldsymbol{1}\sim \bar{\alpha} \lambda\hbar^2 Y^{(1)}\pi_{0,1}\bar{\boldsymbol{h}}\boldsymbol{m}^{2,1}_{0,0}\boldsymbol{1}+\bar{\alpha} \lambda\hbar^2\pi_{0,1}\bar{\boldsymbol{h}}\boldsymbol{m}^{2}_{2,0}\bar{\boldsymbol{h}}\boldsymbol{U}_1\boldsymbol{1}+\mathcal{O}(\lambda^2).
\end{align}
To evaluate the $1$-point function we develop $\bar{\boldsymbol{h}}$ according to \eqref{NHPARTFull}, where in this specific case it reduces to the special case 
\begin{align}
    \bar{\boldsymbol{h}}=\frac{\alpha_j}{\bar{\alpha}}\bar{\boldsymbol{h}}_j.
\end{align}
Let us firstly compute the following contributions of \eqref{NAHTTEXP}
\begin{equation}\label{NAHTTINT1}
    \begin{split}
        &\pi_{0,1}\bar{\boldsymbol{h}}\boldsymbol{m}^{2,1}_{0,0}\boldsymbol{1}=\frac{\alpha_2}{\bar{\alpha}}h_2m^{2,1}_{0,0}=-\frac{\alpha_2}{\bar{\alpha}}h_2f_{2,1}=-\frac{1}{\bar{\alpha}}\int\dd[d]{x}\int\dd[d]{y}\Delta_{2}(x-y)f_{2,0}(y),\\
        &\pi_{2,0}\bar{\boldsymbol{h}}\boldsymbol{U}_1\boldsymbol{1}=\frac{\alpha_1}{\bar{\alpha}}\qty(\id\tp h_1)\int\dd[d]{x}f_{1,0}(x)\tp f_{1,1}(x)=\frac{1}{\bar{\alpha}}\int\dd[d]{x}\int\dd[d]{y}\Delta_1(x-y)f_{1,0}(x)\tp f_{1,0}(y),\\
        &\pi_{0,1}\bar{\boldsymbol{h}}\boldsymbol{m}^{2}_{2,0}\pi_{2,0}\qty(f_{1,0}(x)\tp f_{1,0}(y))=-\frac{\lambda}{2}\frac{\alpha_2}{\bar{\alpha}}h_1\int\dd[d]{z}\delta^d(x-z)\delta^d(y-z)f_{2,1}(z)\\
        &\qquad\qquad\qquad\qquad\qquad\qquad\,\,\,\,\,\,\,\,\,
        =-\frac{\lambda}{2\bar{\alpha}}\int\dd[d]{z}\int\dd[d]{\tau}\delta^d(x-z)\delta^d(y-z)\Delta_2(z-\tau)f_{2,0}(\tau).
    \end{split}
\end{equation}
Thanks to the partial results we can directly compute the counter term contribution
\begin{equation}
    \begin{split}
        \bar{\alpha}\omega_{0,1}(\pi_{0,1}\bar{\boldsymbol{h}}\boldsymbol{m}^{2,1}_{0,0}\boldsymbol{1},f_{2,1}(x_1))=-\int\dd[d]{x}\int\dd[d]{y}\Delta_{2}(x-y)\delta^d(y-x_1)=-\frac{1}{M^2},
    \end{split}
\end{equation}
and the loop contribution
\begin{equation}
    \begin{split}
        &\bar{\alpha}^2\omega_{0,1}(\pi_{0,1}\bar{\boldsymbol{h}}\boldsymbol{m}\bar{\boldsymbol{h}}\boldsymbol{U}\boldsymbol{1},f_{2,1}(x_1))\\
        &=-\frac{\lambda}{2}\int\dd[d]{z}\dd[d]{\omega}\dd[d]{x}\dd[d]{y}\Delta_1(x-y){\omega}\Delta_2(z-\omega)\delta^d(x-z)\delta^d(y-z)\delta^d(x_1-\omega)\\
        &=\frac{\lambda}{M^2}\frac{1}{2}\int\dd[d]{k}\frac{1}{k^2+m^2},
    \end{split}
\end{equation}
The result of $\expval{\Phi(x_1)}$ is similar to \eqref{Tadpole} and reads
\begin{align}\label{2Tadpole}
    \expval{\Phi(x)}=\frac{\lambda\hbar^2}{M^2}\qty{\frac{1}{2}\int\frac{\dd[d]{k}}{(2\pi)^d}\frac{1}{k^2+m^2}+Y^{(1)}}+\mathcal{O}(\lambda^2).
\end{align}
where the $\bar{\alpha}$ and $\alpha_j$ dependency is cancelled by the $\bar{\alpha}^{-1}$ and $\alpha_j^{-1}$ present in the propagators and the the expansions of $\bar{\boldsymbol{h}}$. Therefore the computation of correlators via the homotopy transfer theorem is independent on the choice of $\bar{\alpha},\alpha_j$.
\subsubsection{Two-point function example}
Having two different fields we have two different non vanishing $2$-point functions, namely
\begin{align}
    \expval{\phi(x_1)\phi(x_2)}=\omega_{2,0}(\pi_{2,0}\bar{\boldsymbol{F}}'\boldsymbol{1},f_{1,1}(x_1)\tp f_{1,1}(x_2)),\\
    \expval{\Phi(x_1)\Phi(x_2)}=\omega_{0,2}(\pi_{0,2}\bar{\boldsymbol{F}}'\boldsymbol{1},f_{2,1}(x_1)\tp f_{2,1}(x_2)),
\end{align}
and we proceed to only show how to compute $\expval{\Phi(x_1)\Phi(x_2)}$ at $1$-loop order which without considering the counter-terms. $\expval{\Phi(x_1)\Phi(x_2)}$ has the following non vanishing contributions, modulo counter-terms,
\begin{align}
    \pi_{0,2}\bar{\boldsymbol{F}}'\boldsymbol{1}\sim\bar{\alpha}\hbar\pi_{0,2}\bar{\boldsymbol{h}}\boldsymbol{U}\boldsymbol{1}+\bar{\alpha}^4\hbar^4\pi_{0,2}\bar{\boldsymbol{h}}\boldsymbol{m}\qty(\bar{\boldsymbol{h}}\boldsymbol{m}\bar{\boldsymbol{h}}\boldsymbol{U}+\bar{\boldsymbol{h}}\boldsymbol{U}\bar{\boldsymbol{h}}\boldsymbol{m})\bar{\boldsymbol{h}}\boldsymbol{U}\boldsymbol{1}+\mathcal{O}(\lambda^4).
\end{align}
Using \eqref{NAHTTINT1} we can compute non vanishing intermediate steps
\begin{equation}\label{NAHTTID2}
    \begin{split}
        \bar{\alpha}^2\pi_{4,0}\bar{\boldsymbol{h}}\boldsymbol{U}\bar{\boldsymbol{h}}\boldsymbol{U}\boldsymbol{1}=&\int\dd[d]{x_1}\dd[d]{y_1}\dd[d]{x_2}\dd[d]{y_2}\Delta_1(x_1-y_2)\\
        &\left\{\right.\Delta_1(x_1-y_2)f_{1,0}(x_1)\tp f_{1,0}(y_1)\tp f_{1,0}(x_2)\tp f_{1,0}(y_2)+\\&
        +\Delta_1(y_1-y_2)f_{1,0}(x_1)\tp f_{1,0}(x_2)\tp f_{1,0}(y_1)\tp f_{1,0}(y_2)+\\
        &\left.
        +\Delta_1(y_1-y_2)f_{1,0}(x_2)\tp f_{1,0}(x_1)\tp f_{1,0}(y_1)\tp f_{1,0}(y_2)\right\},
    \end{split}
\end{equation}
\begin{equation}
    \begin{split}
        \bar{\alpha}^2\pi_{0,1}\bar{\boldsymbol{h}}\boldsymbol{m}\bar{\boldsymbol{h}}\boldsymbol{U}\boldsymbol{1}=&-\frac{\lambda}{2}\int\dd[d]{z}\dd[d]{\tau}\dd[d]{x}\dd[d]{y}\Delta_1(x-y){\tau}\Delta_2(z-\tau)\delta^d(x-z)\delta^d(y-z)f_{2,1}(\tau)\\
        =&\int\dd[d]{\tau}{\rm T}(\tau)f_{2,1}(\tau),
    \end{split}
\end{equation}

\begin{equation}\label{NAHTTID1}
    \begin{split}
        \bar{\alpha}\pi_{2,1}\bar{\boldsymbol{h}}\boldsymbol{m}&\qty(f_{1,0}(x_i)\tp f_{1,0}(x_j)\tp f_{1,0}(x_k)\tp f_{1,0}(x_l))=
        -\frac{\lambda}{2 }\int\dd[d]{z}\int\dd[d]{\tau}\Delta_2(z-\tau)\\
        &\left\{\delta^2(x_i-z)\delta^2(x_j-z)f_{1,0}(x_k)\tp f_{1,0}(x_l)+\delta^2(x_j-z)\delta^2(x_k-z)f_{1,0}(x_i)\tp f_{1,0}(x_l)+\right.
        \\&\left.+\delta^2(x_k-z)\delta^2(x_l-z)f_{1,0}(x_i)\tp f_{1,0}(x_j)\right\}\tpt f_{2,0}(\tau),
    \end{split}
\end{equation}
and lastly
\begin{equation}\label{NAHTTID3}
    \begin{split}
        \bar{\alpha}\pi_{0,1}\bar{\boldsymbol{h}}\boldsymbol{m}\qty(f_{1,0}(x_i)\tp f_{1,0}(x_j)\tpt f_{2,0}(\tau))=-\frac{\lambda}{2}\int\dd[d]{z'}\int\dd[d]{\tau'}&\Delta_2(z'-\tau')\delta^d(x_i-z')\delta^d(x_j-z')\\&f_{2,0}(\tau)\tp f_{2,0}(\tau').
    \end{split}
\end{equation}
From the previous intermediate computations it is easy to see that the tree level contribution to $\expval{\Phi(x_1)\Phi(x_2)}$ is the propagator
\begin{align}
    \omega_{0,2}(\bar{\alpha}\hbar\pi_{0,2}\bar{\boldsymbol{h}}\boldsymbol{U}\boldsymbol{1},f_{2,1}(x_1)\tp f_{2,1}(x_2))=\hbar\Delta_2(x_1-x_2),
\end{align}
with the right powers of $\hbar$. Then we have the first disconnected contribution from 
\begin{equation}
    \begin{split}
        \bar{\alpha}^4\hbar^4\omega_{0,2}(\bar{\alpha}\hbar\pi_{0,2}\bar{\boldsymbol{h}}\boldsymbol{m}\bar{\boldsymbol{h}}\boldsymbol{U}&\bar{\boldsymbol{h}}\boldsymbol{m}\bar{\boldsymbol{h}}\boldsymbol{U}\boldsymbol{1},f_{2,1}(x_1)\tp f_{2,1}(x_2))\\&=\qty[\int\dd[d]{\tau_1}{\rm T}(\tau_1)\Delta_2(\tau_1-x_1)]\qty[\int\dd[d]{\tau_2}{\rm T}(\tau_2)\Delta_2(\tau_2-x_2)]\\
        &=\frac{1}{4}\frac{\lambda^2\hbar^4}{M^4}\qty[\int\frac{\dd[d]{k}}{(2\pi)^d}\frac{1}{k^2+m^2}]^2,
    \end{split}
\end{equation}
which at the first loop order reduces to twice the tadpole contribution. Lastly we have the loop correction. In order to compute it we apply \eqref{NAHTTID1} to \eqref{NAHTTID2},  then we apply \eqref{NAHTTID3} to the result giving the following result
\begin{equation}\label{2PFCONN}
    \begin{split}
        &\bar{\alpha}^4\hbar^4\omega_{0,2}(\bar{\alpha}\hbar\pi_{0,2}\bar{\boldsymbol{h}}\boldsymbol{m}\bar{\boldsymbol{h}}\boldsymbol{m}\bar{\boldsymbol{h}}\boldsymbol{U}\bar{\boldsymbol{h}}\boldsymbol{U}\boldsymbol{1},f_{2,1}(x_1)\tp f_{2,1}(x_2))=\frac{3}{4}\frac{\lambda^2\hbar^4}{M^4}\qty[\int\frac{\dd[d]{k}}{(2\pi)^d}\frac{1}{k^2+m^2}]^2+\\&+
        \frac{\lambda^2\hbar^4}{2}\int\frac{\dd[d]{k_1}\dd[d]{k_1}}{(2\pi)^{2d}}\frac{e^{\iota(k_1+k_2)(x_1-x_2)}}{\qty[(k_1+k_2)^2+M^2]^2\qty[k_1^2+m^2]\qty[k_2^2+m^2]}+\\
        &+
        \frac{\lambda^2\hbar^4}{2}\int\frac{\dd[d]{k_1}\dd[d]{k_1}}{(2\pi)^{2d}}\frac{e^{-\iota(k_1+k_2)(x_1-x_2)}}{\qty[(k_1+k_2)^2+M^2]^2\qty[k_1^2+m^2]\qty[k_2^2+m^2]}+
        \\&+
        \frac{\lambda^2\hbar^4}{2}\int\frac{\dd[d]{k_1}\dd[d]{k_1}}{(2\pi)^{2d}}\frac{e^{\iota(k_1-k_2)(x_1-x_2)}}{\qty[(k_1-k_2)^2+M^2]^2\qty[k_1^2+m^2]\qty[k_2^2+m^2]}.
    \end{split}
\end{equation}
At first glance there are too many contributions to the connected part of \eqref{2PFCONN}, but after manipulating the contributions with the following transformations, one for each term,
\begin{equation}
    \begin{split}
        &k_1+k_2=p,\,k_2=q,\Longrightarrow J=\begin{pmatrix}
1 & 1 \\
0 & 1 
\end{pmatrix},\qty|J|=1,\,\,\,k_1=p-q,\\
&k_1+k_2=-p,\,k_2=q,\Longrightarrow J=\begin{pmatrix}
-1 & -1 \\
0 & 1 
\end{pmatrix},\qty|J|=-1,\,\,\,k_1=q-p,\\
&k_1-k_2=p,\,k_2=q,\Longrightarrow J=\begin{pmatrix}
1 & -1 \\
0 & 1 
\end{pmatrix},\qty|J|=1,\,\,\,k_1=p+q,
    \end{split}
\end{equation}
where $J$ is the Jacobian of the transformation, we realize that the first and second connected contributions cancel each other. The result of the $2$-point function for $\Phi$ is
\begin{equation}
    \begin{split}
        \expval{\Phi(x_1)\Phi(x_2)}&=\hbar \Delta_2(x_1-x_2)+\frac{\lambda^2\hbar^4}{M^4}\qty[\int\frac{\dd[d]{k}}{(2\pi)^d}\frac{1}{k^2+m^2}]^2+\\
        &+\frac{\lambda^2\hbar^4}{2}\int\frac{\dd[d]{p}}{(2\pi)^{d}}\frac{e^{p(x_1-x_2)}}{\qty[p^2+M^2]^2}\int\frac{\dd[d]{q}}{(2\pi)^{d}}\frac{1}{\qty[q^2+m^2]\qty[(q+p)^2+m^2]},
    \end{split}
\end{equation}
which corresponds to known literature and faithfully reproduces the symmetry factors for both connected and disconnected diagrams.

\subsection{Schwinger-Dyson equation}
The proof to the Schwinger-Dyson equation provided in \cite{Konosu:2023rkm}\footnote{More specifically to section 4.} can be recycled in order to prove that \ref{NAHTT} satisfies the Schwinger-Dyson equation. In order to recycle the proof in \cite{Konosu:2023rkm} we need to identify the objects used in \ref{NAHTT} with the formulation provided in \cite{Konosu:2023rkm}.\\
\begin{table}[!ht]
\centering
\def\arraystretch{1.5}
\begin{tabular}{ |p{4cm}||p{6cm}|p{2cm}|  }
 \hline
 \multicolumn{3}{|c|}{Table of identifications} \\
 \hline
 Object & Native formulation \ref{NAHTT} & Reference object \cite{Konosu:2023rkm}\\
 \hline
 Hilbert space       &    $\bar{\Hs}:=\displaystyle\bigoplus_{j=1}^{N}\Hs_j$    &    $\Hs$    \\
 Field element       &   $\Phi^i(x):=\displaystyle\sum_{j=1}^N\phi_j(x)\sigma^i_j$   &    $\hat{\Phi}^i$    \\
 Basis element       &   $\bar{f}^i:=\displaystyle\sum_{j=1}^Nf_{j,1}(x)\sigma^i_j$     &    $\hat{\lambda}^i(x)$    \\
 Field derivative    &     $\displaystyle\fdv{\Phi^i(x)}:=\displaystyle\sum_{j=1}^N\sigma^i_j\fdv{\phi_j(x)}$   &    $\displaystyle\fdv{\hat{\Phi}^i(x)}$    \\
 Projector           &   $\pi_{(n)}:=\displaystyle\sum_{n_1+...+n_N=n}\pi_{n_1,...,n_N}$     &    $\pi_n$    \\
 Symplectic form     &     $\bar{\omega}:=\displaystyle\sum_{j=1}^N\omega_j$   &    $\omega$    \\
 Multi-symplectic form     &     \eqref{Nmpf}  &    $\omega_n$    \\
 Co-derivations   & $\boldsymbol{m}_{(n)}:=\displaystyle\sum_{j=1}^N\sum_{n_1+...+n_N=n}\boldsymbol{m}^{j}_{n_1,...,n_N}$    &    $\boldsymbol{m}_n$        \\
 \hline
\end{tabular}
\caption{Table of identifications between \ref{NAHTT} and \cite{Konosu:2023rkm} sec. 4}
\label{table:1}
\end{table}
The identifications provided in table \eqref{table:1} are a $1:1$ map from the definitions present in this paper to the definitions given in \cite{Konosu:2023rkm}. Which implies that the proof given in \cite{Konosu:2023rkm} for the Schwinger-Dyson equation extends to \ref{NAHTT}.\\
Mathematically the identification process provided in table \eqref{table:1} is equivalent to the following isomorphisms 
\begin{equation}
    \begin{split}
        &\tHt:=\tH_1\tpt...\tpt\tH_N\simeq\tH_1\tp...\tp\tH_N:=\mathcal{T}\bar{\Hs},\\
        &\Hom(\tHt,\tHt)\simeq\Hom(\mathcal{T}\bar{\Hs},\mathcal{T}\bar{\Hs}),
    \end{split}
\end{equation}
because of $\tpt\simeq\tp$, provided that all $\Hs_j$ are on the same field $\kk$, which implies that we can always map \ref{NAHTT} back to the results of \cite{Konosu:2023rkm}.
    
\section*{Conclusions}\addcontentsline{toc}{section}{Conclusions}
In this paper we extended the notion of co-algebra to situations where the underlying Fock space/tensor product space has a finite and infinite number of particles/string types and boundaries on world-sheet topologies, as seen in sections \ref{NCOALG:CH} and \ref{INFCOALG:CH}. Thanks to the extended notion of co-algebra developed we demonstrated how the Wess-Zumino-Witten co-algebraic formulation \eqref{WZWCO} and homotopy transfer theorem \ref{HTT} can be extended in order to study theories with more complicated underlying Fock spaces/tensor product spaces.\\
At the same time we provided an axiomatic approach to the definition of Lagrangian field theories, in the form of a CAFT \ref{sec:caft}, using only co-algebraic and homotopy algebraic ingredients, regardless of specific assumptions on the theory. The CAFT approach highlights common features shared between all Lagrangian field theories. In this regard, it has been reinforced that, Lagrangian field theories that satisfy the classical or quantum Batalin–Vilkovisky master equation are built upon an homotopy algebraic structure \eqref{CBV} or loop-algebraic structure \eqref{QBV}, regardless of the structure of the Fock space/tensor product space.\\
The CAFT formulation of QFT/SFT allowed us to formulate the Sphere-Disk Homotopy Algebra \cite{Erbin_2020,Maccaferri:2023gcg} and Open-Closed Homotopy Algebra \cite{Kajiura:2004xu,Kajiura:2005sn,Kajiura:2006mt} in pure Wess-Zumino-Witten co-algebraic formulation \ref{SDHA:SEC}, agreeing with \cite{hoefel2009coalgebra} on the definition of co-derivations \ref{NCODER:SEC}. As a consequence of the CAFT formulation of Sphere-Disk Homotopy Algebra \ref{SDHA:SEC} we were able to derive the open-closed description duality of vertices as a consequence of the cyclical structure of the homotopy algebraic structure of the Sphere-Disk Homotopy Algebra. Furthermore the Open-Closed Homotopy Algebra \ref{OCHA:sec} can be interpreted as the result of the breaking of cyclicity in the Sphere-Disk Homotopy Algebra.\\
Thanks to the CAFT formulation, we were able to prove that MRV's formulation of quantum bosonic open-closed SFT \cite{Maccaferri:2023gcg} is the proper Wess-Zumino-Witten co-algebraic formulation of the theory \ref{QOCSFT}. Furthermore the linear operators formulated in \cite{Maccaferri:2023gcg} are in fact fully fledged co-derivations \ref{INFCODER:SEC}. Lastly, similarly to the Sphere-Disk Homotopy Algebra, we were able to derive the open-closed description duality \eqref{QOCSFTD1} and boundary equivalence relations of vertices \eqref{QOCSFTD2} as a consequence of the cyclical structure of the homotopy algebraic structure of the quantum open-closed SFT.\\
As a consequence of the extension of the homotopy transfer theorem to more generalized homotopy algebras, we were able to extend the methods defined in \cite{Okawa:2022sjf,Konosu:2023pal,Konosu:2023rkm} to compute scattering amplitudes to QFTs with $N$ different scalar fields \ref{NAHTT}.
\\\\
The natural continuation of this work will be to investigate the extension of the co-algebraic formulation from bosonic SFT to supersymmetric SFT in a way to reproduce the Susy Open-Closed Homotopy Algebra relations \cite{Kunitomo:2022qqp} and try to define the full Susy open-closed SFT.\\
Another possible continuation to this work is to actively extend the co-algebraic amplitude computation methods \cite{Okawa:2022sjf,Konosu:2023pal,Konosu:2024zrq,Chiaffrino:2023wxk,Jurco:2019yfd} to many particles QFTs with spin $\frac{1}{2},1$ degrees of freedom and possibly Gauge symmetries. Lastly it might prove noteworthy to investigate the possible connection between iterated integrals and the systematic nature of computing correlators with the homotopy transfer theorem, in order to formulate a more efficient method to compute correlators with the homotopy transfer theorem.
\\\\
To conclude, it is the author's opinion that SFT is giving rise to vast number of advanced mathematical methods, that hopefully will bring forward many more results applicable to both SFT and QFT, therefore this line of research should continue to be an exciting and fruitful research topic for many years to come.

\section*{Acknowledgments}
\addcontentsline{toc}{section}{Acknowledgments}
First and foremost, I would like to thank my master advisor and mentor Carlo Maccaferri for the constant academic support, valuable advice and guidance over the course of the writing of this paper.\\
I would like to thank Alberto Ruffino for his collaboration on the project and for the many fruitful discussions entertained.\\
I would also like to thank Daniele Migliorati for his help on the more formal mathematical aspects of this paper
.\\
I would like to thank Jackub Vo\v{s}mera, Cristhiam Lopez Arcos, Hyungrok Kim and Christian Sämann for pointing out relevant literature to the paper and fruitful discussions.
\\
Lastly I would like to sincerely thank my PhD advisor Oliver Schlotterer for his support on finishing this work and his mentorship on writing papers.
\\\\
I would like to thank the University of Turin, for hosting the first stages of this work, started during my master degree.\\
I would also like to thank the University of Uppsala for hosting the final stages of this work during the start of my PhD.
\\\\
This work was partially funded by the European Union (ERC Synergy Grant MaScAmp 101167287). Views and opinions expressed are however those of the author(s) only and do not necessarily reflect those of the European Union or the European Research Council. Neither
the European Union nor the granting authority can be held responsible for them.

    \appendix
    
 \renewcommand{\thesection}{\Alph{section}}

\section{ Homotopy transfer theorem proofs}

\subsection{Derivation of F and F'}\label{Appendix_Hom}
To derive the form of $\boldsymbol{F}$ and $\boldsymbol{F}'$ discussed in \ref{HTT} \cite{Okawa:2022sjf,Erbin_2020,Bonezzi:2023xhn}  we start with the following ansatz: let
there be a linear operator $\boldsymbol{A}$ such that
\begin{align}
    \boldsymbol{A}:\tH\longrightarrow \tH,\,\,d(\boldsymbol{A}).
\end{align}
Morphisms $\boldsymbol{F}$ and $\boldsymbol{F}'$ are defined as the formal power series
\begin{align}
    \boldsymbol{F}:=\boldsymbol{P}\sum_{n=0}^\infty\alpha_n\boldsymbol{A}^n,\,\boldsymbol{F}':=\sum_{n=0}^\infty\tilde{\alpha}_n\boldsymbol{A}^n\boldsymbol{P},\,\alpha,\tilde{\alpha}\in\cc.
\end{align}
The coefficients $\alpha,\tilde{\alpha}$ are fixed by the right invertibility condition
\begin{align}
    \boldsymbol{F}\boldsymbol{F}'=\id \,\Rightarrow\,0=\boldsymbol{P}\qty{\sum_{n,m=0}^\infty\alpha_n\tilde{\alpha}_m\boldsymbol{A}^{n+m}-1}\boldsymbol{P}.
\end{align}
By rearranging the sums and factoring common powers of $\boldsymbol{A}^n$ we arrive at the recursive relations
\begin{align}\label{APPA1}
    \alpha_0\tilde{\alpha}_0=1,\,\sum_{l=0}^{n\geq1}\alpha_l\tilde{\alpha}_{n-l}=0.
\end{align}
The two simplest solutions to \eqref{APPA1} are
\begin{align}
    & \alpha_n=1\,\forall n\geq0,\,\tilde{\alpha}_0=1,\,\tilde{\alpha}_1=-1,\,\tilde{\alpha}_m=0\,\forall m\geq 2\Longrightarrow\,\boldsymbol{F}=\boldsymbol{P}\frac{1}{1-\boldsymbol{A}},\,\boldsymbol{F}'=(1-\boldsymbol{A})\boldsymbol{P},\\
    &\tilde{\alpha}_n=1\,\forall n\geq0,\,\alpha_0=1,\,\alpha_1=-1,\,\alpha_m=0\,\forall m\geq 2\Longrightarrow\,\boldsymbol{F}=(1-\boldsymbol{A})\boldsymbol{P},\,\boldsymbol{F}'=\boldsymbol{P}\frac{1}{1-\boldsymbol{A}},
\end{align}
where we formally collapsed the geometric series to $\frac{1}{1-\boldsymbol{A}}$.\\
The next step in deriving $\boldsymbol{A}$ is to check the morphism condition in both cases
\begin{align}
    \partial'=\boldsymbol{F}\partial\boldsymbol{F}',\,\,\,{\rm remembering\,\,\,that}\,\,\,\,\partial'\boldsymbol{P}=\boldsymbol{P}\partial,
\end{align}
which implies two side conditions on $\boldsymbol{A}$ depending on the choice of $\alpha,\tilde{\alpha}$
\begin{align}
    &\partial'=\boldsymbol{P}\frac{1}{1-\boldsymbol{A}}\partial(1-\boldsymbol{A})\boldsymbol{P}\Rightarrow\,\boldsymbol{P}\frac{1}{1-\boldsymbol{A}}\partial\boldsymbol{A}\boldsymbol{P}=0\,\Rightarrow\,\boldsymbol{AP}=0,\,\frac{1}{1-\boldsymbol{A}}\boldsymbol{P}=\boldsymbol{P},\\
    &\partial'=\boldsymbol{P}(1-\boldsymbol{A})\partial\frac{1}{1-\boldsymbol{A}}\boldsymbol{P}\Rightarrow\,\boldsymbol{P}\boldsymbol{A}\partial\frac{1}{1-\boldsymbol{A}}\boldsymbol{P}=0\,\Rightarrow\,\boldsymbol{PA}=0,\,\boldsymbol{P}\frac{1}{1-\boldsymbol{A}}=\boldsymbol{P}.
\end{align}
Before deriving the form of $\boldsymbol{A}$ we require some identities which will simplify computations. The first identity is the action of a graded operator $\boldsymbol{X}$ on $\frac{1}{1-\boldsymbol{A}}$
\begin{align}
    \boldsymbol{X}\frac{1}{1-\boldsymbol{A}}=\frac{1}{1-\boldsymbol{A}}\boldsymbol{X}+\frac{1}{1-\boldsymbol{A}}\commutator{\boldsymbol{X}}{\boldsymbol{A}}\frac{1}{1-\boldsymbol{A}},
\end{align}
where $\commutator{\boldsymbol{X}}{\boldsymbol{A}}$ is the graded commutator. The second identity is specific for the side condition $\boldsymbol{PA} = 0 $
and reads
\begin{align}
    \boldsymbol{PX} = \boldsymbol{P}\frac{1}{1-\boldsymbol{A}}\boldsymbol{X}=\boldsymbol{P}\frac{1}{1-\boldsymbol{A}}\boldsymbol{X}+\boldsymbol{P}\frac{1}{1-\boldsymbol{A}}\boldsymbol{X}\frac{1}{1-\boldsymbol{A}}\boldsymbol{A},
\end{align}
and similarly for the side condition $\boldsymbol{AP} = 0 $
\begin{align}\label{APPA3}
    \boldsymbol{XP}=\frac{1}{1-\boldsymbol{A}}\boldsymbol{XP}-\boldsymbol{A}\frac{1}{1-\boldsymbol{A}}\boldsymbol{X}\frac{1}{1-\boldsymbol{A}}\boldsymbol{P}.
\end{align}
The functional form of $\boldsymbol{A}$ directly depends on the side condition and the choice of signs in the Hodge-
Kodaira decomposition. To derive specific form of $\boldsymbol{A}$ we chose to express the Hodge-Kodaira decomposition as
\begin{align}
    \boldsymbol{P}=\id+\alpha\commutator{\boldsymbol{\partial}}{\boldsymbol{h}},
\end{align}
where by choosing $\alpha = +1$ we recover the morphisms featured in \cite{Erbin_2020,Bonezzi:2023xhn} and by choosing $\alpha = -1$ we recover the one featured in \cite{Okawa:2022sjf} provided the right choice of side conditions. To derive the form of A we impose that $\boldsymbol{D}'=\boldsymbol{FDF}'$ is nilpotent
\begin{align}\label{APPA2}
    (\boldsymbol{D}')^2=0\,\Longrightarrow\, \boldsymbol{\partial}'\boldsymbol{B}'+\boldsymbol{B}'\boldsymbol{\partial}'+\boldsymbol{B}'\boldsymbol{B}'=0,\,\,\boldsymbol{D}'=\boldsymbol{\partial}'+\boldsymbol{B}'.
\end{align}
Let us derive $\boldsymbol{A}$ with side conditions $\boldsymbol{AP} = 0$ where $\boldsymbol{F}' = (1 - \boldsymbol{A})\boldsymbol{P}$ simplifies to $\boldsymbol{F}' = \boldsymbol{P}$. 
Let us start
by computing the single entries of \eqref{APPA2}
\begin{align}
    &\boldsymbol{\partial}'\boldsymbol{B}'=\boldsymbol{P}\frac{1}{1-\boldsymbol{A}}\boldsymbol{\partial B P}+\boldsymbol{P}\frac{1}{1-\boldsymbol{A}}\commutator{\boldsymbol{\partial}}{\boldsymbol{A}}\frac{1}{1-\boldsymbol{A}}\boldsymbol{BP},\\
    &\boldsymbol{B}'\boldsymbol{\partial}'=\boldsymbol{P}\frac{1}{1-\boldsymbol{A}}\boldsymbol{B\partial P},\\
    &\boldsymbol{B}'\boldsymbol{B}'=\boldsymbol{P}\frac{1}{1-\boldsymbol{A}}\boldsymbol{B}(1+\alpha\commutator{\boldsymbol{\partial}}{\boldsymbol{h}})\frac{1}{1-\boldsymbol{A}}\boldsymbol{BP}.
\end{align}
We now use \eqref{D_rel} and \eqref{APPA3} to further manipulate
\begin{equation}
    \boldsymbol{\partial}'\boldsymbol{B}'+\boldsymbol{B}'\boldsymbol{\partial}'=\boldsymbol{P}\frac{1}{1-\boldsymbol{A}}\qty{-\boldsymbol{B}+\boldsymbol{BA}+\commutator{\boldsymbol{\partial}}{\boldsymbol{A}}}\frac{1}{1-\boldsymbol{A}}\boldsymbol{P}.
\end{equation}
By substituting our results into \eqref{APPA2} we get
\begin{align}
    0=\boldsymbol{P}\frac{1}{1-\boldsymbol{A}}\qty{\boldsymbol{B}-\boldsymbol{B}+\boldsymbol{BA}+\commutator{\boldsymbol{\partial}}{\boldsymbol{A}}+\alpha\commutator{\boldsymbol{B\partial}}{\boldsymbol{h}}}\frac{1}{1-\boldsymbol{A}}\boldsymbol{P},
\end{align}
therefore
\begin{align}
    (\boldsymbol{D}')^2=0\,\Longrightarrow \boldsymbol{BA}+\commutator{\boldsymbol{\partial}}{\boldsymbol{A}}+\alpha\commutator{\boldsymbol{B\partial}}{\boldsymbol{h}}=0
\end{align}
By unpacking the commutators and using \eqref{D_rel} we get to 
\begin{align}
    0=(\alpha\boldsymbol{Bh}-\boldsymbol{A} )\boldsymbol{\partial}+\boldsymbol{\partial}(\boldsymbol{A}-\alpha\boldsymbol{Bh})+\boldsymbol{B}(\boldsymbol{A}-\alpha\boldsymbol{Bh}),
\end{align}
and we conclude that
\begin{align}
    \boldsymbol{A}=\alpha\boldsymbol{Bh},\,\,\boldsymbol{F}=\boldsymbol{P}\frac{1}{1-\alpha\boldsymbol{Bh}},\,\,\boldsymbol{F}'=(1-\alpha\boldsymbol{Bh})\boldsymbol{P}.
\end{align}
The derivation with the side condition $\boldsymbol{PA} = 0$ follows the same logic and it implies that
\begin{align}
    \boldsymbol{A}=\alpha\boldsymbol{hB},\,\,\boldsymbol{F}=\boldsymbol{P}(1-\alpha\boldsymbol{hB}),\,\,\boldsymbol{F}'=\frac{1}{1-\alpha\boldsymbol{hB}}\boldsymbol{P}.
\end{align}
\subsection{ Proof of co-algebraic extension of $h$ }\label{h-action}
To extend the action of the contacting homotopy map $h$ from the vector space $\Hs$ to the full tensor
algebra $\tH$ we will rely upon properties of $\boldsymbol{P}$ and the Hodge-Kodaira decomposition
\begin{align}
    \boldsymbol{P}\pi_N=(\underbrace{P\tp...\tp P}_n)\pi_n,\,\,\,\,P=\id+\alpha\commutator{\partial}{h}.
\end{align}
To define $\boldsymbol{h}$ we take the Hodge-Kodaira decomposition on the tensor algebra
\begin{align}\label{APPAh1}
    \boldsymbol{P}=\id+\alpha\commutator{\boldsymbol{\partial}}{\boldsymbol{h}},
\end{align}
and restrict its action on the $\Hs^{\tp n}$ subspace of $\tH$, namely
\begin{align}
    \boldsymbol{P}\pi_n=(\underbrace{P\tp...\tp P}_n)\pi_n=\id\pi_n+\alpha\commutator{\boldsymbol{\partial}}{\boldsymbol{h}}\pi_n,
\end{align}
where the right side of the equation will provide the connection between $h$ and $\boldsymbol{h}$. Now we rearrange the expression to
\begin{align}\label{APPAh2}
    \alpha\commutator{\boldsymbol{\partial}}{\boldsymbol{h}}\pi_n=(\underbrace{P\tp...\tp P}_n)\pi_n-\pi_n.
\end{align}
By remembering that $\pi_n$ satisfies
\begin{align}
    \pi_n=\pi_i\tp\pi_{n-i},\,\,\,\pi_n=\pi_{j-1}\tp\underbrace{\pi_1}_{j-th}\tp\pi_{n-j},\,\,\,\pi_{j<0}=0,\,\,\,\pi_0=1,
\end{align}
we can choose to isolate a specific $P$ element and manipulate it as follows
\begin{equation}
    \begin{split}
        (\underbrace{P\tp...\tp P}_n)\pi_n & =(\underbrace{P\tp...\tp P}_{j-1})\pi_n\tp P\pi_{j-1} \tp(\underbrace{P\tp...\tp P}_{n-j})\pi_{n-1}\\
        &=\boldsymbol{P}\pi_{j-1}\tp\qty(\id+\alpha\commutator{\partial}{h})\tp\boldsymbol{P}\pi_{n-j}\\
        &=\alpha \boldsymbol{P}\pi_{j-1}\tp\commutator{\partial}{h}\pi_1\tp\boldsymbol{P}\pi_{n-j}+\qty(\id+\alpha\commutator{\boldsymbol{\partial}}{\boldsymbol{h}})\pi_{j-1}\tp\pi_1\tp\boldsymbol{P}\pi_{n-j}\\
        &=\alpha \boldsymbol{P}\pi_{j-1}\tp\commutator{\partial}{h}\pi_1\tp\boldsymbol{P}\pi_{n-j}+\alpha\commutator{\boldsymbol{\partial}}{\boldsymbol{h}}\pi_{j-1}\tp\pi_1\tp\boldsymbol{P}\pi_{n-j}\\&+\alpha\pi_j\tp\commutator{\boldsymbol{\partial}}{\boldsymbol{h}}\pi_{n-j}+\pi_n,
    \end{split}
\end{equation}
or equivalently by switching the order of operations on $\boldsymbol{P}$
\begin{equation}
    \begin{split}
        (\underbrace{P\tp...\tp P}_n)\pi_n 
        &=\alpha \boldsymbol{P}\pi_{j-1}\tp\commutator{\partial}{h}\pi_1\tp\boldsymbol{P}\pi_{n-j}+\alpha\boldsymbol{P}\pi_{j-1}\tp\pi_1\tp\commutator{\boldsymbol{\partial}}{\boldsymbol{h}}\pi_{n-j}\\&+\alpha\commutator{\boldsymbol{\partial}}{\boldsymbol{h}}\pi_{j-1}\tp\pi_{n-j+1}+\pi_n.
    \end{split}
\end{equation}
Because $\boldsymbol{\partial}$ commutes with $\boldsymbol{P}$ and is a co-derivation we can pull out the graded commutator
\begin{align}
    \boldsymbol{P}\pi_n=\commutator{\boldsymbol{\partial}}{\alpha\boldsymbol{P}\pi_{j-1}\tp h \tp \boldsymbol{P}\pi_{n-j}+\alpha \boldsymbol{h}\pi_{j-1}\tp\pi_1\tp\boldsymbol{P}\pi_{n-j}+\alpha\pi_j\tp\boldsymbol{h}\pi_{n-j}  }+\pi_n.
\end{align}
Substituting into \eqref{APPAh2} we reach the result 
\begin{align}\label{APPAh3}
    \boldsymbol{h}\pi_n=\boldsymbol{P}\pi_{j-1}\tp h \tp \boldsymbol{P}\pi_{n-j}+ \boldsymbol{h}\pi_{j-1}\tp\pi_1\tp\boldsymbol{P}\pi_{n-j}+\pi_j\tp\boldsymbol{h}\pi_{n-j},
\end{align}
or equivalently by switching the order of operations on $\boldsymbol{P}$
\begin{align}\label{APPAh4}
    \boldsymbol{h}\pi_n=\boldsymbol{P}\pi_{j-1}\tp h \tp \boldsymbol{P}\pi_{n-j}+ \boldsymbol{P}\pi_{j-1}\tp\pi_1\tp\boldsymbol{h}\pi_{n-j}+\boldsymbol{h}\pi_{j-1}\tp\pi_{n-j+1}.
\end{align}
Finally, by taking \eqref{APPAh4} and setting $j=n$ we derive the first definition of \eqref{lr-h-def}
\begin{align}
    \boldsymbol{h}\pi_n=\boldsymbol{P}\pi_{n-1}\tp h+\boldsymbol{h}\pi_{n-1}\tp\pi_{1},
\end{align}
and by taking \eqref{APPAh3} and setting $j=1$ we derive the second definition of \eqref{lr-h-def}
\begin{align}
    \boldsymbol{h}\pi_n= h \tp \boldsymbol{P}\pi_{n-1}+\pi_1\tp\boldsymbol{h}\pi_{n-1}.
\end{align}
\section{Proof of short hand result for repeated derivation}\label{Appendix_CAFT}
To prove \eqref{princ-result} and \eqref{princ-result2} we start by taking the CAFT action
\begin{align}
    S[\gl]:=\int_0^1\dd{t}\omega(\pi_1\boldsymbol{\partial}_t\gl,\pi_1\boldsymbol{m}\gl),
\end{align}
and perform a degree zero cyclic field redefinition generated by the co-derivation $\boldsymbol{\delta}$
\begin{align}
    S'[\gl]:=\int_0^1\dd{t}\omega(\pi_1\boldsymbol{\partial}_t e^{\varepsilon\boldsymbol{\delta}}\gl,\pi_1\boldsymbol{m}e^{\varepsilon\boldsymbol{\delta}}\gl):=\int_0^1\dd{t}\omega(\pi_1\boldsymbol{\partial}_t \boldsymbol{F}_\varepsilon\gl,\pi_1\boldsymbol{m}\boldsymbol{F}_\varepsilon\gl).
\end{align}
A co-homomorphism defined via the exponentiation of a co-derivation always allows for the inverse co-homomorphisms
\begin{align}
    \boldsymbol{F}_\varepsilon:=e^{\varepsilon\boldsymbol{\delta}}\,\Longrightarrow\,\boldsymbol{F}^{-1}_\varepsilon:=e^{-\varepsilon\boldsymbol{\delta}}.
\end{align}
It's always possible to rewrite the integrand of $S'[\gl]$ in the following form
\begin{align}
    \omega(\pi_1\boldsymbol{\partial}_t\boldsymbol{F}_\varepsilon\mathcal{G},\pi_1\boldsymbol{m}\boldsymbol{F}_\varepsilon\mathcal{G}) =&\, \omega(\pi_1\boldsymbol{\partial}_t\mathcal{G},\pi_1\boldsymbol{F}_\varepsilon^{-1}\boldsymbol{m}\boldsymbol{F}_\varepsilon\mathcal{G})+\omega(\pi_1\boldsymbol{F}_\varepsilon^{-1}\commutator{\boldsymbol{\partial}_t}{\boldsymbol{F}_{\varepsilon}}\mathcal{G},\pi_1\boldsymbol{m}\mathcal{G})
            \\
            &+\omega(\pi_1\commutator{\boldsymbol{\partial}_t}{\boldsymbol{F}_{\varepsilon}}\mathcal{G},\pi_1\commutator{\boldsymbol{m}}{\boldsymbol{F}_\varepsilon}\mathcal{G}).
\end{align}
Let us explicitly write the field redefinition as a shift of the base field $\Psi$ 
\begin{align}\label{diff_delta}
    \pi_1\boldsymbol{\delta}=\delta:=\delta\Psi^a\overset{\rightarrow}{\pdv{}{\Psi^a}},\,\,\,\,\overset{\rightarrow}{\pdv{}{\Psi^a}}\Psi=f_a,\,\,\,\overset{\rightarrow}{\pdv{}{\Psi^a}}\gl=\boldsymbol{f}_a\gl.
\end{align}
We require that the $\delta\Psi$ is independent of the WZW parametrization 
\begin{align}
    \partial_t\delta\Psi=0,
\end{align}
and that
\begin{align}
    \commutator{\partial_t}{\overset{\rightarrow}{\pdv{}{\Psi^a}}}=0\,\Longrightarrow\,\commutator{\boldsymbol{\partial}_t}{\boldsymbol{\delta}}=0.
\end{align}
Thanks to this we can write the variation of the action as follows
\begin{align}
    \delta_\varepsilon S[\mathcal{G}]=S[\mathcal{G}]'-S[\mathcal{G}]=\sum_{n=1}^{\infty}\frac{1}{n!}\int_0^1\dd{t}\omega(\pi_1\boldsymbol{\partial}_t\mathcal{G},\pi_1\commutator{\commutator{\commutator{\boldsymbol{n}}{\varepsilon\boldsymbol{\delta}}}{\varepsilon\boldsymbol{\delta}}...}{\varepsilon\boldsymbol{\delta}}\mathcal{G})
\end{align}
To find the repeated differentiation of the action we connect the variation of the action with the variational Taylor expansion around $\varepsilon$
\begin{align}
        \delta_\varepsilon S[\mathcal{G}]:=\sum_{n=0}^{\infty}\frac{\varepsilon^n}{n!}\delta\Psi^{a_1}...\delta\Psi^{a_n}\overset{\rightarrow}{\pdv{}{\Psi^{a_n}}}...\overset{\rightarrow}{\pdv{}{\Psi^{a_1}}}S[\mathcal{G}].
\end{align}
Using the cyclicity of $\boldsymbol{\delta}$, the fact that $\pi_1\boldsymbol{\delta}^{n\geq2}\mathcal{G}=0$, $\pi_1\boldsymbol{\partial}_t\boldsymbol{\delta}\mathcal{G}=0$  we find that for the $n$-th power in $\varepsilon$ 
    \begin{equation}
        \begin{split}
            \delta_\varepsilon^n S[\mathcal{G}]=\frac{\varepsilon^n}{n!}\int_0^1\dd{t}\omega(\pi_1\boldsymbol{\partial}_t\mathcal{G},\pi_1\boldsymbol{n}\boldsymbol{\delta}^{n}\mathcal{G}),
        \end{split}
    \end{equation}
which connected to the $n$-th power of $\varepsilon$ in the Taylor expansion and writing  $\boldsymbol{\delta}$ in terms of \eqref{diff_delta} gives us the result \eqref{princ-result}
\begin{align}
    \overset{\rightarrow}{\pdv{}{\Phi^{a_n}}}...\overset{\rightarrow}{\pdv{}{\Phi^{a_1}}}S[\gl]=\int_0^1\dd{t}\omega(\pi_1\boldsymbol{\partial}_t\gl,\pi_1\boldsymbol{n}\boldsymbol{f}_{a_n}...\boldsymbol{f}_{a_1}\gl)\,\,\,\forall\,n\geq0,\,\,\,\,\,\boldsymbol{\partial}_t\boldsymbol{f}_{a_i}=0,
\end{align}
For the alternative formulation of the case $n=1$ we just need to smartly add zero in the form of
\begin{align}
    \omega(\pi_1\boldsymbol{\partial}_t\boldsymbol{\delta}\mathcal{G},\pi_1\boldsymbol{n}\mathcal{G})=0,
\end{align}
and by using the cyclicity we can render the expression independent from the WZW parametrization
\begin{align}
    \overset{\rightarrow}{\pdv{S}{\Psi^a}}=(-1)^{d(\Psi^a)}\omega(\pi_1\boldsymbol{f}_a\mathcal{G},\pi_1\boldsymbol{n}\mathcal{G}).
\end{align}
\section{$N$ component co-algebra pedagogical example}\label{B1}
Let us provide a specific example of the proof of \eqref{N-uplift}. Let us work on a 2 component tensor co-algebra
and let us want to uplift to co-derivation the map
\begin{align}
    c^1_{1,1}:\Hs_1\tpt\Hs_2\longrightarrow\Hs_1.
\end{align}
Let us recall the properties of a co-derivation
\begin{equation}
    \begin{split}
        & \bDelta_j\boldsymbol{c}^1_{1,1}=\qty{\boldsymbol{c}^1_{1,1}\tp_j'\id_j+\id_j\tp_j'\boldsymbol{c}^1_{1,1}}\bDelta_j,\,\,j\in\qty[1,2],\\
        &\pi_{1,0}\boldsymbol{c}^1_{1,1}=c^1_{1,1},\,\,\,\boldsymbol{c}^1_{1,1}\pi_{1,1}=c^1_{1,1}.
    \end{split}
\end{equation}
As an example we feed the co-derivation $\boldsymbol{c}^1_{1,1}$ with an element of $\Hs^{\tp 2}_1\tpt\Hs^{\tp 2}_2$, which is the first non trivial application of a co-derivation. Let us then apply first $\bDelta_1$, where using \eqref{NCOLEIB} and \eqref{NCOPROJ} leads
to
\begin{align}\label{BPPA1}
    \bDelta_1\boldsymbol{c}^1_{1,1}\pi_{2,2}=\qty{\boldsymbol{c}^1_{1,1}\pi_{1,2}\tp_1'\pi_{1,0}+\pi_{1,0}\tp_1'\boldsymbol{c}^1_{1,1}\pi_{1,2}}\bDelta_1.
\end{align}
Now that we isolated $\pi_{1,2}$ any further split introduced by $\bDelta_1$ is unnecessary (gives only trivial results). Let us now isolate the first element of \eqref{BPPA1} and split it using $\bDelta_2$ in the following way
\begin{equation}\label{BPPA2}
    \begin{split}
        (\bDelta_2\tp_1'\id_1)\qty{\boldsymbol{c}^1_{1,1}\pi_{1,2}\tp_1'\pi_{1,0}}\bDelta_1=
        \qty[\qty{\boldsymbol{c}^1_{1,1}\pi_{1,1}\tp_2'\pi_{0,1}+\pi_{0,1}\tp_2'\boldsymbol{c}^1_{1,1}\pi_{1,1}}\tp_1'\pi_{1,0}](\bDelta_2\tp_1'\id_1)\bDelta_1.
    \end{split}
\end{equation}
We can perform the same process seen in \eqref{BPPA2} to the second element of \eqref{BPPA1} leading us to
\begin{equation}\label{BPPA3}
    \begin{split}
        (\id_1\tp_1'\bDelta_2)\qty{\boldsymbol{c}^1_{1,1}\pi_{1,2}\tp_1'\pi_{1,0}}\bDelta_1=
        \qty[\pi_{1,0}\tp_1'\qty{\boldsymbol{c}^1_{1,1}\pi_{1,1}\tp_2'\pi_{0,1}+\pi_{0,1}\tp_2'\boldsymbol{c}^1_{1,1}\pi_{1,1}}](\id_1\tp_1'\bDelta_2)\bDelta_1.
    \end{split}
\end{equation}
Note that due to co-associativity  $(\id_1\tp_1'\bDelta_2)$ and $(\bDelta_2\tp_1'\id_1)$ are the same operator.\\
Now that we have exhausted all non trivial applications of $\bDelta_j$ on $\boldsymbol{c}^1_{1,1}\pi_{2,2}$ we can identify the co-derivations with the multilinear
products using $\boldsymbol{c}^1_{1,1}\pi_{1,1}=c^1_{1,1}$ resulting in
\begin{equation}
    \begin{split}
        (\bDelta_2\tp_1'\id_1)\bDelta_1\boldsymbol{c}^1_{1,1}\pi_{2,2}&=\qty[\qty{c^1_{1,1}\pi_{1,1}\tp_2'\pi_{0,1}+\pi_{0,1}\tp_2'c^1_{1,1}\pi_{1,1}}\tp_1'\pi_{1,0}](\bDelta_2\tp_1'\id_1)\bDelta_1\\
        &+\qty[\pi_{1,0}\tp_1'\qty{c^1_{1,1}\pi_{1,1}\tp_2'\pi_{0,1}+\pi_{0,1}\tp_2'c^1_{1,1}\pi_{1,1}}](\id_1\tp_1'\bDelta_2)\bDelta_1.
    \end{split}
\end{equation}
By now applying $\qty(\id_1\tp_1'\bnabla_2)$ we recover
\begin{equation}
    \begin{split}
        \bDelta_1\boldsymbol{c}^1_{1,1}\pi_{2,2}&=\qty[\qty{c^1_{1,1}\pi_{1,1}\tp_2\pi_{0,1}+\pi_{0,1}\tp_2c^1_{1,1}\pi_{1,1}}\tp_1'\pi_{1,0}]\bDelta_1\\
        &+\qty[\pi_{1,0}\tp_1'\qty{c^1_{1,1}\pi_{1,1}\tp_2\pi_{0,1}+\pi_{0,1}\tp_2c^1_{1,1}\pi_{1,1}}]\bDelta_1.
    \end{split}
\end{equation}
Lastly if we apply $\bnabla_1$ we recover
\begin{equation}
    \begin{split}
        \boldsymbol{c}^1_{1,1}\pi_{2,2}&=\qty[\qty{c^1_{1,1}\pi_{1,1}\tp_2\pi_{0,1}+\pi_{0,1}\tp_2c^1_{1,1}\pi_{1,1}}\tp_1\pi_{1,0}]\\
        &+\qty[\pi_{1,0}\tp_1\qty{c^1_{1,1}\pi_{1,1}\tp_2\pi_{0,1}+\pi_{0,1}\tp_2c^1_{1,1}\pi_{1,1}}],
    \end{split}
\end{equation}
which can be compactly written as
\begin{align}
    \boldsymbol{c}^1_{1,1}\pi_{2,2}=\sum_{i,j=0}^{2}\bar{\id}^{i,j}\tpb c^1_{1,1}(\bar{\id}^{1,1})\tpb \bar{\id}^{1-i,1-j},
\end{align}
which is exactly \eqref{N-uplift} for $n_1=n_2=2$. By induction it is possible to extend the proof and prove that \eqref{N-uplift} is the only consistent way to extend a multilinear product to a co-derivation.   
    
    %%%%%%%%%%%%%%%%%%%%%%%%%

 %   \endgroup
    
%    \bibliography{bibliography.bib}  
%    \bibliographystyle{utphys}

\providecommand{\href}[2]{#2}\begingroup\raggedright\endgroup

%    \printbibliography
\end{document}